\documentclass[12pt,twoside]{book}
\usepackage{a4,amssymb,epsfig,fancyheadings}
\advance \headheight by 3.0truept

\catcode`@=11

\def\@citex[#1]#2{\if@filesw\immediate\write\@auxout{\string\citation{#2}}\fi
  \def\@citea{}\@cite{\@for\@citeb:=#2\do
    {\@citea\def\@citea{,\penalty\@m}\@ifundefined
      {b@\@citeb}{{\bf ?}\@warning
       {Citation `\@citeb' on page \thepage \space undefined}}%
\hbox{\csname b@\@citeb\endcsname}}}{#1}}

\def\citer{\@ifnextchar [{\@tempswatrue\@citexr}{\@tempswafalse\@citexr[]}}

\def\@citexr[#1]#2{\if@filesw\immediate\write\@auxout{\string\citation{#2}}\fi
  \def\@citea{}\@cite{\@for\@citeb:=#2\do
    {\@citea\def\@citea{--\penalty\@m}\@ifundefined
       {b@\@citeb}{{\bf ?}\@warning
       {Citation `\@citeb' on page \thepage \space undefined}}%
\hbox{\csname b@\@citeb\endcsname}}}{#1}}

\begin{document}

% ===== title page ========================================
\pagestyle{empty}
\renewcommand{\baselinestretch}{1.3}

\begin{flushright}
{\ttfamily http://tumb1.biblio.tu-muenchen.de/publ/diss/ph/2002/bosch.html}\\
MPI-PHT-2002-35\\
June 2002
\end{flushright}
\vspace*{-1.5cm}
\begin{center}
\epsfig{figure=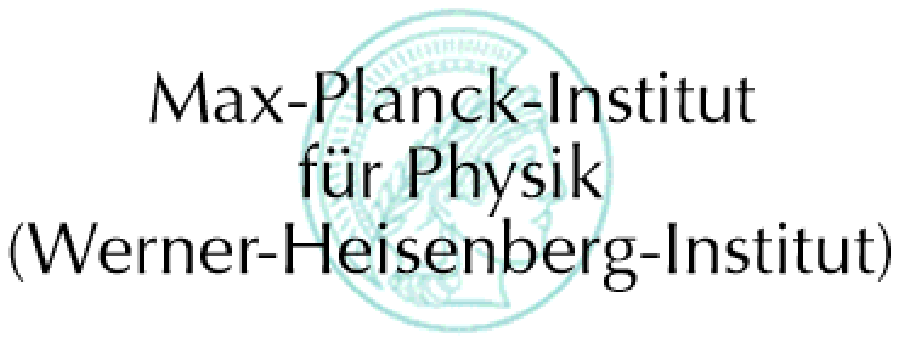}

\vfill
{\huge\bf Exclusive Radiative Decays of \\$B$ Mesons in QCD Factorization}

\vfill
{\Large\sc Stefan W. Bosch}\\[0.3cm]

Max-Planck-Institut f\"ur Physik\\
(Werner-Heisenberg-Institut)\\
F\"ohringer Ring 6\\
D-80805 Munich, Germany\\
Email: bosch@mppmu.mpg.de
\vfill

\parbox{3.2cm}{\begin{flushleft}\epsfig{figure=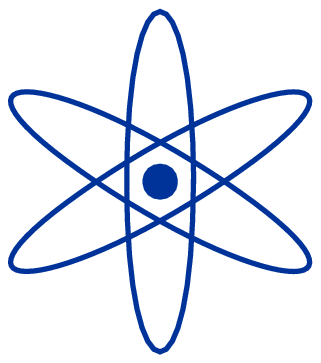}\end{flushleft}}\hfill\parbox{3.2cm}{\begin{flushright}\epsfig{figure=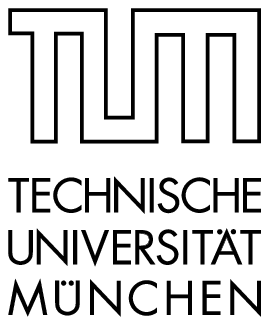}\end{flushright}}

\end{center}

\cleardoublepage
\markboth{}{}

\thispagestyle{empty}
\begin{center}

\begin{center}{\large\bf Physik-Department}\\ {\bf Technische Universit\"at M\"un\-chen}\\Institut f\"ur Theoretische Physik\\Lehrstuhl Univ.-Prof. A.J. Buras\end{center}

\vspace{4cm}
{\huge\bf Exclusive Radiative Decays of \\$B$ Mesons in QCD Factorization}

\vfill
{\Large\sc Stefan W. Bosch}

\vfill
\end{center}
Vollst\"andiger Abdruck der von der Fakult\"at f\"ur Physik der Technischen Universit\"at M\"unchen zur Erlangung des akademischen Grades eines
\begin{center}
\bf Doktors der Naturwissenschaften (Dr. rer. nat.)
\end{center}
genehmigten Dissertation.

\begin{center}
\begin{tabular}{lll}
Vorsitzender:              &    & Univ.-Prof. Dr. Stephan Paul\\[0.3cm]
Pr\"ufer der Dissertation: & 1. & Univ.-Prof. Dr. Andrzej J. Buras\\[0.1cm]
		  	   & 2. & Univ.-Prof. Dr. Gerhard Buchalla,\\
                               && Ludwig-Maximilians-Universit\"at M\"unchen\\
\end{tabular}
\end{center}

\noindent Die Dissertation wurde am 10. Juni 2002 bei der Technischen Universit\"at M\"unchen eingereicht und durch die Fakult\"at f\"ur Physik am 22. Juli 2002 angenommen.

% ===== abstract ===================== ====================
\frontmatter
\renewcommand{\thepage}{\roman{page}}
\pagestyle{plain}
\noindent
{\Large\bf Abstract}\bigskip

\noindent
We discuss exclusive radiative decays in QCD factorization within the Standard Model. In particular, we consider the decays $B\to V\gamma$, with a vector meson $K^*$ or $\rho$ in the final state, and the double radiative modes $B_s^0\to\gamma\gamma$ and $B_d^0\to\gamma\gamma$. At quark level, all these decays are governed by the flavour-changing neutral-current $b\to s\gamma$ or $b\to d\gamma$ transitions, which appear at the one-loop level in the Standard Model. Such processes allow us to study CP violation and the interplay of strong and electroweak interactions, to determine parameters of the CKM matrix, and to search for New Physics. The exclusive decays are experimentally better accessible, but pose more problems for the theoretical analysis. The heavy-quark limit $m_b\gg\Lambda_{QCD}$, however, allows to systematically separate perturbatively calculable hard scattering kernels from nonperturbative form factors and universal light-cone distribution amplitudes.\medskip

\noindent
The main results of this work are the following:
\begin{itemize}
  \item We apply QCD factorization methods based on the heavy-quark limit to the hadronic matrix elements of the exclusive radiative decays $B\to V\gamma$ and $B\to\gamma\gamma$. A power counting in $\Lambda_{QCD}/m_b$ implies a hierarchy among the possible transition mechanisms and allows to identify leading and subleading contributions. In particular, effects from quark loops are calculable in terms of perturbative hard-scattering functions and universal meson light-cone distribution amplitudes rather than being generic, uncalculable long-distance contributions. Our approach is model independent and similar in spirit to the treatment of hadronic matrix elements in two-body non-leptonic $B$ decays formulated by Beneke, Buchalla, Neubert, and Sachrajda.
  \item For the $B\to V\gamma$ decays we evaluate the leading $\Lambda_{QCD}/m_b$ contributions complete to next-to-leading order in QCD. We adopt existing results for the hard-vertex corrections and calculate in addition hard-spectator corrections, including also QCD penguin operators.
  \item Weak annihilation topologies in $B\to V\gamma$ are shown to be power suppressed. We prove to one-loop order that they are nevertheless calculable within QCD factorization. Because they are numerically enhanced we include the ${\cal O}(\alpha_s^0)$ annihilation contributions of current-current and QCD-penguin operators in our analysis.
  \item The double radiative $B\to\gamma\gamma$ decays are analyzed with leading-logarithmic accuracy. In the heavy quark limit the dominant contribution at leading power comes from a single diagram. The contributions from one-particle irreducible diagrams are power suppressed but still calculable within QCD factorization. We use these corrections, including QCD penguins, to estimate CP asymmetries in $B\to\gamma\gamma$ and so-called long-distance contributions in $B$ and $D\to\gamma\gamma$.
  \item We predict branching ratios, CP and isospin asymmetries, and estimate U-spin breaking effects for $B\to K^*\gamma$ and $B\to\rho\gamma$. For the $B\to\gamma\gamma$ decays we give numerical results for branching ratios and CP asymmetries. Varying the individual input parameters we estimate the error of our predictions. The dominant uncertainty comes from the poorly known nonperturbative input parameters.
\end{itemize}

% ===== table of contents =================================
\renewcommand{\baselinestretch}{1.1}
\tableofcontents

% =========================================================
% =     main part                                         =
% =========================================================

\mainmatter
\renewcommand{\baselinestretch}{1.1}
\renewcommand{\thepage}{\arabic{page}}
\pagestyle{fancyplain}
\renewcommand{\chaptermark}[1]{\markboth{\thechapter.  #1}{}}
\renewcommand{\sectionmark}[1]{\markright{\thesection\  #1}}
\rhead[\sl \leftmark]{\fancyplain{} {\bf \thepage}}
\lhead[\bf \thepage]{\sl \rightmark}
\cfoot{}

% ===== introduction ======================================
\chapter{Introduction}
\label{ch:intro}
Elementary particle physics represents man's effort to answer the basic question: ``What is the World Made of?'' In search of the fundamental building blocks of matter physicists penetrated to smaller and smaller constituents, which later turned out to be divisible. The primary matter of Anaximenes of Miletus, the periodic table of Mendeleev, Rutherford's alpha scattering experiments, the detection of the neutron by Chadwick and that of the positron by Anderson, the discovery of nuclear fission by Hahn and Meitner, or Reines finding the neutrino were just some of the milestones on this way. What underlies our current theoretical understanding of nature is quantum field theory in combination with a gauge principle. Electromagnetism, weak and strong nuclear forces, and their interaction with quarks and leptons are described by the {\em Standard Model} of particle physics \cite{GSWSM,QCD}. Combined with general relativity, this theory is so far consistent with virtually all physics down to the scales probed by particle accelerators, roughly $10^{-16}\,$cm, and also passes a variety of indirect tests that probe even shorter distances.

In spite of its impressive successes, the Standard Model is believed to be not complete. For a really final theory it is too arbitrary, especially considering the large number of sometimes even ``unnatural'' parameters in the Lagrangian. Examples for such parameters, that are largely different from what one naively expects them  to be, are the weak scale compared with the Planck scale or the small value of the strong CP-violation parameter $\theta_\mathrm{QCD}$. Questions like: ``Why are there three particle generations?'', ``Why is the gauge structure with the assignment of charges as it is?'', or ``What is the origin of the mass spectrum?'' demand an answer by a really fundamental theory, but the Standard Model gives no replies. Furthermore, the union of gravity with quantum theory yields a nonrenormalizable quantum field theory, indicating that New Physics should show up at very high energies.

The ideas of grand unification, extra dimensions, or supersymmetry were put forward to find a more complete theory. But applying these ideas has not yet led to theories that are substantially simpler or less arbitrary than the Standard Model. To date, string theory \cite{string}, the relativistic quantum theory of one-dimensional objects, is a promising, and so far the only, candidate for such a ``Theory of Everything''.\medskip

Within this thesis we work exclusively in the Standard Model, where particularly in the flavour sector beside the aforementioned problems some unsolved issues remain. Among them are the mechanism of electroweak symmetry breaking and the hierarchy problem that comes along with it, the generation of fermion masses, quark mixing, or the violation of the discrete symmetries C, P, CP, and T. Especially CP violation is of particular interest as it is one of the three vital ingredients to generate a cosmological matter-antimatter asymmetry \cite{AS}. In 1964, James Cronin, Val Fitch, and collaborators discovered that the decays of neutral kaons do not respect CP symmetry \cite{CCFT}. Only recently CP violation was established undoubtfully also in the neutral $B$-meson system \cite{BaBeCP}.

We will deal with the bottom quark system, which is an ideal laboratory for studying flavour physics. The history of $B$ physics started 1977 with the ``observation of a dimuon resonance at $9.5\,\mathrm{GeV}$ in $400\,\mathrm{GeV}$ proton-nucleon collisions'' at Fermilab \cite{ups}. It was baptized the ``$\Upsilon$ resonance'' and its quark content is $\bar b b$. With the start of {\sc BaBar} \cite{BaBar} and Belle \cite{KEK-B} in 1999, dedicated $B$ factories add a wealth of data to the results of CLEO \cite{CLEO-B} and the CERN \cite{CERN-B} and Fermilab \cite{Fermilab-B} experiments. The upcoming $B$-physics experiments at the Tevatron Run II \cite{BTeV} and LHC \cite{LHCb} will bring us ever closer to the main goal of $B$ physics, which is a precision study of the flavour sector with its phenomenon of CP violation to pass the buck of being the experimentally least constrained part of the Standard Model. This is not only to pin down the parameters of the Standard Model, but in particular to reveal New Physics effects via deviations of measured observables from the Standard Model expectation. Such an indirect search for New Physics is complementary to the direct search at particle accelerators. It invites both experimenters and theoreticians to work with precision. We need accurate and reliable measurements and calculations. The calculational challenge we will meet for this thesis are exclusive radiative decays of $B$ mesons.\medskip

But why investigate {\em $B$ meson} decays? Due to confinement quarks appear in nature not separately, but have to be bound into colourless hadrons. Considering constituent quarks only, the simplest possible of such objects consists of a quark and an antiquark only and is called a meson. The bound states with a $b$ quark and a $\bar d$ or $\bar u$ antiquark are referred to as the $\bar B^0$ and $B^-$ meson, respectively. Those $B$ mesons containing an $s$ or $c$ quark are denoted $B_s$ and $B_c$, respectively. So ``meson'' decays because mesons are the simplest hadrons. But why of all mesons ``$B$'' mesons? Apart from the $\Upsilon$ resonances, the $B$ mesons are the heaviest mesons, because the top quark decays before it can hadronize. The fact that $B$ mesons are heavy has two weighty consequences: $B$ decays show an extremely rich phenomenology and theoretical techniques using an expansion in the heavy mass allow for model-independent predictions. The rich phenomenology is based on the one hand on the large available phase space allowing for a plethora of possible final states and on the other hand on the possibility for large CP-violating asymmetries in $B$ decays. The latter feature is in contrast to the Standard Model expectations for decays of $K$ and $D$ mesons. In $D$ decays only the comparably light $d$, $s$, and $b$ quarks can enter internal loops which leads to a strong GIM suppression of CP-violating phenomena. CP violation in $K\to\pi\pi$ is small due to flavour suppression and not because the CP violating phase itself is small. Actually, the $\sin 2\beta$ measurement in $B\to J/\psi K_S$ showed that the CP-violating phase is large. Furthermore, especially the $B\to J/\psi K_S$ decay mode is theoretically extremely clean as opposed to the large theoretical uncertainties in the kaon system. The pattern of CP violation in the $K$ and $B$ system just represents the hierarchy of the CKM matrix. The $B$ meson system offers an excellent laboratory to quantitatively test the CP-violating sector of the Standard Model, determine fundamental parameters, study the interplay of strong and electroweak interactions, or search for New Physics.

We will concentrate on a subgroup of $B$ decays: exclusive radiative $B$ decays, i.e. we are interested in the exclusive decay products of a $B$ meson containing at least one photon. The quark level decay is a flavour-changing-neutral-current process $b\to s\gamma$ or $b\to d\gamma$. Such decays are rare, i.e. they come with small exclusive branching ratios of ${\cal O}(10^{-5})$, because they arise only at the loop level in the Standard Model. Thereby they test the detailed structure of the theory at the level of radiative corrections and provide information on the masses and couplings of the virtual SM or beyond-the-SM particles participating. Among the rare $B$ decays the $b\to s\gamma$ modes are the most prominent ones because they are already experimentally measured.

Primarily we are interested in the underlying decay of the heavy quark, which is governed by the weak interaction. But it is the strong force that is responsible for the formation of the hadrons that are observed in the detectors. If we want to do our experimental colleagues a favour, we let them measure the easier accessible exclusive decays, i.e. those where all decay products are detected. Then, however, we impose the burden of a more difficult theoretical treatment on ourselves. The ``easier'' option for theorists is to consider the inclusive decay, where e.g. for $B\to X_s\gamma$ all final states with a photon and strangeness content $-1$ have to be summed over. But quark-hadron duality allows us to consider instead of all the $B$ decays only the parton-level $b\to s\gamma$ decay, which amounts to a fully perturbative calculation. Much effort was put into the inclusive mode to achieve a full calculation at next-to-leading order in renormalization-group-improved perturbation theory. Yet, for the exclusive decays we have to dress the $b$ quark with the ``brown muck'', the light quark and gluon degrees of freedom inside the $B$ meson, and have to keep struggling with hadronization effects. Despite the more complicated theoretical situation of the exclusive channels it is worthwhile to better understand them. Especially in the difficult environment of hadron machines, like the Fermilab Tevatron or the LHC at CERN, they are easier to investigate experimentally. In any case the systematic uncertainties, both experimental and theoretical, are very different for inclusive and exclusive modes. A careful study of the exclusive modes can therefore yield valuable complementary information in testing the Standard Model.

 The field-theoretical tool kit at our disposal for this analysis includes operator product expansion and renormalization group equations in the framework of an effective theory. Herewith the problem of calculating transition amplitudes can be separated into the perturbatively calculable short distance Wilson coefficients and the long distance operator matrix elements. In principle, the latter ones have to be calculated by means of a nonperturbative method like lattice QCD or QCD sum rules. For the exclusive decays of $B$ mesons, however, one can use additionally the fact that the $b$-quark mass $m_b$ is large compared to $\Lambda_{QCD}$, the typical scale of QCD. This in turn allows one to establish factorization formulas for the evaluation of the relevant hadronic matrix elements of local operators in the weak Hamiltonian. Herewith a further separation of long-distance contributions to the process from a perturbatively calculable short-distance part, that depends only on the large scale $m_b$, is achieved. The long-distance contributions have to be computed non-perturbatively or determined from experiment. However, they are much simpler in structure than the original matrix element. This QCD factorization technique was developed by Beneke, Buchalla, Neubert, and Sachrajda for the non-leptonic two-body decays of $B$ mesons \cite{BBNS}. We apply similar arguments based on the heavy-quark limit $m_b\gg \Lambda_\mathrm{QCD}$ to the decays $B\to K^*\gamma$, $B\to\rho\gamma$, and $B\to\gamma\gamma$. This allows us to separate perturbatively calculable contributions from the nonperturbative form factors and universal meson light-cone distribution amplitudes in a systematic way. With power counting in $\Lambda_\mathrm{QCD}/m_b$ we can identify leading and subleading contributions.\medskip

We have organized the subsequent 105 pages as follows: In the first part we fill our toolbox with the necessary ingredients. After a mini review of the Standard Model we present the basic equipment: operator product expansion, effective theories, and renormalization group improved perturbation theory. A discussion of the effective $b\to s\gamma$ Hamiltonian and a short survey of its theoretical status quo concludes this chapter. Our best tool so far to treat the tough nut of exclusive $B$ decays is QCD factorization. We devote Chapter \ref{ch:fact} to the description of this useful technique. In order to be able to appreciate its merits we first present its predecessors ``naive factorization'' and generalizations. We then discuss rather in detail the QCD factorization approach, give a sample application to the calculation of the pion form factor, and mention limitations of QCD factorization. Finally, we comment on other attempts to treat hadronic matrix elements in exclusive non-leptonic $B$ decays.

Part II and III contain the main subject of this work: the treatment of $B\to V\gamma$ ($V=K^*$ or $\rho$) and $B\to\gamma\gamma$ in QCD factorization. For both types of decays we first present the necessary formulas and basic expressions and then give numerical results and phenomenological applications.

We derive for $B\to V\gamma$ the decay amplitude complete at next-to-leading order in QCD and leading power in $\Lambda_\mathrm{QCD}/m_b$. Hard-vertex and hard-spectator corrections are discussed separately. Annihilation contributions turn out to be power suppressed, but nevertheless calculable. As they are factorizable and numerically important we include them for our phenomenological analysis. Doing so we become sensitive to the charge of the light spectator quark inside the $B$ meson and can estimate isospin breaking effects. These turn out to be large. The most important phenomenological quantity we predict is the branching ratio. The NLL value is considerably larger than both the leading logarithmic prediction when the same form factor is used and the experimental value. Our calculation also allows us to estimate CP-asymmetries and U-spin breaking effects. We want to stress that this thesis is the first totally complete next-to-leading-logarithmic presentation of exclusive $B\to V\gamma$ decays, and it is a model-indpendent one.

The double radiative $B\to\gamma\gamma$ decays are analyzed with leading logarithmic accuracy in QCD factorization based on the heavy-quark limit $m_b\gg\Lambda_\mathrm{QCD}$. The dominant effect arises from the one-particle-irreducible magnetic-moment type transition $b\to s(d)\gamma$ where an additional photon is emitted from the light quark. The contributions from one-particle irreducible diagrams are power suppressed but still calculable within QCD factorization. They are used to compute the CP-asymmetry in $B\to\gamma\gamma$ and to estimate so-called long-distance contributions in $B$ and $D\to\gamma\gamma$. Numerical results are presented for branching ratios and CP asymmetries.

We give our conclusions and an outlook in chapter \ref{ch:conclusions}.

Some more technical details are discussed in the Appendices. We give the explicit formulas for the Wilson coefficients, transformation relations between the two operator bases for $b\to s\gamma$, and a one-loop proof that weak annihilation contributions to $B\to V\gamma$ are calculable within QCD factorization.

% =========================================================
% =     part I: Fundamentals                              =
% =========================================================
\part{Fundamentals}

% ===== Basic Concepts ====================================
\chapter{The Basic Concepts}
\label{ch:SM}

In this chapter we briefly introduce the basic concepts needed for doing calculations in elementary particle physics. We present the Standard Model in a nutshell and introduce the concepts of renormalization, renormalization group, operator product expansion, and effective theories. We assume that the reader is familiar with quantum field and gauge theories and refer to the pertinent textbooks \cite{qft}.

% =========================================
% =      The Standard Model               =
% =========================================

\section{The Standard Model}
\label{sec:SM}

As mentioned already in the introduction, the Standard Model (SM) is a comprehensive theory of particle interactions. Its success in giving a complete and correct description of all non-gravitational physics tested so far is unprecedented. In the following we give a short introduction into this beautiful theory. We want to introduce the ``Old Standard Model,'' i.e. the one where neutrinos are massless. The recent evidence for neutrino masses, coming from the observation of neutrino oscillations \cite{SK}, has no direct consequences for our work.

The Standard Model is made up of the Glashow-Salam-Weinberg Model \cite{GSWSM} of electroweak interaction and Quantum Chromodynamics (QCD) \cite{QCD}. It is based on the principle of gauge symmetry. The Lagrangian of a gauge theory is invariant under local ``gauge'' transformations of a symmetry group. Such a symmetry can be used to generate dynamics - the {\em gauge interactions.} The prototype gauge theory is quantum electrodynamics (QED) with its Abelian $U(1)$ local symmetry. It is believed that all fundamental interactions are described by some form of gauge theory.

The gauge group of strong interactions is the non-Abelian group $SU(3)_C$ which has eight generators. These correspond to the gluons that communicate the strong force between objects carrying colour charge - therefore the ``C'' as subscript. Since the gluons themselves are coloured, they can directly interact with each other, which leads to the phenomena of {\em ``asymptotic freedom''} and {\em ``confinement.''} At short distances, the coupling constant $\alpha_s$ becomes small. This allows us to compute colour interactions using perturbative techniques and turns QCD into a quantitative calculational scheme. For long distances on the other hand, the coupling gets large, which causes the quarks to be ``confined'' into colourless hadrons. In the words of Yuri Dokshitzer: ``QCD, the marvellous theory of the strong interactions, has a split personality. It embodies `hard' and `soft' physics, both being hard subjects, the softer ones being the hardest.'' \cite{YD}

Electroweak interaction is based on the gauge group $SU(2)_L \otimes U(1)_Y$ - ``L'' stands for ``left'' and ``Y'' denotes the hypercharge - which is spontaneously broken to $U(1)_{QED}$. This is achieved through the non vanishing vacuum expectation value of a scalar isospin doublet {\em Higgs} field \cite{Higgs}
\begin{equation}
\label{higgs}
  \phi=\left(\begin{array}{c}\phi^+\\ \phi^0\end{array}\right)
\end{equation}
Three of the four scalar degrees of freedom of the Higgs field give masses to the $W$ and $Z$ bosons. The remaining manifests itself in a massive neutral spin zero boson, the physical Higgs boson. It is the only particle of the Standard Model which lacks direct experimental detection. The current lower limit on its mass is 114.1 GeV at the $95\%$ confidence level \cite{mH}. From electroweak precision data there is much evidence for a light Higgs. But as soon as such a light Higgs is found, this gives birth to the {\em hierarchy problem.} A scalar (Higgs) mass is not protected by gauge or chiral symmetries so we expect $m_H \approx \Lambda \approx 10^{16}\,\mbox{GeV}$ if we do not want to fine-tune the bare Higgs mass against the mass aquired from quantum effects. Why should $m_H$ be much smaller than $\Lambda$?

Fermions are the building blocks of matter. In the SM they appear in three generations which differ only in their masses. The two species of fundamental fermions are leptons and quarks. With regard to the gauge group $SU(2)_L$ the quarks and leptons can be classed in left-handed doublets and right-handed singlets.
\begin{displaymath}
  \begin{array}{p{2.3cm}ccccc}
    \mbox{Quarks:}&
    \left(\begin{array}{c} u\\ d\:\!' \end{array} \right)_{\!\!\!L} & \qquad \qquad &
    \left(\begin{array}{c} c\\ s\:\!' \end{array} \right)_{\!\!\!L} & \qquad \qquad &
    \left(\begin{array}{c} t\\ b\:\!' \end{array} \right)_{\!\!\!L} \\
     & u_R & \qquad \qquad & c_R & \qquad \qquad & t_R \\
     & d_R & \qquad \qquad & s_R & \qquad \qquad & b_R \\
    \\
    \mbox{Leptons:}&
    \left(\begin{array}{c} \nu_e\\ e^- \end{array} \right)_{\!\!\!L}
    & \qquad \qquad &
    \left(\begin{array}{c} \nu_\mu\\ \mu^- \end{array}
    \right)_{\!\!\!L} & \qquad \qquad &
    \left(\begin{array}{c} \nu_\tau\\ \tau^- \end{array}
    \right)_{\!\!\!L} \\
     & e_R & \qquad \qquad & \mu_R & \qquad \qquad & \tau_R
   \end{array}
\end{displaymath}
Within the Old Standard Model there are no right-handed neutrinos. The quarks carry colour charge and transform as $SU(3)_C$ triplets whereas the colourless leptons are $SU(3)_C$ singlets.

Fermion masses are generated via a Yukawa interaction $\overline \psi(x)\phi(x)\psi(x)$ with the Higgs field (\ref{higgs}). Using global unitary transformations in flavour space, the Yukawa interaction can be diagonalized to obtain the physical mass eigenstates
\begin{equation}
\label{defCKM}
  \left( \begin{array}{c} d\:\!' \\ s\:\!' \\ b\:\!' \end{array} \right) =
  \underbrace{\left( \begin{array}{ccc} V_{ud} & V_{us} & V_{ub} \\
      V_{cd} & V_{cs} & V_{cb} \\
      V_{td} & V_{ts} & V_{tb} \end{array} \right)}_{\displaystyle \equiv V_\mathrm{CKM}} \cdot
  \left( \begin{array}{c} d \\ s \\ b \end{array} \right)
\end{equation}
The non-diagonal elements of the Cabibbo-Kobayashi-Maskawa matrix \cite{CKM} allow for transitions between the quark generations in the charged quark current
\begin{equation}
\label{cc}
  J^+_\mu = \left( \overline{u}, \overline{c}, \overline{t} \right)_L
  \gamma_\mu V_\mathrm{CKM} \left( \begin{array}{c}
      d\\s\\b\end{array}\right)_{\!\!\!L}
\end{equation}
Unitarity of $V_\mathrm{CKM}$ guaranties the absence of flavour-changing-neutral-current (FCNC) processes at tree level. This Glashow-Iliopoulos-Maiani (GIM) mechanism \cite{GIM} would forbid FCNC transitions even beyond the tree level if we had exact horizontal flavour symmetry which assures the equality of quark masses of a given charge. Such a symmetry is in nature obviously broken by the different quark masses so that at the one-loop level effective $b\to s$, $b\to d$, or $s\to d$ processes like $B\to X_{s}\gamma$ or $K\to\pi\nu\bar\nu$ can appear.

\begin{figure}
  \begin{center}
    \psfig{figure=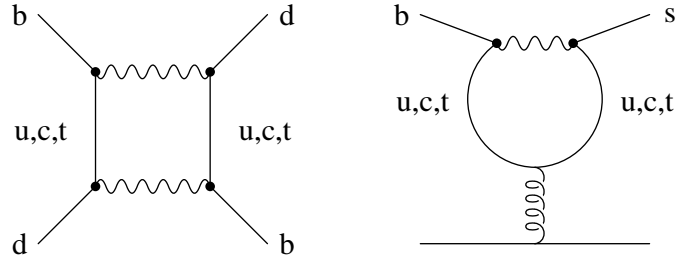}
  \end{center}
  \caption{Example of a box and penguin diagram in the full theory. CP violating effects are possible because quarks of all three generations are involved in the processes.\label{fig:boxpen}}
\end{figure}

A unitary complex $N\times N$ matrix can be described by $N^2$ real parameters. If this matrix mixes $N$ families each with two quarks one can remove $2N-1$ phases through a redefinition of the quark states, which leaves the Lagrangian invariant. Because an orthogonal $N\times N$ matrix has $N(N-1)/2$ real parameters (angles) we are left with $N^2-(2N-1)-N(N-1)/2=(N-1)(N-2)/2$ independent physical phases in the quark mixing matrix. Therefore, it is real if it mixes two generations only. The three-generation CKM matrix, however, has to be described by three angles and one complex phase. The latter one is the only source of CP violation within the Standard Model if we desist from the possibility that $\theta_\mathrm{QCD}\neq 0$. But these CP-violating effects can show up only if really all three generations of the Standard Model are involved in the process. Typically this is the case if one considers loop contributions of weak interaction, like box or penguin diagrams as in Fig.~\ref{fig:boxpen}.
%%%%%%%%%%%%%%%%%%%%%%%%%%%%%%%%%%%%%%%%%%%%%%%%%%%%%
%%%%%%%%%%%%%%%%%%%%%%%%%%%%%%%%%%%%%%%%%%%%%%%%%%%%%

In principle there are many different ways of parametrizing the CKM matrix. For practical purposes most useful is the so called Wolfenstein parametrization \cite{Wolfen}
\begin{equation}
\label{wolfen}
V_\mathrm{CKM}=\left(
  \begin{array}{ccc} 
    1-\frac{\lambda^2}{2} & \lambda & A\lambda^3 (\rho-i\eta) \\
    -\lambda & 1-\frac{\lambda^2}{2} & A\lambda^2 \\
    A \lambda^3 (1-\rho-i\eta) & -A\lambda^2 & 1
  \end{array}\right)
\end{equation}
which is an expansion to ${\cal O}(\lambda^3)$ in the small parameter $\lambda=|V_{us}|\approx 0.22$. It is possible to improve the Wolfenstein parametrization to include higher orders of $\lambda$ \cite{impwolf}. In essence $\rho$ and $\eta$ are replaced with $\bar\rho=\rho(1-\lambda^2/2)$ and $\bar\eta=\eta(1-\lambda^2/2)$, respectively.

\begin{figure}
  \begin{center}
    \input{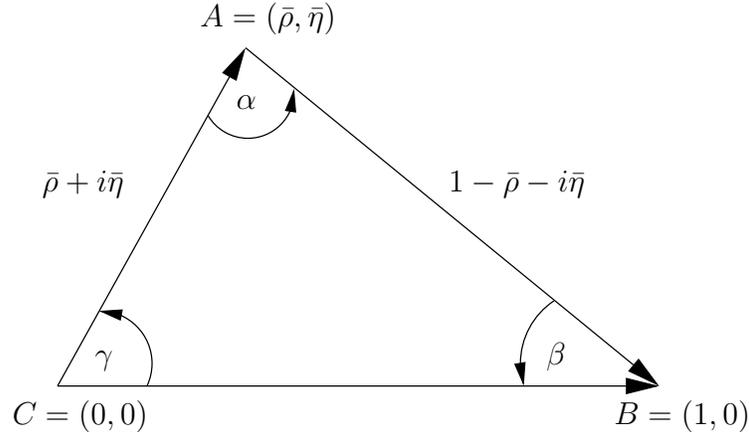}
  \end{center}
  \caption{Unitarity Triangle\label{fig:UT}}
\end{figure}

\begin{figure}
   \centerline{\epsffile{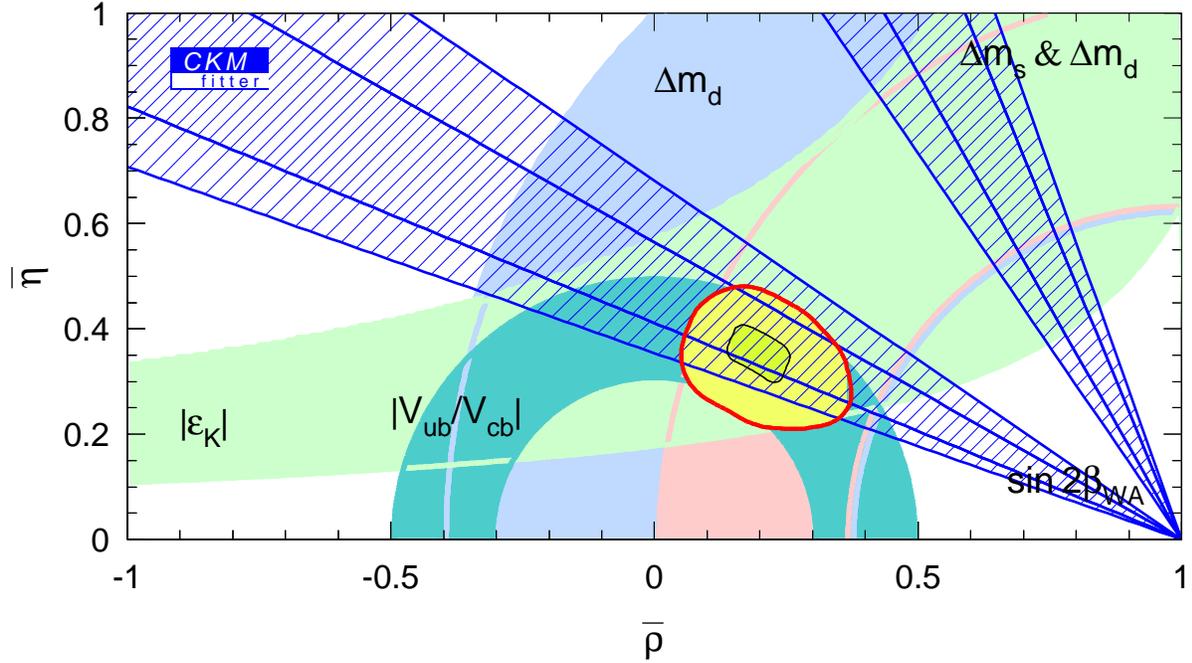}}
\caption{The Range-fit result of the global CKM fit from CKMfitter (status May 2002) \cite{CKMfitter}.\label{fig:ckmfit}}
\end{figure}

For phenomenological studies of CP-violating effects, the so called standard unitarity triangle (UT) plays a special role. It is a graphical representation of one of the six unitarity relations, namely
\begin{equation}
\label{ut}
V_{ud}V_{ub}^*+V_{cd}V_{cb}^*+V_{td}V_{tb}^*=0
\end{equation}
in the complex $(\rho,\eta)$ plane. This unitarity relation involves simultaneously the elements $V_{ub}$, $V_{cb}$, and $V_{td}$ which are under extensive discussion at present. The area of this and all other unitarity triangles equals half the absolute value of $J_{CP}=\mathrm{Im}(V_{us}^{}V_{cb}^{}V_{ub}^*V_{cs}^*)$, the Jarlskog measure of CP violation \cite{JCP}. Usually, one chooses a phase convention where $V_{cd}^{}V_{cb}^*$ is real and rescales the above equation with $|V_{cd}^{}V_{cb}^*|=A\lambda^3$. This leads to the triangle in Figure \ref{fig:UT} with a base of unit length and the apex $(\bar\rho,\bar\eta)$.
%%%%%%%%%%%%%%%%%%%%%%%%%%%%%%%%%%%%%%%%%%%%%%%%%%%%%
%%%%%%%%%%%%%%%%%%%%%%%%%%%%%%%%%%%%%%%%%%%%%%%%%%%%%
A phase transformation in (\ref{ut}) only rotates the triangle, but leaves its form unchanged. Therefore, the angles and sides of the unitarity triangle are physical observables and can be measured. Much effort was and is put into the determination of the UT parameters. One tries to measure as many parameters as possible. The consistency of the various measurements tests the consequences of unitarity in the three generation Standard Model. Any discrepancy with the SM expectations would imply the presence of new channels or particles contributing to the decay under consideration. So far, all experimental results are consistent with the Standard Model picture \cite{CKMfitter,UTfit}. The state-of-the-art ``frequentists'' result for the unitarity triangle from the 2002 Winter conferences is displayed in Fig.~\ref{fig:ckmfit} \cite{CKMfitter}.
%%%%%%%%%%%%%%%%%%%%%%%%%%%%%%%%%%%%%%%%%%%%%%%%%%%%%%%%%%%%%%%%%%%
%%%%%%%%%%%%%%%%%%%%%%%%%%%%%%%%%%%%%%%%%%%%%%%%%%%%%%%%%%%%%%%%%%%
Actually the good agreement of measurements with the Kobayashi-Maskawa mechanism gives rise to some theoretical puzzles: the KM mechanism for example does explain neither the cosmic baryon asymmetry nor the smallness of $\theta_\mathrm{QCD}$ and basically all extensions of the Standard Model introduce a large number of new CP-violating phases.

% =========================================
% =      Renormalization                  =
% =========================================

\section{Renormalization and Renormalization Group}
\label{sec:ren}

Given the Lagrangian of a theory one can deduce the Feynman rules by means of which amplitudes of the processes occuring in this theory can be calculated in perturbation theory. In Feynman diagrams with internal loops, however, one often encounters ultraviolet divergences. This is because the momentum variable of the virtual particle in the loop integration ranges from zero to infinity. The theory of renormalization is a prescription which allows us to consistently isolate and remove all these infinities from the physically measurable quantities. A two-step procedure is necessary.

First, one {\em regulates} the theory. That is, one modifies it in a way that observable quantities are finite and well defined to all orders in perturbation theory. We are then free to manipulate formally these quantities, which are divergent only when the regularization is removed. The most straightforward way to make the integrals finite is to introduce a momentum cutoff. But this violates for example Lorentz invariance or the Ward identities. A regularization method that preserves all symmetries of a gauge theory is {\em dimensional regularization} \cite{HV,dimreg}. The basic idea is to compute the Feynman diagram as an analytic function of the dimensionality of space time $D=4-2\epsilon$. For sufficiently small $D$, any loop-momentum integral will converge. The singularities are extracted as poles for $\epsilon\to 0$.

Potential problems of dimensional regularization concern the treatment of $\gamma_5$ in $D\neq 4$ dimensions. The definition
\begin{equation}
\label{gamma5}
\gamma_5=\frac{i}{4!}\varepsilon_{\kappa\lambda\mu\nu}\gamma^\kappa \gamma^\lambda \gamma^\mu \gamma^\nu
\end{equation}
with $\epsilon_{\kappa\lambda\mu\nu}$ the completely antisymmetric tensor in four dimensions, cannot straightforwardly be translated to $D\neq 4$ dimensions. In the so called ``naive dimensional regularization'' (NDR) scheme \cite{NDR} the metric tensor is generalized to $D$ dimensions: $g_\mu^\mu=D$, and the $\gamma$ matrices obey the same anticommuting rules as in four dimensions. Even if these rules are algebraically inconsistent \cite{BM}, the NDR scheme gives correct results provided one can avoid the calculation of traces like ${\rm tr}(\gamma_5 \gamma_\kappa \gamma_\lambda \gamma_\mu \gamma_\nu)$ \cite{BW}.

The scheme originally proposed by 't Hooft and Veltman (HV scheme) \cite{HV} allows a consistent formulation of dimensional regularization even when $\gamma_5$ couplings are present \cite{BM}. Besides the $D$- and 4-dimensional metric tensors $g$ and $\tilde g$ one introduces the $-2\varepsilon$-dimensional tensor $\hat g$. One can then split the $D$-dimensional Dirac matrix $\gamma_\mu$ into a 4- and a $-2\varepsilon$-dimensional part $\tilde\gamma_\mu$ and $\hat\gamma_\mu$ which separately obey anticommutation relations with the appropriate metric tensors. A $\gamma_5$ can be introduced which anticommutes with $\tilde\gamma$ but commutes with $\hat\gamma$. The price one has to pay for a consistent dimensional regularization scheme is a substantial increase in the complexity of calculations. 

The second step in our programme to eliminate the infinities from a theory is {\em renormalization}. This is the process of relating the unphysical {\em (bare)} and physical {\em (renormalized)} parameters like couplings $g$ or masses $m$ and rewrite observables as functions of the physical quantities. The renormalization procedure hides all divergences in a redefinition of the fields and parameters in the Lagrangian, i.e.
\begin{equation}
  \label{ren}
  \begin{array}{lcl}
   g_0=Z_g g \mu^\varepsilon & \qquad & m_0=Z_m m\\
   q_0=Z_q^{1/2}q            & \qquad & A_0^\mu=Z_3^{1/2}A^\mu
  \end{array}
\end{equation}
and thus guaranties that measurable quantities stay finite. The index ``0'' indicates bare quantities. Introducing the parameter $\mu$ with dimension of mass in (\ref{ren}) is necessary to keep the coupling dimensionless. The factors $Z$ are the renormalization constants. The renormalization process is performed recursively in powers of the coupling constant $g$. If at every order of perturbation theory all divergences are reabsorbed in $Z$'s, the theory is called ``renormalizable''. Theories with gauge symmetries, like the Standard Model, are renormalizable. This is true even if the gauge symmetry is spontaneously broken via the Higgs mechanism because gauge invariance of the Lagrangian is conserved \cite{GtH}.

Renormalization can be straightforwardly implemented via the counter-term method. According to (\ref{ren}) the unrenormalized quantities are reexpressed through the renormalized ones in the original Lagrangian. Thus
\begin{equation}
\label{lcounter}
{\cal L}_0={\cal L}+{\cal L}_\mathrm{counter}
\end{equation}
The counter terms ${\cal L}_\mathrm{counter}$ are proportional to $(Z-1)$ and can be treated as new interaction terms. For these new interactions Feynman rules can be derived and the renormalization constants $Z_i$ are determined such that the contributions from these new interactions cancel the divergences in the Green functions. This fixes the renormalization constants only up to an arbitrary subtraction of finite parts. Different finite parts define different {\em renormalization schemes.} In the Minimal Subtraction (MS) scheme only the divergences and no finite parts are subtracted \cite{MS}. The modified MS scheme ($\overline{\rm MS}$) \cite{BBDM} defines the finite parts such that terms $\ln 4\pi-\gamma_E$, the artifacts of dimensional regularization, vanish. This can be achieved if one calculates with
\begin{equation}
\label{mumsbar}
  \mu_{\overline{\mathrm{MS}}}=\frac{\mu \, e^{\gamma_E/2}}{\sqrt{4\pi}}
\end{equation}
instead of $\mu$ and performs minimal subtraction afterwards. We will exclusively work with the $\overline{\rm MS}$ scheme in the following.

Every renormalization procedure necessitates to introduce a dimensionful parameter $\mu$ into the theory. Even after renormalization the theoretical predictions depend on this {\em renormalization scale} $\mu$. At this momentum scale the renormalization prescriptions, which the parameters of a renormalized field theory depend on, are applied. One ``defines the theory at the scale $\mu$.'' The bare parameters are $\mu$-independent. To determine the renormalized parameters from experiment, a specific choice of $\mu$ is necessary: $g\equiv g(\mu)$, $m\equiv m(\mu)$, $q\equiv q(\mu)$. Different values of $\mu$ define different parameter sets $g(\mu)$, $m(\mu)$, $q(\mu)$. The set of all tranformations that relates parameter sets with different $\mu$ is called {\em renormalization group} (RG).

The scale dependence of the renormalized parameters can be obtained from the $\mu$-independence of the bare ones. In QCD we get from (\ref{ren}) the {\em renormalization group equations} (RGE) for the {\em running coupling} and the {\em running mass}
\begin{eqnarray}
  \label{RGEg}
  \frac{dg(\mu)}{d\ln \mu} &=& \beta\!\left(g(\mu),\varepsilon\right)\\
  \label{RGEm}
  \frac{dm(\mu)}{d\ln \mu} &=&-\gamma_m\left(g(\mu)\right) m(\mu)
\end{eqnarray}
with the $\beta$-function
\begin{equation}
\label{beta}
  \beta\!\left(g(\mu),\varepsilon\right)=-\varepsilon g \underbrace{-\frac{1}{Z_g}\frac{dZ_g}{d \ln\mu}}_{\displaystyle =: \beta(g)}
\end{equation}
and the anomalous dimension of the mass operator
\begin{equation}
\label{gamma}
  \gamma_m\left(g(\mu)\right)=\frac{1}{Z_m}\frac{dZ_m}{d \ln\mu}
\end{equation}
Calculating to two-loop accuracy we get
\begin{eqnarray}
  \beta(g)&=& -\frac{g^3}{16\pi^2}\beta_0 -\frac{g^5}{(16\pi^2)^2}\beta_1\\
  \gamma_m(\alpha_s) &=& \frac{\alpha_s}{4\pi}\gamma_m^{(0)} +\left(\frac{\alpha_s}{4\pi}\right)^2\gamma_m^{(1)}
\end{eqnarray}
where
\begin{eqnarray}\label{betadef}
  \begin{array}{p{4cm}p{6.5cm}}
    $\displaystyle \beta_0=\frac{11N-2f}{3}$ & $\displaystyle \beta_1=\frac{34}{3}N^2 -\frac{10}{3}N f-2C_F f$
  \end{array}\\ \label{gammadef}
  \begin{array}{p{4cm}p{6.5cm}}
    $\displaystyle \gamma_m^{(0)}=6C_F$ & $\displaystyle \gamma_m^{(1)}=C_F\left(3C_F+\frac{97}{3}N-\frac{10}{3}f\right)$
  \end{array}\\ \label{alphasCFdef}
  \begin{array}{p{4cm}p{6.5cm}}
    $\displaystyle \alpha_s(\mu)=\frac{g^2(\mu)}{4\pi}$ & $\displaystyle C_F=\frac{N^2-1}{2N}$
  \end{array}
\end{eqnarray}
with $N$ the number of colours and $f$ the number of active flavours. The solutions for $\alpha_s(\mu)$ and $m(\mu)$ then are \cite{BBDM}
\begin{eqnarray}
\label{runalpha}
  \alpha_s(\mu) &=&\frac{4\pi}{\beta_0 \ln(\mu^2/\Lambda_{\overline{MS}}^2)}\left[1-\frac{\beta_1}{\beta_0^2}\frac{\ln\ln(\mu^2/\Lambda_{\overline{MS}}^2)}{\ln(\mu^2/\Lambda_{\overline{MS}}^2)}\right]\\
\label{runm}
  m(\mu) &=& m(\mu_0)\left[\frac{\alpha_s(\mu)}{\alpha_s(\mu_0)}\right]^{\frac{\gamma_m^{(0)}}{2\beta_0}}\left[1+\left(\frac{\gamma_m^{(1)}}{2\beta_0}-\frac{\beta_1\gamma_m^{(0)}}{2\beta_0^2}\right)\frac{\alpha_s(\mu)-\alpha_s(\mu_0)}{4\pi}\right]
\end{eqnarray}
Here $\Lambda_{\overline{MS}}$ is a characteristic scale both for QCD and the used $\overline{MS}$ scheme and depends also on the number of effective flavours present in $\beta_{0}$ and $\beta_1$. An $\alpha_s^{(5)}(M_Z)=0.118\pm0.005$ corresponds to $\Lambda^{(5)}_{\overline{MS}}=225^{+70}_{-57}$ MeV in NLO. It is interesting to note that such a mass scale $\Lambda$ emerges without making reference to any dimensional quantity and would be present also in a theory with completely massless particles. In QCD with three colours, even for six active flavours, both $\beta_0=7$ and $\gamma_m^{(0)}/2\beta_0=4/7$ are positive. This leads to asymptotic freedom as the coupling tends to zero with increasing $\mu$. The pole at $\Lambda_{\overline{MS}}$ signals the breakdown of perturbation theory but gives a plausible argument for confinement. Similarly, $m(\mu)$ decreases with $\mu$ getting larger.

A particularly useful application of the renormalization group is the summation of large logarithms. To see this we reexpress $\alpha_s$ of (\ref{runalpha}) as
\begin{equation}
\label{runalphamu0}
  \alpha_s(\mu)=\frac{\alpha_s(\mu_0)}{v(\mu)}\left[1-\frac{\beta_1}{\beta_0}\frac{\alpha_s(\mu_0)}{4\pi}\frac{\ln v(\mu)}{v(\mu)}\right]
\end{equation}
with
\begin{equation}
  v(\mu)=1-\beta_0 \frac{\alpha_s}{4\pi} \ln\frac{\mu_0^2}{\mu^2}
\end{equation}
If we expand the leading order term of (\ref{runalphamu0}) in $\alpha_s(\mu_0)$ we get
\begin{equation}
\label{RGIalpha}
  \alpha_s(\mu)=\alpha_s(\mu_0)\sum_{m=0}^\infty\left(\beta_0\frac{\alpha_s(\mu_0)}{4\pi}\ln\frac{\mu_0^2}{\mu^2}\right)^m
\end{equation}
Thus the solution of the RGE automatically sums the logarithms $\ln(\mu_0^2/\mu^2)$ which get large for $\mu\ll\mu_0$. Generally, solving the RGE to order $n$ sums in $\alpha_s(\mu)$ all terms of the form
\begin{equation}
  \alpha_s(\mu_0)^{m+1}\left(\alpha_s(\mu_0)\ln\frac{\mu_0^2}{\mu^2}\right)^k,\quad 0\le m\le n, \quad k \in \mathbb{N}_{\;\!0}
\end{equation}
This is particularly useful if, though $\alpha_s(\mu_0)$ is smaller than one, the combination $\alpha_s(\mu_0)\ln(\mu_0^2/\mu^2)$ is close to or even larger than one. Then the large logarithms would spoil the convergence of the perturbation series.

% =========================================
% =      Operator Product Expansion       =
% =========================================

\section{Operator Product Expansion}
\label{sec:OPE}

Up to now we treated processes of strong and electroweak interaction separately. But all weak processes involving hadrons receive QCD corrections, which can be substantial especially for non-leptonic and rare decays. The underlying quark level decay of a hadron is governed by the electroweak scale given by $M_{W,Z}={\cal O}(100 \,{\rm GeV})$. On the other hand, the available energy inherent in a $B$ meson decay is of ${\cal O}(m_B)$. In dimensional regularization for example we encounter logarithms of the ratio of either of these scales with the renormalization scale $\mu$. If the scales involved are widely separated it is not possible to make all the logarithms small by a suitable choice of the renormalization scale. As we have seen in the last section these large logarithms can be summed systematically using renormalization group techniques.

But we have yet another energy scale in the problem. A priori we cannot consider the decay of free quarks. Due to confinement quarks appear in colourless bound systems only. The binding of the quarks inside the hadron via strong interaction is characterized by a typical hadronic scale of ${\cal O}(1\,{\rm GeV})$. Here, even without large logarithms the strong coupling $\alpha_s$ is too large for perturbation theory to make sense. Unfortunately, in many cases the non-perturbative methods we have at hand nowadays are not yet developed enough to give accurate results.

Coming back to the typical energy $m_B$ in a $B$ decay. Do we have to know at all what is {\em really} going on at energies of ${\cal O}(100 \,{\rm GeV})$ or the corresponding extremely short distances? In fact we do not. We also do not bother general relativity to calculate the trajectory of an apple falling from a tree or QED and QCD to learn something about the properties of condensed matter. Instead, we employ Newtonian mechanics or the laws of chemistry and solid state physics, respectively. They are nonrelativistic approximations or {\em effective theories} appropriate for the low energy scale under consideration. This is exactly what we want to achieve for the weak interaction of quarks as well. The theoretical tool for this purpose is {\em operator product expansion} (OPE) \cite{OPE} which we shall introduce in the following.

For small separations, the product of two field operators $A(x)$ and $B(y)$ can be expanded in local operators $Q_i$ with potentially singular coefficient functions $c_i$ as
\begin{equation}
\label{OPE}
  A(x)B(y)=\sum_i c_i(x-y)Q_i(x)
\end{equation}
The {\em Wilson coefficient} $c_1(z)$ accompanying the operator with lowest energy dimension is the most singular one for $z\to 0$ and the degree of divergence of the $c_i$ decreases for increasing operator dimension. Furthermore, for dimensional reasons, contributions of operators with higher dimension are suppressed by inverse powers of the heavy mass (small distance) scale. In principle, we have to consider all operators compatible with the global symmetries of the operator product $AB$. The physical picture is that a product of local operators should appear as one local operator if their distance is small compared to the characteristic length of the system. One can systematically approximate the behaviour of an operator product at short distances with a finite set of local operators. This is exactly what is done for the theory of weak decays.

\begin{figure}
  \begin{center}
    \psfig{figure=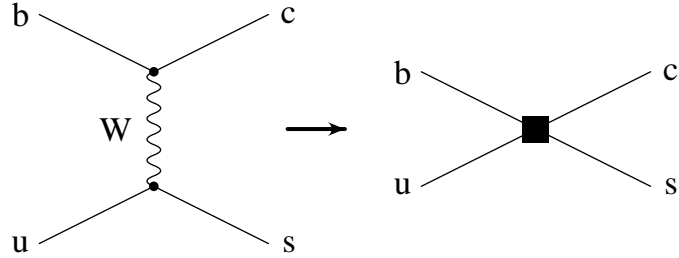}
  \end{center}
  \caption[The effective vertex]{Replacing a $W$ propagator with an effective four-fermion vertex.\label{fig:effvertex}}
\end{figure}

Here, the mass of the $W$ boson $M_W\approx 80\,{\rm GeV}$ is very large compared to a typical hadronic scale. Therefore, the $W$ propagator $\Delta^{\mu\nu}(x,y)$ is of very short range only. In the amplitude of weak decays it connects two charged currents $J_\mu^-(x)$ and $J_\nu^+(y)$, which hence interact almost locally so that we can perform an OPE. Let us consider the quark level decay $b\to cs\bar u$ for definiteness. The tree-level $W$-exchange amplitude for this decay is given by
\begin{eqnarray}
\label{Abtocsu}
  A(b \to cs\bar u)&=& -\frac{G_F}{\sqrt{2}} V^{}_{cb}
  V^*_{us}\frac{M_W^2}{k^2-M_W^2}(\bar s u)_{V-A}(\bar c b)_{V-A} 
  \nonumber \\
  &=&\frac{G_F}{\sqrt{2}} V^{}_{cb}V^*_{us}\underbrace{(\bar s u)_{V-A}
  (\bar c b)_{V-A}}_{\mbox{local operator}} +
  {\cal O}\!\left(\frac{k^2}{M_W^2}\right) 
\end{eqnarray}
Since the momentum transfer $k$ through the $W$ propagater is small as compared to $M_W$, we can safely neglect the terms ${\cal O}(k^2/M_W^2)$. The $W$ propagator then quasi shrinks to a point (see fig. \ref{fig:effvertex})
%%%%%%%%%%%%%%%%%%%%%%%%%%%%%%%%%%%%%%%%%%%%%%%%%%%%%
%%%%%%%%%%%%%%%%%%%%%%%%%%%%%%%%%%%%%%%%%%%%%%%%%%%%%
and we obtain an effective four-fermion interaction. This is the modern formulation of the classical Fermi theory of weak interaction with $G_F=1.166\cdot 10^{-5}\,{\rm GeV}^{-2}$ the Fermi constant. The Wilson coefficient in this example is simply one. The notation $(\overline{q}_1 q_2)_{V-A}$ in (\ref{Abtocsu}) is a practical shorthand for a left-handed charged quark current with the chiral vector minus axialvector structure
\begin{equation}
  (\overline{q}_1 q_2)_{V-A}:=\overline{q}_1\gamma_\mu (1-\gamma_5)q_2
\end{equation}

% =========================================
% =      Effective Theories               =
% =========================================

\section{Effective Theories}
\label{sec:ET}

The result (\ref{Abtocsu}) can also be derived from an effective Hamiltonian
\begin{equation}
  {\cal H}_{\rm eff}=\frac{G_F}{\sqrt{2}} V_{cb}V^*_{us}
  (\bar s u)_{V-A}(\bar c b)_{V-A}+\mbox{operators of higher dimension}
\end{equation}
where the operators of higher dimensions correspond to the terms ${\cal O}(k^2/M_W^2)$ in (\ref{Abtocsu}) and can likewise be neglected.  In the effective theory the $W$ boson is removed as an explicit, dynamical degree of freedom. It is ``integrated out'' or ``contracted out'' using the language of the path integral or canonical operator formalism, respectively. One can proceed in a completely analougous way with the heavy quarks. This leads to {\em effective $f$ quark theories} where $f$ denotes the ``active'' quarks, i.e. those that have not been integrated out.

If we include also short distance QCD or electroweak corrections more operators have to be added to the effective Hamiltonian which we generalize to
\begin{equation}
  \label{Heffgen}
  {\cal H}_\mathrm{eff}=\frac{G_F}{\sqrt{2}}\sum_i V_\mathrm{CKM}^i C_i(\mu)Q_i(\mu)
\end{equation}
Here the factor $V^i_\mathrm{CKM}$ denotes the CKM structure of the particular operator. If we want to calculate the amplitude for the decay of a meson $M=K,\,D,\,B,\,\ldots$ into a final state $F$ we just have to project the Hamilton operator onto the external states
\begin{eqnarray}
  A(M\to F)&=&\langle F| {\cal H}_\mathrm{eff} |M\rangle
  \nonumber \\
  &=&\frac{G_F}{\sqrt{2}}\sum_i V_\mathrm{CKM}^i C_i(\mu) \langle
  F|Q_i(\mu) |M\rangle\,.
\end{eqnarray}
The Wilson coefficients $C_i(\mu)$ can be interpreted as the coupling constants for the effective interaction terms $Q_i(\mu)$. They are calculable functions of $\alpha_s$, $M_W$, and the renormalization scale $\mu$. To any order in perturbation theory the Wilson coefficients can be obtained by {\em matching} the full theory onto the effective one. This simply is the requirement that the amplitude in the effective theory should reproduce the corresponding amplitude in the full theory. Hence, we first have to calculate the amplitude in the full theory and then the matrix elements $\langle Q_i\rangle$. In this second step the resulting expressions may, even after quark field renormalization, be still divergent. Consequently we have to perform an {\em operator renormalization}
\begin{equation}
  Q_i^{(0)}=Z_{ij} Q_j
\end{equation}
where $Q_i^{(0)}$ denotes the unrenormalized operator. This notation is somewhat sloppy and misleading. What actually is renormalized is not the operator but the operator matrix elements, or, even more exactly, the amputated Green functions $\langle Q_i\rangle$. Then we have to include the renormalization constant $Z_q^{1/2}$ for each of the four external quark fields:
\begin{equation}\label{opren}
  \langle Q_i\rangle^{(0)} =Z_q^{-2} Z_{ij} \langle Q_j\rangle
\end{equation}
In general, the renormalization constant $Z_{ij}$ is a matrix so that operators carrying the same quantum numbers can {\em mix under renormalization}. Operators of a given dimension mix only into operators of the same or of lower dimension. Again, the divergent parts of the renormalization constant are determined from the requirement that the amplitude in the effective theory is finite. The finite part in $Z_{ij}$ on the other hand defines a specific renormalization scheme. In a third step we extract the Wilson coefficients by comparing the full and the effective theory amplitude. These are the Wilson coefficients at some fixed scale $\mu_0$. A caveat here is that the external states in the full and the effective theory have to be treated in the same manner. Especially the same regularization and renormalization schemes have to be used on both sides. 

As the Wilson coefficients appear already at the level of the effective Hamiltonian, they are independent of the external states this Hamiltonian is projected onto to obtain the complete amplitude. When determining the Wilson coefficients, any external, even unphysical, state can be used. The coefficient functions represent the short-distance structure of the theory. Because they depend for example on the masses of the particles that were integrated out, they contain all information about the physics at the high energy scale. The long-distance contribution, on the other hand, is parametrized by the process-dependent matrix elements of the local operators. This {\em factorization} of SD and LD dynamics is one of the salient features of OPE. We can calculate the Wilson coefficients in perturbation theory and the hadronic matrix elements by means of some non-perturbative technique like $1/N$ expansion, sum rules, or lattice gauge theory. Especially to use the latter one, a separation of the SD part is essential for today's lattice sizes. The factorization can be visualized with large logarithms $\ln(M_W^2/m_q^2)$ being split into $\ln(M_W^2/\mu^2)+\ln(\mu^2/m_q^2)$. In doing so, the first logarithm will be retrieved in the Wilson coefficients and the second one in the matrix elements. From this point of view the renormalization scale $\mu$ can be interpreted as the {\em factorization scale} at which the full contribution is separated into a low energy and a high energy part.

A typical scale at which to calculate the hadronic matrix elements of local operators is low compared to $M_W$. For $B$ decays we would choose $\mu={\cal O}(m_B)$. Therefore, the logarithm $\ln(M_W^2/\mu^2)$ contained in the Wilson coefficient is large. So why not use the powerful technique of summing large logarithms developped in section \ref{sec:ren}? In order to do so we have to find the renormalization group equations for the Wilson coefficients and solve them. But so far the Wilson coefficients were not renormalized at all. If we remember, however, that in the effective Hamiltonian the operators, which have to be renormalized, are accompanied always by the appropriate Wilson coefficent we can shuffle the renormalization as well to the Wilson coefficients. Let us start with the Hamiltonian of the effective theory with fields and coupling constants as bare quantities, which are renormalized according to
\begin{eqnarray}
  q^{(0)} &=& Z_q^{1/2}q\\
  C_i^{(0)} &=& Z_{ij}^c C_j
\end{eqnarray}
Then the Hamiltonian (\ref{Heffgen}) is in essence
\begin{eqnarray}
  {\cal H}_\mathrm{eff} &\propto& C_i^{(0)} Q_i(q^{(0)})\nonumber\\
    &\equiv& Z_{ij}^c C_j Z_q^2 Q_i\nonumber\\ \label{Heffcounter}
    &\equiv& C_i Q_i +(Z_q^2 Z_{ij}^c -\delta_{ij}) C_j Q_i
\end{eqnarray}
i.e. it can be written in terms of the renormalized couplings $C_i$ and fields $Q_i$ plus counterterms. The $q^{(0)}$ indicates that the interaction term $Q_i$ is composed of unrenormalized fields. If we calculate the amplitude with the Hamiltonian (\ref{Heffcounter}) including the counterterms, we get the finite renormalized result
\begin{equation}
  Z_q^2 Z_{ij}^c C_j \langle Q_i\rangle^{(0)} = C_j\langle Q_j\rangle
\end{equation}
Comparing with (\ref{opren}) we read off the renormalization constant for the Wilson coefficients
\begin{equation}
  Z_{ij}^c = Z_{ji}^{-1}
\end{equation}
So we can think of the operator renormalization in terms of the completely equivalent renormalization of the coupling constants $C_i$, as in any field theory. If we again demand the unrenormalized Wilson coefficients not to depend on $\mu$ we obtain the renormalization group equation
\begin{equation}
  \label{RGEC}
  \frac{dC_i(\mu)}{d\ln\mu}=\gamma_{ji}(\mu)C_j(\mu)\,,
\end{equation}
with the anomalous dimension matrix for the operators
\begin{equation}
  \label{admop}
  \gamma_{ij}(\mu)=Z^{-1}_{ik}\frac{dZ_{kj}}{d\ln\mu}
\end{equation}
Let us simply state here that the numerical values for the $\gamma_{ij}$ can be determined directly from the divergent parts of the renormalization constants $Z_{ij}$. In (\ref{RGEC}) the transposed of this anomalous dimension matrix appears. It is only the sign and the fact that the anomalous dimension is a matrix instead of a single number that distinguishes the RGE for the Wilson coefficients from that of the running mass in (\ref{RGEm}). Therefore, we could use the solution (\ref{runm}) with the appropriate changes. To leading order this is in fact possible. But if we want to go to next-to-leading-order accuracy we run into problems, because the matrices $\gamma^{(0)}_{ij}$ and $\gamma^{(1)}_{ij}$ in the perturbative expansion
\begin{equation}
\label{adm}
  \gamma_{ij} = \gamma^{(0)}_{ij}\frac{\alpha_s}{4\pi} +\gamma^{(1)}_{ij}\left(\frac{\alpha_s}{4\pi}\right)^2 +{\cal O}(\alpha_s^3)
\end{equation}
do not commute with each other. Let us instead formally write the solution for the Wilson coefficients with an evolution matrix $U(\mu,\mu_0)$
\begin{equation}\label{cevolve}
  C_i(\mu)=U_{ij}(\mu,\mu_0) C_j(\mu_0)
\end{equation}
The leading order evolution matrix can be read off from (\ref{runm})
\begin{eqnarray}
  U^{(0)}(\mu,\mu_0) &=& \left[\frac{\alpha(\mu)}{\alpha(\mu_0)}\right]^{-\frac{{\gamma^{(0)}}^T}{2\beta_0}}\\ \nonumber
  \label{ulo}
  &=& V\left(\left[\frac{\alpha(\mu_0)}{\alpha(\mu)}\right]^\frac{\vec\gamma^{(0)}}{2\beta_0}\right)_{\!\!\!\!D} V^{-1}
\end{eqnarray}
where $V$ is the matrix that diagonalizes $\gamma^{(0)T}$
\begin{equation}
  \gamma^{(0)}_D=V^{-1} \gamma^{(0)T} V
\end{equation}
and $\vec\gamma^{(0)}$ is the vector containing the eigenvalues of $\gamma^{(0)}$. For the next-to-leading order solution we make the clever ansatz
\begin{equation}
\label{unlo}
  U(\mu,\mu_0)=\left[1+\frac{\alpha_s(\mu)}{4\pi}J\right] U^{(0)}(\mu,\mu_0)\left[1-\frac{\alpha_s(\mu_0)}{4\pi} J\right]
\end{equation}
which proves to solve (\ref{RGEC}) if \cite{RGEC}
\begin{equation}
  J=V H V^{-1}
\end{equation}
where the elements of $H$ are
\begin{equation}
  H_{ij}=\delta_{ij} \gamma_i^{(0)} \frac{\beta_1}{2\beta_0^2} -\frac{G_{ij}}{2\beta_0+\gamma_i^{(0)}-\gamma_j^{(0)}}
\end{equation}
with
\begin{equation}
  G=V^{-1} \gamma^{(1)T} V
\end{equation}
As we have mentioned in section \ref{sec:ren}, the procedure of renormalization allows to subtract arbitrary finite parts along with the ultraviolet singularities. Whereas physical quantities must clearly be independent of the renormalization scheme chosen, at NLO unphysical quantities, like the Wilson coefficients and the anomalous dimensions, depend on the choice of the renormalization scheme. To ensure a proper cancellation of this scheme dependence in the product of Wilson coefficients and matrix elements the same scheme has to be used for both. In order to uniquely define a renormalization scheme it is not sufficient to quote only the regularization and renormalization procedure but one also has to choose a specific form for the so-called {\em evanescent operators}. These are operators which exist in $D\neq 4$ dimensions but vanish in $D=4$ \cite{BW,DG,HN}.

So what do we have achieved so far? We have determined the Wilson coefficients at a scale $\mu_0$ via a matching procedure. These are the initial conditions for the evolution from $\mu_0$ down to an appropriate low energy scale $\mu$ via $U(\mu,\mu_0)$ which sums large logarithms. Herefore, we had to determine the anomalous dimensions of the operators and solve the renormalization group equation for the Wilson coefficients. We thus arrive at a RG improved perturbation theory and officially don't speak any more of ``leading'' (LO) and ``next-to-leading order'' (NLO) but rather of ``leading'' (LL) and ``next-to-leading-logarithmic order'' (NLL). Yet, we might carelessly use the terms synonymously. In our task to evaluate weak decay amplitudes involving hadrons in the framwork of a low energy effective theory we then only lack the calculation of the hadronic matrix elements $\langle Q_i(\mu)\rangle$. This, however, is a highly non-trivial problem which this and many other works are devoted to.

Looking at the complicated NLO formulas one might ask why at all going to next-to-leading order accuracy. After all these calculations imply the evaluation of two or even more loop diagrams which are technically very challenging. But they are very important. First of all we can test the validity of the renormalization group improved perturbation theory. Then, of course, we hope that the theoretical uncertainties get reduced. One particular issue is the residual renormalization scale dependence of the result. The scale $\mu$ enters for example in $\alpha_s(\mu)$ or the running quark masses, in particular $m_t(\mu)$, $m_b(\mu)$, and $m_c(\mu)$. In principle, a physical quantity cannot depend on the renormalization scale. But as we have to truncate the perturbative series at some fixed order, this property is broken. The renormalization scale dependence of Wilson coefficients and operator matrix elements cancels only to the order of perturbation theory included in the calculation. Therefore, one can use the remaining scale ambiguity as an estimate for the neglected higher order corrections. Usually one varies $\mu$ between half and twice the typical scale of the problem, i.e. $m_b/2<\mu<2m_b$ for $B$ decays. Going to NLO significantly reduces these scale ambiguities. Furthermore, the renormalization scheme dependence of the Wilson coefficients appears at NLO for the first time. Only if we properly match the long distance matrix elements, obtained for example from lattice calculations, to the short distance contributions, these unphysical scheme dependences will cancel. Another issue is that the QCD scale $\Lambda_{\overline{\mathrm{MS}}}$, which can be extracted from various high energy processes, cannot be used meaningfully in weak decays without going to NLO.

% =========================================
% =      The Effective Hamiltonian        =
% =========================================

\section{The Effective $b\to s\gamma$ Hamiltonian}
\label{sec:EH}

\begin{figure}
  \begin{center}
    \psfig{figure=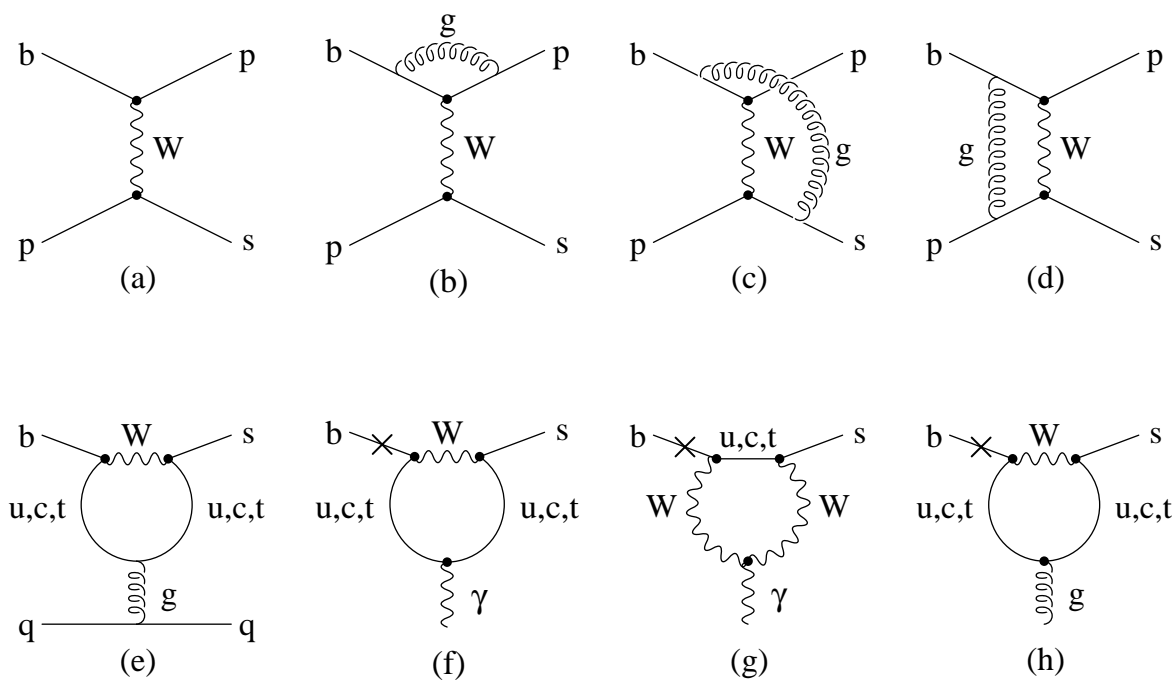}
  \end{center}
  \caption{The diagrams where the operator basis for $b\to s\gamma$ originates from: Current-current diagram (a) with QCD corrections (b), (c), (d); gluon penguin diagram (e), magnetic photon penguin diagrams (f), (g), and magnetic gluon penguin diagram (h). The cross in (f), (g), and (h) denotes that the mass of the external $b$ quark has to be kept.\label{fig:opdiag}}
\end{figure}

In this section we want to discuss the effective Hamiltonian necessary for the calculations to follow. For $b\to s\gamma$ transitions it reads
\begin{equation}\label{heff}
{\cal H}_\mathrm{eff}=\frac{G_F}{\sqrt{2}}\sum_{p=u,c}\lambda_p^{(s)}
\left[ C_1 Q^p_1 + C_2 Q^p_2 +\sum_{i=3,\ldots ,8} C_i Q_i\right]
\end{equation}
where
\begin{equation}\label{lamps}
\lambda_p^{(s)}=V^*_{ps}V_{pb}
\end{equation}
The operators originate from the diagrams in Fig.~\ref{fig:opdiag}
%%%%%%%%%%%%%%%%%%%%%%%%%%%%%%%%%%%%%%%%%%%%%%%%%%%%%
%%%%%%%%%%%%%%%%%%%%%%%%%%%%%%%%%%%%%%%%%%%%%%%%%%%%%
and are given by
\begin{eqnarray}
  \label{q1def}
    Q^p_1 &=& (\bar sp)_{V-A}(\bar pb)_{V-A} \\
  \label{q2def}
    Q^p_2 &=& (\bar s_i p_j)_{V-A}(\bar p_j b_i)_{V-A} \\
  \label{q3def}
    Q_3 &=& (\bar sb)_{V-A} \sum_q (\bar qq)_{V-A} \\
  \label{q4def}
    Q_4 &=& (\bar s_i b_j)_{V-A} \sum_q (\bar q_j q_i)_{V-A} \\
  \label{q5def}
    Q_5 &=& (\bar sb)_{V-A} \sum_q (\bar qq)_{V+A} \\
  \label{q6def}
    Q_6 &=& (\bar s_i b_j)_{V-A} \sum_q (\bar q_j q_i)_{V+A} \\
  \label{q7def}
    Q_7 &=& \frac{e}{8\pi^2}m_b\,\bar s_i\sigma^{\mu\nu}(1+\gamma_5)b_i\, F_{\mu\nu}\\
  \label{q8def}
    Q_8 &=& \frac{g_s}{8\pi^2}m_b\,\bar s_i\sigma^{\mu\nu}(1+\gamma_5)T^a_{ij} b_j\, G^a_{\mu\nu}
\end{eqnarray}
with $e$ and $g_s$ the coupling constants of electromagnetic and strong interaction and $F_{\mu\nu}$ and $G_{\mu\nu}$ the photonic and gluonic field strength tensors, respectively. In $Q_7$ and $Q_8$ we neglected the very small $m_s(1-\gamma_5)$ contribution. The $i,j$ are colour indices. If no colour index is given the two operators are assumed to be in a colour singlet state. The operator basis (\ref{q1def}--\ref{q8def}) consists of all possible gauge invariant operators with energy dimension six with the following properties: they have the correct quantum numbers to contribute to $b\to s\gamma$, they are compatible with the symmetries of electroweak interaction, and they cannot be transformed into each other by applying equations of motion. As a consequence, our operator basis is only the correct one if all external states are taken on-shell \cite{HP}. Additional operators have to be considered if an off-shell calculation is performed. There is yet another operator basis used in the literature. It was introduced by Chetyrkin, Misiak, and M\"unz (CMM) \cite{CMM,CMM2} because then no Dirac traces containing $\gamma_5$ arise in effective theory calculations, wich allows to use fully anticommuting $\gamma_5$ in dimensional regularization. It reads
\begin{eqnarray}
  \label{p1def}
    P^p_1 &=& (\bar s_L \gamma_\mu T^a p_L)(\bar p_L \gamma^\mu T^a b_L) \\
  \label{p2def}
    P^p_2 &=& (\bar s_L \gamma_\mu p_L)(\bar p_L \gamma^\mu b_L) \\
  \label{p3def}
    P_3 &=& (\bar s_L \gamma_\mu b_L) \sum_q (\bar q\gamma^\mu q) \\
  \label{p4def}
    P_4 &=& (\bar s_L \gamma_\mu T^a b_L) \sum_q (\bar q\gamma^\mu T^a q) \\
  \label{p5def}
    P_5 &=& (\bar s_L \gamma_{\mu_1}\gamma_{\mu_2}\gamma_{\mu_3} b_L) \sum_q (\bar q \gamma^{\mu_1}\gamma^{\mu_2}\gamma^{\mu_3} q) \\
  \label{p6def}
    P_6 &=& (\bar s_L \gamma_{\mu_1}\gamma_{\mu_2}\gamma_{\mu_3} T^a b_L) \sum_q (\bar q \gamma^{\mu_1}\gamma^{\mu_2}\gamma^{\mu_3} T^a q) \\
  \label{p7def}
    P_7 &=& \frac{e}{16\pi^2}m_b(\bar s_L\sigma^{\mu\nu}b_R)F_{\mu\nu}\\
  \label{p8def}
    P_8 &=& \frac{g_s}{16\pi^2}m_b(\bar s_L\sigma^{\mu\nu}T^a b_R)G^a_{\mu\nu}
\end{eqnarray}
where $T^a$ stand for $SU(3)_\mathrm{color}$ generators and $L=(1-\gamma_5)/2$ and $R=(1+\gamma_5)/2$ for the left and right-handed projection operators. We denote the corresponding Wilson coefficients with $Z_i$. The operator basis (\ref{p1def}--\ref{p8def}) is in principle the more natural one as the operators appear exactly in this form when calculating the diagrams of Fig.~\ref{fig:opdiag}. In our normal operator basis (\ref{q1def}--\ref{q8def}), the following four-dimensional identity was used to ``simplify'' $P_5$ and $P_6$
\begin{equation}
  \gamma_\mu \gamma_\nu \gamma_\rho = g_{\mu\nu} \gamma_\rho + g_{\nu\rho} \gamma_\mu -g_{\mu\rho} \gamma_\nu +i\epsilon_{\sigma\mu\nu\rho} \gamma^\sigma \gamma_5 \qquad\mbox{in }D=4
\end{equation}
This step, however, requires to introduce several more evanescent operators and leads to problematic traces with $\gamma_5$ in two-loop calculations. The choice of the operator's colour structure is more natural in the CMM basis, too, as it is the one emerging from the diagrams in Fig.~\ref{fig:opdiag}.

In the common nomenclature $Q_1$ and $Q_2$ are called {\em current-current operators}, $Q_{3\ldots 6}$ {\em QCD penguin operators} and $Q_7$ and $Q_8$ {\em electromagnetic} and {\em chromomagnetic penguin operator}, respectively. For historical reasons the numbering of $Q^p_{1,2}$ is sometimes reversed in the literature \cite{BBL}. The sign conventions for the electromagnetic and strong couplings correspond to the covariant derivative $D_\mu=\partial_\mu +ie Q_f A_\mu + i g T^a A^a_\mu$. The coefficients $C_{7}$ and $C_8$ then are negative in the Standard Model, which is the choice generally adopted in the literature. The effective Hamiltonian for $b\to d\gamma$ is obtained from equations (\ref{heff}--\ref{p8def}) by the replacement $s\to d$.

Let us summarize the status quo in determining the effective Hamiltonian for $b\to s\gamma$. At leading order only $Q_7$ contributes to $b\to s\gamma$. The corresponding Wilson coefficient was first calculated by Inami and Lim in 1980 \cite{IL}. For the leading logarithmic RG improved calculation the $8\times 8$ anomalous dimension matrix was obtained bit by bit. Mixing of the four-quark operators $Q_{1\ldots 6}$ among each other was considered already earlier for the analysis of nonleptonic decays \cite{6x6ADM}. The $2\times 2$ submatrix for the magnetic penguins was obtained by Grinstein, Springer, and Wise \cite{GSWbsg}. Because the mixing of $Q_{1\ldots 6}$ into $Q_7$ and $Q_8$ vanishes at the one-loop level, one has to perform two-loop calculations if the LL result for $C_7(\mu_b)$ is wanted. This technical complication delayed the first completely correct result for the $8\times 8$ anomalous dimension matrix until 1993 \cite{CFMRS} followed by independent confirmations \cite{confLL}.

Going to next-to-leading order accuracy was highly desirable, because the leading logarithmic expression for the branching ratio $B(B\to X_s\gamma)$ suffers from sizable renormalization scale uncertainties at the $\pm 25\%$ level. The step from leading order to leading logarithmic order already increased the branching ratio $B(B\to X_s\gamma)$ by almost a factor of three \citer{BBM,GOSN}. The high-flying NLL enterprise was a joint effort of many groups. QCD corrections affect both Wilson coefficients and operator matrix elements. The ${\cal O}(\alpha_s)$ matching for the current-current operators was calculated already a long time ago \cite{ACMP}, the one for the QCD penguins more than ten years later \cite{BJLW}. Two-loop matching is necessary for $Q_7$ and $Q_8$, which was achieved first by Adel and Yao \cite{AY} and subsequently checked thoroughly \cite{checkAY,CDGG}. The ${\cal O}(\alpha_s^2)$ contributions to the $6\times 6$ anomalous dimension matrix were calculated in \cite{BW,ACMP,BJLW,CFMR}, the submatrix for $Q_7$ and $Q_8$ in \cite{MM}. Chetyrkin, Misiak, and M\"unz finally succeeded in determining the three-loop mixing of four-quark operators into magnetic penguin operators \cite{CMM}. The explicit formulas for the Wilson coefficients needed subsequently can be found in Appendix~\ref{app:WC}, and in Appendix~\ref{app:opbase} we comment on the transformation properties of the two operator bases used in this work.

For the {\em inclusive} decay $B\to X_s\gamma$, i.e. the radiative decay of a $B$ meson into the sum of final states with strangeness $S=-1$, the operator matrix elements can be computed perturbatively employing the {\em heavy-quark expansion} (HQE) \cite{HQE}. This method consists of an OPE in inverse powers of the large dynamical scale of energy release $\sim m_b$ followed by a nonrelativistic expansion for the $b$ field. Using the optical theorem, the inclusive $B$-decay rate can be written in terms of the absorptive part of the forward scattering amplitude $\langle B|{\cal T}|B\rangle$. The transition operator ${\cal T}$ is the absorptive part of the time-ordered product of ${\cal H}_\mathrm{eff}(x){\cal H}_\mathrm{eff}(0)$
\begin{equation}\label{Tdef}
  {\cal T} = {\rm Im}\!\left[ i\!\int\! d^4x \,T{\cal H}_\mathrm{eff}(x){\cal H}_\mathrm{eff}(0)\right]
\end{equation}
If we insert a complete set of states inside the time-ordered product we see that this was just a fancy way of writing the standard expression for the decay rate
\begin{equation}
  \Gamma(B\to X)=\frac{1}{2m_B}\sum_X (2\pi)^4 \delta^4(p_B-p_X)\left|\langle X|{\cal H}_\mathrm{eff}|B\rangle\right|^2
\end{equation}
However, the formulation in terms of the $T$ product allows for a direct evaluation using Feynman diagrams. Because of the large mass of the $b$ quark, we can construct an OPE in which ${\cal T}$ is represented as a series of local operators containing the heavy-quark fields. The operator of lowest dimension is $\bar b b$. Its matrix element is simplified by a nonrelativistic expansion in powers of $1/m_b$ starting with unity. In the limit of an infinitely heavy $b$ quark the $B$ meson decay rate is therefore given by the $b$ quark decay rate. This {\em quark-hadron duality} is of great use for many inclusive calculations and justifies for example the spectator model. Corrections to this relation appear only at ${\cal O}(1/m_b^2)$ because the potential dimension 4 operator $\bar b D\hspace{-0.65em}/\hspace{0.15em} b$ can be reduced to $m_b \bar b b$. 
However, the OPE only converges for sufficiently inclusive observables. But for $B\to X_s\gamma$ decays experimental cuts are necessary to reduce the background from charm production. This restricts the available phase space considerably leading to complications of the OPE based analysis.

The virtual corrections to the matrix elements $\langle s\gamma|Q_{1,7,8}|b\rangle$ were first calculated by Greub, Hurth, and Wyler \cite{GHW} and checked by Buras, Czarnecki, Misiak, and Urban using another method \cite{BCMU}. The latter authors presented recently also the last missing item in the NLO analysis of $B\to X_s\gamma$, namely the two-loop matrix elements of the QCD-penguin operators \cite{BCMU2}. Yet, their contribution is numerically very small because the corresponding Wilson coefficients almost vanish. The Bremsstrahlung corrections, i.e. the process $b\to s\gamma g$, influence especially the photon energy spectrum. At leading order this spectrum is a $\delta$ function, smeared out by the Fermi motion of the $b$ quark inside the $B$ meson, whereas it is broadened substantially at NLO \cite{Brems}.

Even higher electroweak corrections have been calculated for $B\to X_s\gamma$ \cite{EW}. Finally, there are non-perturbative contributions which can be singled out in the framework of HQE. The $1/m_b^2$ corrections mainly account for the fact that in reality a $B$ meson and not a $b$ quark is decaying. Additionally, one has to consider long distance contributions originating in the photon coupling to a virtual $\bar c c$ loop. This {\em Voloshin effect} is proportional to $1/m_c^2$ and enhances the decay rate by 3\% \citer{Voloshin,BIR}.

All this effort was taken to reduce the theoretical error and account for the ever increasing experimental precision. The current experimental world average for the inclusive branching fraction is
\begin{equation}
\label{bsgamexp}
  {\cal B}(B\to X_s\gamma)^\mathrm{exp}=(3.23\pm 0.42)\cdot 10^{-4}
\end{equation}
combining the results of \citer{CHEN,ABE}. For the theoretical prediction \cite{BCMU2}
\begin{equation}
\label{bsgamth}
  {\cal B}(B\to X_s\gamma)_{E_\gamma >1.6\,\mathrm{GeV}}^\mathrm{th}=(3.57\pm0.30)\cdot 10^{-4}
\end{equation}
there is an ongoing discussion about quark mass effects. The matrix elements of $Q_1$ depend at two-loop level on the mass ratio $m_c^2/ m_b^2$. Gambino and Misiak \cite{GM} argue that the running charm quark mass should be used instead of the pole mass, because the charm quarks in the loop are dominantly off-shell. Strictly speaking, this is a NNLO issue and could be used to estimate the sensitivity to NNLO corrections. Numerically, however, it increased the branching ratio by 11\%. A more conservative error estimate would rather add this shift to the theoretical error.

As $b\to s\gamma$ decays appear at one-loop level for the first time, they play an important role in indirect searches for physics beyond the Standard Model. Effects from new particles in the loops could easily be of the same order of magnitude than the Standard Model contributions. The excellent agreement of experimental measurement (\ref{bsgamexp}) and theoretical prediction (\ref{bsgamth}) therefore places severe bounds on the parameter space of New Physics scenarios, like multi-Higgs models \cite{CDGG,2HDM}, Technicolor \cite{techcol}, or the MSSM \cite{MSSM}.

The treatment of the matrix elements for {\em exclusive} $b\to s\gamma$ decays as for example $B\to K^*\gamma$ is in general more complicated. In this case, bound-state effects are essential and need to be described by non-perturbative hadronic quantities like form factors. Exactly these exclusive radiative decays of $B$ mesons are the subject of this work and will be dealt with in detail in part II and III.

% ===== QCD Factorization =================================
\chapter{Factorization}
\label{ch:fact}

The idea of factorization in hadronic decays of heavy mesons is already quite old. Let us in the following sections describe ``naive factorization'' and its extensions, state the problems and shortcomings of these models, and then introduce QCD factorization, which allows for a systematic and model-independent treatment of two-body $B$ decays. In the last section of this chapter we will shortly present some other approaches used to tackle the difficulties with these decays.

%===================================================
%=      Naive Factorization and its Offspring      =
%===================================================

\section{Naive Factorization and its Offspring}
\label{sec:naive}

For leptonic and semi-leptonic two-body decays, the amplitude can be {\em factorized} into the product of a leptonic current and the matrix element of a quark current, because gluons cannot connect quark and lepton currents. Pictorially spoken, the Feynman diagram falls apart into two simpler separate diagrams if we cut the $W$ propagator. For non-leptonic decays, however, we also have non-factorizable contributions, because gluons can connect the two quark currents and additional diagrams can contribute.

In highly energetic two-body decays, hadronization of the decay products takes place not until they have separated already. If the quarks have arranged themselves into colour-singlet pairs, low-energetic (soft) gluons cannot affect this arrangement ({\em colour transparency}) \cite{CT,DGfact}. In the {\em naive factorization} approach, the matrix element of a four-fermion operator in a heavy-quark decay is assumed to separate (``factorize'') into two factors of matrix elements of bilinear currents with colour-singlet structure \cite{fact,BSW}, e.g.
\begin{eqnarray}
\label{nfact}
  \langle D^+ \pi^-|(\bar c b)_{(V-A)}(\bar d u)_{(V-A)}|\bar B_d\rangle &\to& \langle\pi^-|(\bar d u)_{(V-A)}|0\rangle \langle D^+|(\bar c b)_{(V-A)}|\bar B_d\rangle\nonumber\\
  &\propto& f_\pi F^{B\to D}
\end{eqnarray}
In general, the complicated non-leptonic matrix elements $\langle M_1 M_2|Q_i|B\rangle$ are decomposed into a {\em form factor} $F^{B\to M_1}$ and a meson {\em decay constant} $f_{M_2}$. The factorized matrix elements of $Q_1$ and $Q_2$ are dressed with the parameters
\begin{equation}
  a_{1,2} = C_{1,2}(\mu)+\frac{C_{2,1}(\mu)}{N}
\end{equation}
respectively, to give the amplitude. In the literature one distinguishes three classes of non-leptonic two-body decays. The first class contains only $a_1$ such that the meson generated from the colour-singlet current is charged as in (\ref{nfact}). The second class involves only $a_2$ and therefore consists of those decays in which the meson generated directly from the current is neutral. The third class finally covers decays in which the $a_1$ and $a_2$ amplitudes interfere.

Already two of the prominent proponents of factorization in heavy meson decays, Dugan and Grinstein, admitted that ``at first sight, factorization is a ridiculous idea.'' \cite{DGfact}. In (\ref{nfact}) the exchange of ``non-factorizable''\footnote{We put quotes on ``non-factorizable'' if we mean the corrections to naive factorization to avoid confusion with the meaning of factorization in the context of hard processes in QCD.} gluons between the $\pi^-$ and the $(B_d D^+)$ system was completely neglected, which consequently does not allow for rescattering in the final state and for the generation of a strong phase shift between different amplitudes. Yet, the main problem of naive factorization is that it reduces the renormalization scale dependent matrix element to the form factor and decay constant, which have a rather different scale dependence. This destroys the cancellation of the scale dependence in the amplitude and is therefore unphysical. Naive factorization cannot be correct exactly and gives at most an approximation for one single suitable {\em factorization scale} $\mu_f$. The value of this particular scale is not provided by the model itself, but usually expected to be ${\cal O}(m_b)$ and ${\cal O}(m_c)$ for $B$ and $D$ decays, respectively. Another problem with naive factorization arises beyond the leading logarithmic level. Here, the Wilson coefficients become renormalization scheme dependent whereas the factorized matrix elements are renormalization scheme independent such that no cancellation of the scheme dependence in the amplitude can take place. This is unphysical again.

The concept of {\em ``generalized factorization''} tries to solve these problems by introducing non-perturbative hadronic parameters, which shall quantify the ``non-factorizable'' contributions and herewith cancel the scale and scheme dependence of the Wilson coefficients \citer{JS,genfact}. Neubert and Stech \cite{NS} replace $a_1$ and $a_2$ by
\begin{eqnarray}
  a_1^\mathrm{eff}&=&\left(C_1(\mu) +\frac{C_2(\mu)}{N}\right)[1+\varepsilon_1^{(BD,\pi)}(\mu)] +C_2(\mu)\varepsilon_8^{(BD,\pi)}(\mu)\nonumber\\ \label{a12eff}
  a_2^\mathrm{eff}&=&\left(C_2(\mu) +\frac{C_1(\mu)}{N}\right)[1+\varepsilon_1^{(B\pi,D)}(\mu)] +C_1(\mu)\varepsilon_8^{(B\pi,D)}(\mu)
\end{eqnarray}
Due to the aforementioned renormalization scheme dependence, however, it is possible to find for any chosen scale $\mu_f$ a renormalization scheme in which the non-perturbative parameters $\varepsilon_{1} (\mu_f)$ and $\varepsilon_{8} (\mu_f)$ vanish simultaneously \cite{BS1}. This leads back to naive factorization instead of describing the ``non-factorizable'' contributions to non-leptonic decays properly. Another variant of generalized factorization achieves the scale and scheme independence via effective Wilson coefficient functions and an effective number of colours \cite{genfact}:
\begin{equation}
  a_{1,2}^\mathrm{eff}=C_{1,2}^\mathrm{eff}+\frac{1}{N^\mathrm{eff}}C_{2,1}^\mathrm{eff}
\end{equation}
Again, these effective parameters are nicely renormalization scale and scheme independent. Yet, Buras and Silvestrini \cite{BS1} showed that they depend on the gauge and the infrared regulator and are thus unphysical.

All these shortcomings are resolved in the {\em QCD factorization} approach.

%================================
%=      QCD Factorization       =
%================================

\section{QCD Factorization}
\label{sec:fact}

The typical scale for a $B$ meson decay is of order $m_B$ and therefore much larger than $\Lambda_\mathrm{QCD}$, the long-distance scale where non-perturbative QCD takes over. QCD factorization, introduced by Beneke, Buchalla, Neubert, and Sachrajda,  uses exactly this fact \cite{BBNS}. This time, something is factorized in the general sense of QCD applications: Namely the long-distance dynamics in the matrix elements and the short-distance interactions that depend only on the large scale $m_B$. And again, the short-distance contributions can be computed in a perturbative expansion in the strong coupling $\alpha_s\left({\cal O}(m_b)\right)$. The long-distance part has still to be computed non-perturbatively or determined experimentally. However, these non-perturbative parameters are mostly simpler in structure than the original matrix element and they are process independent.

%=      The factorization formula       =
\subsection{The factorization formula}
\label{subsec:factform}
\begin{figure}
  \begin{center}  
    \input{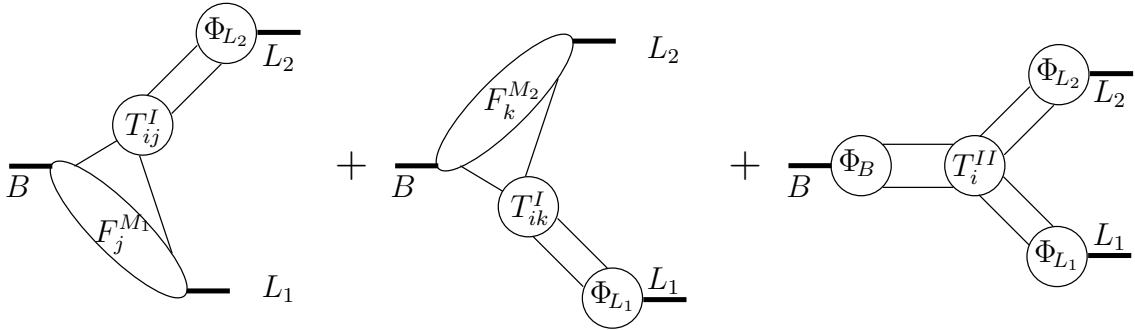}
  \end{center}
\caption{Graphical representation of the factorization formula for a $B$ meson decaying into two light mesons, e.g. $B^-\to\pi^0 K^-$. At leading order in $\Lambda_\mathrm{QCD}/m_b$ there are no long-distance interactions between the system of $B$ meson and the meson that picks up the spectator quark, and the other final state meson.\label{fig:fact}}
\end{figure}

Let us here explain QCD factorization for exclusive non-leptonic two-meson $B$ decays $B\to M_1 M_2$ following \cite{BBNS}. Applications of these methods to exclusive radiative decays will be the subject of the following two parts. We consider $B\to M_1 M_2$ in the heavy-quark limit and differentiate between decays into final states containing a heavy and a light meson $H_1$ and $L_2$ or two light mesons $L_1$ and $L_2$. A meson is called ``light'' if its mass $m$ remains finite and ``heavy'' if its mass scales with $m_b$ such that $m/m_b$ stays fixed in the heavy-quark limit $m_b\to\infty$. Up to power corrections of order $\Lambda_\mathrm{QCD}/m_b$, for the transition matrix element of an operator $Q_i$ in the weak effective Hamiltonian the following {\em factorization formula} holds:
\begin{eqnarray}\label{ffll}
  \langle L_1 L_2|Q_i|\bar{B}\rangle &=& \;\;\;\,\sum_j F_j^{B\to L_1}(m_2^2)\,\int_0^1 \!du\,T_{ij}^I(u)\,\Phi_{\!L_2}(u)\nonumber\\
  &&+ \sum_k F_k^{B\to L_2}(m_1^2)\,\int_0^1 \!dv\,T_{ik}^I(v)\,\Phi_{\!L_1}(v)\nonumber\\
  &&+\,\int_0^1 \!d\xi du dv \,T_i^{II}(\xi,u,v)\,\Phi_{\!B}(\xi)\,\Phi_{\!L_1}(v)\,\Phi_{\!L_2}(u)\\[0.2cm] \label{ffhl}
  \langle H_1 L_2|Q_i|\bar{B}\rangle &=&  \;\;\;\,\sum_j F_j^{B\to H_1}(m_2^2)\,\int_0^1 \!du\,T_{ij}^I(u)\,\Phi_{\!L_2}(u)
\end{eqnarray}
$F_j^{B\to M}$ denotes a $B\to M$ {\em form factor} and $\Phi_M$ is the {\em light-cone distribution amplitude} (LCDA) for the quark-antiquark Fock state of meson $M$. These non-perturbative quantities are much simpler than the original non-leptonic matrix element. The LCDA $\Phi_M$ reflects universal properties of a single meson state and the form factors refer only to a relatively simple $B\to$ meson transition matrix element of a local current. Both can be calculated using some non-perturbative technique, like lattice QCD or QCD sum rules, or they can be obtained from experimental results. The hard-scattering functions $T_{ij}^I(u)$, $T^I_{ik}(v)$ and $T_i^{II}(\xi,u,v)$ are perturbatively calculable functions of the light-cone momentum fractions $u$, $v$ and $\xi$ of the quarks inside the final state mesons and the $B$ meson, respectively. We distinguish ``type I'' or ``hard vertex'' and ``type II'' or ``hard spectator'' contributions. The idea that the decay amplitude can be expressed as a convolution of a hard-scattering factor with light-cone wave functions of the participating mesons is analogous to more familiar applications of this method to hard exclusive reactions involving only light hadrons \cite{BL}. A graphical representation of (\ref{ffll}) is given in Fig.~\ref{fig:fact}.
%%%%%%%%%%%%%%%%%%%%%%%%%%%%%%%%%%%%%%%%%%%%%%%%%%%%%%%%%%%%%%%%%%%
%%%%%%%%%%%%%%%%%%%%%%%%%%%%%%%%%%%%%%%%%%%%%%%%%%%%%%%%%%%%%%%%%%%

When the spectator quark in the $B$ meson goes to a heavy meson as in $\bar B_d\to D^+ \pi^-$, the spectator interaction is power suppressed in the heavy-quark limit and we arrive at the simpler equation (\ref{ffhl}). For the opposite situation where the spectator quark goes to a light meson, but the other meson is heavy, factorization does not hold, because the heavy meson is neither fast nor small and cannot be factorized from the $B\to M_1$ transition. Annihilation topologies do not contribute at leading order in the heavy-quark expansion.

The hard spectator interactions appear at ${\cal O}(\alpha_s)$ for the first time. Since at ${\cal O}(\alpha_s^0)$ the functions $T^I_{ij/k}$ are independent of $u$ and $v$, the convolution integral results in a meson decay constant and we see that (\ref{ffll}) and (\ref{ffhl}) reproduce naive factorization. We now, however, can systematically compute radiative corrections to naive factorization. This immediately solves the problem of scale and scheme dependence in the conventional factorization approaches. Because the form factors are real quantities, all strong rescattering phases are either generated perturbatively or are power suppressed.

In principle, QCD factorization is nothing else but a consistent formalization and generalization of Bjorken's colour transparency argument \cite{CT}. This is most obvious for the decay $\bar B_d\to D^+ \pi^-$. The spectator quark and the other light degrees of freedom inside the $B$ meson can easily form a $D$ meson after the weak $b\to c$ decay. This $B\to D$ transition can be parametrized by a set of form factors. The other two light quarks are very energetic. To form a pion they must be highly collinear and in a colour-singlet configuration, the probability of which is described by the leading-twist pion light-cone distribution amplitude. Such an energetic ``colour-transparent'' compact object can leave the decay region without interfering with the $D$ meson formation because soft interactions decouple.

%=      The non-perturbative input       =
\subsection{The non-perturbative input}
\label{subsec:nonpert}
We will now discuss the necessary non-perturbative input, namely form-factors and light-cone distribution amplitudes, in more detail.

\subsubsection*{Form factors}
A {\em form factor} is a function of scalar variables accompanying the independent terms in the most general decomposition of the matrix element of a current consistent with Lorentz and gauge invariance. In the literal meaning a form factor goes with the matrix element of some current where initial and final state are one and the same particle. In the non-relativistic limit a particle has two electromagnetic form factors, which are simply the Fourier transforms of its charge and magnetic moment distributions. They therefore indeed give information on the {\em form} of the particle under consideration.

In the context of QCD factorization we often need the matrix element of the vector current which is conventionally parametrized by two scalar form factors
\begin{equation}
  \langle P(k)|\bar{q}\gamma^\mu b|\bar{B}(p)\rangle = F_+^{B\to P}(q^2)\,(p^\mu+{k}^\mu) + \Big[F_0^{B\to P}(q^2)-F_+^{B\to P}(q^2)\Big]\,\frac{m_B^2-m_P^2}{q^2}\,q^\mu
\end{equation}
where $q=p-k$. Strictly speaking $F^{B\to P}_{0/+}$ are {\em transition} form factors, which do not describe the form of neither the $B$ meson nor the pseudoscalar meson $P$, but rather their overlap during the weak decay. For $q^2=0$ the two form factors coincide, $F_+^{B\to P}(0)=F_0^{B\to P}(0)$. The scaling behaviour of the form factors is
\begin{eqnarray}
  F_{0/+}^{B\to D}(0)              &\sim& 1\\[0.2cm]
  F_{0/+;\:\mathrm{hard}}^{B\to\pi}(0) &\sim& \alpha_s\left(\sqrt{m_b\Lambda_\mathrm{QCD}}\right)\left(\frac{\Lambda_\mathrm{QCD}}{m_b}\right)^{3/2}\\
  F_{0/+;\:\mathrm{soft}}^{B\to\pi}(0) &\sim& \left(\frac{\Lambda_\mathrm{QCD}}{m_b}\right)^{3/2}
\end{eqnarray}
for the $B\to D$ and $B\to\pi$ transition, respectively \cite{BBNS,CZ}. Both the heavy-to-heavy and heavy-to-light $B\to M$ form factors receive a leading contribution from soft gluon exchange. This is why the form factor has to enter the factorization formula as a non-perturbative input.

\subsubsection*{Light-cone distribution amplitudes for light mesons}
Let us now consider the other non-perturbative ingredients which are the meson light-cone distribution amplitudes. We define {\em light-cone} components
\begin{equation}\label{vlcc}
  k_\pm =\frac{k^0\pm k^3}{\sqrt{2}}
\end{equation}
for any four vector $k^\mu=(k_+,k_-,\vec k_{\!\perp})$. These variables naturally distinguish between a particle's longitudinal and transverse degrees of freedom. Let us construct a light pseudoscalar meson out of the on-shell constituent quarks in a spin singlet state and with no net transverse momentum:
\begin{equation}
  |P(k)\rangle = \int\!\frac{dv}{\sqrt{v\bar v}} \frac{d^2l_{\perp}}{16\pi^3} \frac{1}{\sqrt{2}}\left(a^\dagger_{l_1\uparrow} b^\dagger_{l_2\downarrow}-a^\dagger_{l_1\downarrow} b^\dagger_{l_2\uparrow}\right)|0\rangle \,\psi(v,l_{\perp})
\end{equation}
where $a^\dagger_{l\uparrow}$ ($b^\dagger_{l\uparrow}$) creates a (anti)quark with momentum $l$ and spin up. Their momentum shall be a fraction $v$ and $\bar v$ of the meson momentum $k$: $l_1^+=v k^+$ and $l_2^+=\bar v k^+$ plus a momentum $l_\perp$ perpendicular to the meson momentum direction, which adds up to zero for both quarks. Here and in the following we use the short-hand notation
\begin{equation}
  \bar v\equiv 1-v
\end{equation}
As long as $v\bar v={\cal O}(1)$, i.e. not at the endpoints of the spectrum, $l_\perp^2/v\bar v$ can be neglected compared to $E^2$ for a meson with large energy $E$. If we are interested in the leading twist contributions only, we can perform the $l_\perp$ integration and define, with $f_P$ being the meson decay constant, via
\begin{equation}\label{phiPdef}
  \int\!\frac{d^2l_{\perp}}{16\pi^3} \frac{1}{\sqrt{2N}} \psi(v,l_\perp) \equiv -\frac{i f_P}{4N} \Phi_P(v)
\end{equation}
the {\em light-cone wave function} $\Phi_P$. The latter is normalized as $\int_0^1\!dv \Phi_P(v)=1$ and has the asymptotic form $\Phi_P(v)=6v\bar v$. Using the explicit expressions for Dirac spinors we can then write a light pseudoscalar meson via its {\em light-cone distribution amplitude} in position space for light-like separations of the constituent quarks as
\begin{equation}\label{PLCDA}
  \langle P(k)|q'_i(z)_\alpha \bar q_j(0)_\beta |0\rangle =\frac{i f_P}{4N}\delta_{ij}(\gamma_5\!\not\! k)_{\alpha\beta} \int_0^1\! dv e^{i\bar vk\cdot z}\Phi_P(v)
\end{equation}
with $i$ and $j$ colour and $\alpha$ and $\beta$ spinor indices. In the following we will need also the leading twist LCDA for a vector meson with polarization vector $\eta$ \cite{BB96,BB98}:
\begin{eqnarray}\label{VLCDA}
  \langle V(k,\eta)|q'_i(z)_\alpha \bar q_j(0)_\beta |0\rangle &=& -\frac{f^\perp_V}{8N}\delta_{ij}\left[\not\!\eta,\not\!k\right]_{\alpha\beta} \int_0^1\! dv\, e^{i\bar vk\cdot z}\Phi_V^\perp(v)\\
  &&-\frac{f_V m_V}{4N}\delta_{ij}\left[i\!\not\!k_{\alpha\beta}\, \eta\!\cdot\! z\int_0^1\! dv\,e^{i\bar vk\cdot z}\Phi_V^\parallel(v)\right.\nonumber\\
  &&\qquad\qquad\quad\, +\not\!\eta_{\alpha\beta} \int_0^1\! dv e^{i\bar vk\cdot z} g_V^{\perp\,(v)}(v)\nonumber\\
  &&\qquad\qquad\quad\, \left.-\frac{1}{4} \left(\varepsilon^{\kappa\lambda\mu\nu}\eta_\kappa k_\lambda z_\mu \gamma_\nu \gamma_5\right)_{\alpha\beta} \int_0^1\! dv  e^{i\bar vk\cdot z} g_V^{\perp\,(a)}(v)\right]\nonumber
\end{eqnarray}
The light-cone wave function $\Phi_V^\perp$ has an expansion in terms of Gegenbauer polynomials $C^{3/2}_n(2v-1)$
\begin{equation}\label{phiperpgbp}
  \Phi_V^\perp(v)=6v(1-v)\left[ 1+\sum^\infty_{n=1} \alpha^V_n(\mu) C^{3/2}_n(2v-1)\right]
\end{equation}
where $C^{3/2}_1(x)=3x$, $C^{3/2}_2(x)=\frac{3}{2}(5x^2-1)$, etc. The Gegenbauer moments $\alpha^\perp_n(\mu)$ are multiplicatively renormalized. They vanish logarithmically as the scale $\mu\to\infty$. In this limit, $\Phi_V^\perp$ reduces to its asymptotic form $\Phi_V^\perp(v)=6v\bar v$, which often is a reasonable first approximation. The remaining leading-twist light-cone wave functions for light vector mesons, $\Phi_V^\parallel$, $g_V^{\perp\,(v)}$, and $g_V^{\perp\,(a)}$ do not contribute at leading power if the mesons are transversely polarized. An example of how to use the light-cone distribution amplitudes in an actual calculation will be given in the following subsection~\ref{subsec:Fggpi}.

We conclude the discussion of light meson LCDAs with the counting rules for the wave functions. Using the asymptotic form $\Phi_X(v)=6v\bar v$ for $X=P$ and $X=\perp$ we count the {\em endpoint region,} where $v$ or $\bar v$ is of order $\Lambda_\mathrm{QCD}/m_b$, as order $\Lambda_\mathrm{QCD}/m_b$. Away from the endpoint the wave function is ${\cal O}(1)$:
\begin{equation}\label{philightcount}
  \Phi_X(v)\sim\left\{\begin{array}{cl} \Lambda_\mathrm{QCD}/m_b & \mbox{for }v,\,\bar v\sim \Lambda_\mathrm{QCD}/m_b\\
   1 & \mbox{for }v\mbox{ away from endpoint}\end{array}\right.
\end{equation}

\subsubsection*{$B$ meson light-cone distribution amplitudes}
The $B$-meson light-cone distribution amplitude appears in the hard-spectator interaction term in (\ref{ffll}), because a hard gluon can probe the momentum distribution of $b$ and spectator quark inside the $B$ meson. Compared to the light meson LCDAs we now have one very heavy quark as constituent: The $b$ quark carries the largest part of the $B$ mesons momentum $p$: $p_b^+ =\bar\xi p^+ \approx p^+$, whereas for the spectator quark we have $l^+ =\xi p^+$ with $\xi ={\cal O}(\Lambda_\mathrm{QCD}/m_b)$. From the explicit expression for a heavy quark and light antiquark spinor we can pull out a factor $(\not\! p+m_B)\gamma_5$. In the $B$-meson rest frame the remainder is essentially proportional to $\not\! l$ since the spectator quark is neither energetic nor heavy, and thus no restrictions on the components of its spinor exist. However, we still want to perform the $l_\perp$ integration and absorb it into the $B$-meson light-cone wave functions. To this end we decompose $\not\! l$ into a part proportional to $\not\! n_+$ and $\not\! n_-$ where $n$ is an arbitrary light-like vector which can be chosen for example in the direction of one of the final state particle momenta. Our standard choice is $n=n_-=(1,0,0,-1)$. For this choice of $n$ we define two scalar wave functions $\Phi_{B1/2}$ in the $B$ meson light-cone distribution amplitude as \cite{BBNS,GN,BF}:
\begin{equation}\label{BLCDA}
  \langle 0|b_i(0)_\alpha \bar q_j(z)_\beta|\bar B(p)\rangle = \frac{i f_B}{4N} \delta_{ij} \left[(\not\! p + m_b)\gamma_5\right]_{\alpha\gamma} \int^1_0 d\xi\, e^{-i\xi p_+ z_-}[\Phi_{B1}(\xi)+\not\! n \Phi_{B2}(\xi)]_{\gamma\beta}
\end{equation}
This is the most general decomposition of the leading-power LCDA only if the transverse momentum of the spectator quark $l_\perp$ can be neglected. The wave functions describe the distribution of light-cone momentum fraction $\xi=l_+/p_+$ of the spectator quark with momentum $l$ inside the $B$ meson. They depend on the choice of $n$ and are normalized as
\begin{equation}\label{phi12norm}
  \int^1_0 d\xi\, \Phi_{B1}(\xi)=1 \qquad \int^1_0 d\xi\, \Phi_{B2}(\xi)=0
\end{equation}
Unfortunately, the $B$-meson wave functions are poorly known, even theoretically. At scales much larger than $m_b$ they should approach a symmetric form as for light mesons. At scales of order $m_b$ and smaller, however, one expects the distribution to be highly asymmetric with $\xi={\cal O}(\Lambda_\mathrm{QCD}/m_b)$. 

What we will need later on besides the normalization conditions is the first negative moment  of $\Phi_{B1}(\xi)$, which we parametrize by a quantity $\lambda_B={\cal O}(\Lambda_{QCD})$, i.e.
\begin{equation}\label{lambdef}
  \int^1_0 d\xi \frac{\Phi_{B1}(\xi)}{\xi}=\frac{m_B}{\lambda_B}
\end{equation}

The counting for $B$ mesons is different from the counting for light mesons. We use the normalization condition (\ref{phi12norm}) to get
\begin{equation}\label{phiBcount}
  \Phi_{B1}(\xi)\sim\left\{\begin{array}{cl} m_b/\Lambda_\mathrm{QCD} & \mbox{for }\xi\sim \Lambda_\mathrm{QCD}/m_b\\
  0 & \mbox{for }\xi\sim 1\end{array}\right.
\end{equation}
This represents the fact that it is practically impossible to find the light spectator quark with momentum of order $m_b$, because there is only a small probability for hard fluctuations that transfer a large momentum to the spectator. The same counting applies to other heavy mesons.

%=      A simple application       =
\subsection{A simple application: $F_{\gamma\gamma\pi^0}$}
\label{subsec:Fggpi}
\begin{figure}
  \begin{center}  
    \input{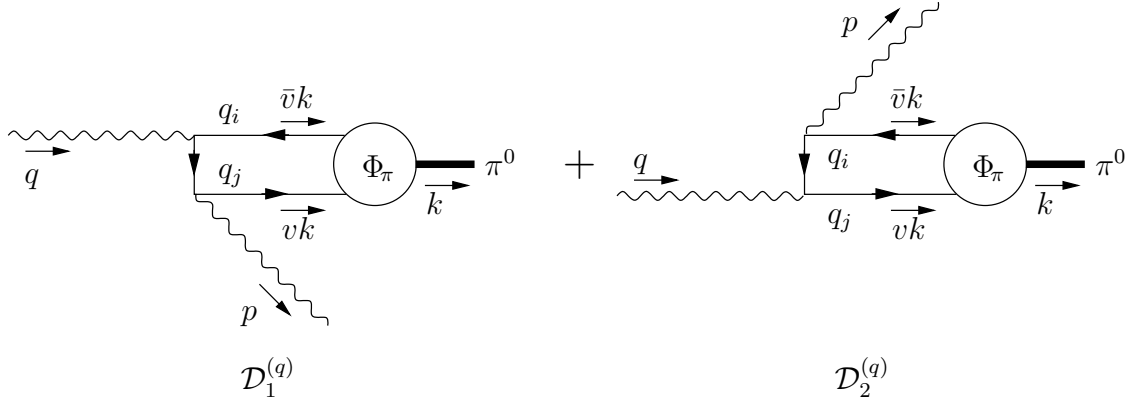}
  \end{center}
\caption{The Feynman diagrams contributing to the $\gamma^* \to \pi^0 +\gamma$ form factor.\label{fig:Fggpi}}
\end{figure}

We now want to apply the pion light-cone distribution amplitude in the calculation of the $\gamma^* \to \pi^0 +\gamma$ form factor. At lowest order in the hard scattering this is a pure QED process where the Feynman diagrams in Fig.~\ref{fig:Fggpi} contribute.
%%%%%%%%%%%%%%%%%%%%%%%%%%%%%%%%%%%%%%%%%%%%%%%%%%%%%%%%%%%%%%%%%%%
%%%%%%%%%%%%%%%%%%%%%%%%%%%%%%%%%%%%%%%%%%%%%%%%%%%%%%%%%%%%%%%%%%%
We assign momenta $vk$ and $\bar vk$ to the quark and antiquark in the outgoing pion, respectively. The Feynman rules in momentum space give for on-shell final-state particles
\begin{eqnarray}
  {\cal D}^{(q)} &=&\bar q_j(vk)_\beta\left[(-ieQ_q \!\not\!\epsilon)\frac{i\delta_{ji}}{\not\! q -\bar v \!\not\! k -m_q}(-ie Q_q\gamma_\nu) \right. \nonumber\\
  &&\left.\qquad\quad +(-ieQ_q \gamma_\nu)\frac{i\delta_{ji}}{v \!\not\!k-\not\!q-m_q} (-ieQ_q \!\not\!\epsilon)\right]_{\beta\alpha}q_i(\bar vk)_\alpha\nonumber\\ \label{Dq}
  &=& iQ_q^2 e^2 \delta_{ji} q_i(\bar vk)_\alpha \bar q_j(vk)_\beta\left[\frac{\not\!\epsilon (\not\! q-\bar v \!\not\! k)\gamma_\nu}{vq^2} +\frac{\gamma_\nu(v \!\not\! k-\not\! q)\not\!\epsilon}{\bar v q^2}\right]_{\beta\alpha}
\end{eqnarray}
where we neglected the quark and pion mass and simply rearranged the quark fields to enable the replacement
\begin{equation}\label{replacepion}
  q_i(\bar vk)_\alpha\bar q_j(vk)_\beta \longrightarrow \frac{i f_\pi}{4N}\delta_{ij} \int_0^1\! dv (\gamma_5\!\not\! k)_{\alpha\beta} \Phi_\pi(v)
\end{equation}
according to (\ref{PLCDA}) for $q=u$ or $q=d$. We perform the trace in (\ref{Dq}) and use the phase convention
\begin{equation}
  \pi^0 =\frac{u\bar u-d \bar d}{\sqrt{2}}
\end{equation}
for the neutral pion. This leads to
\begin{equation}
  {\cal D}=\frac{{\cal D}^{(u)} -{\cal D}^{(d)}}{\sqrt{2}} =-i e^2 \varepsilon^{\kappa\lambda\mu\nu}k_\kappa q_\lambda \epsilon_\mu \gamma_\nu \underbrace{\frac{f_\pi}{3\sqrt{2}} \frac{1}{-q^2} \int_0^1\! dv \frac{\Phi_\pi(v)}{v \bar v}}_{\displaystyle \equiv F_{\gamma\gamma\pi^0}}
\end{equation}
which defines the neutral pion form factor $F_{\gamma\gamma\pi^0}$. For the asymptotic form $\Phi_\pi(v)=6v\bar v$ of the pion wave function we obtain its asymptotic value
\begin{equation}
  F_{\gamma\gamma\pi^0} =\sqrt{2} f_\pi \frac{1}{-q^2}
\end{equation}
If one measures the form factor experimentally one can get information on the pions valence quark distribution.

%=      One-loop proof of factorization       =
\subsection{One-loop proof of factorization in $B\to D\pi$}\label{subsec:1loopproof}
In order to prove the validity of the factorization formulas (\ref{ffll}) and (\ref{ffhl}), one has to show that the hard scattering amplitudes $T^I_{ij/k}$ and $T_i^{II}$ are free of infrared singularities to all orders in perturbation theory. For heavy-light final states such a proof exists at two loops \cite{BBNS}. It has subsequently been extended to all orders \cite{BPS}. We here only want to sketch the proof of factorization for $B\to D\pi$ at the one-loop order as given in the second reference in \cite{BBNS}. 

\subsubsection*{The leading order}
\begin{figure}
  \begin{center}
    \psfig{figure=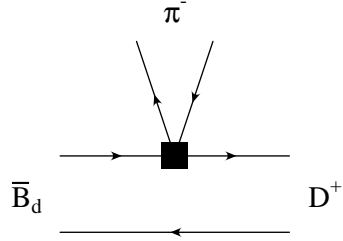}
  \end{center}
  \caption{The leading-order contribution to the hard-scattering kernel $T^I_{ij}$ in $\bar B_d\to D^+\pi^-$. The pion's momentum is $k$.\label{fig:factLO}}
\end{figure}

The relevant quark level decay for $\bar B_d\to D^+\pi^-$ is $b\to c\bar u d$ so that no penguin operators or diagrams can contribute. At lowest order there is only a single diagram with no hard gluon interactions, shown in Fig.~\ref{fig:factLO}.
%%%%%%%%%%%%%%%%%%%%%%%%%%%%%%%%%%%%%%%%%%%%%%%%%%%%%
%%%%%%%%%%%%%%%%%%%%%%%%%%%%%%%%%%%%%%%%%%%%%%%%%%%%%
The spectator quark is soft and absorbed as a soft quark by the recoiling $D$ meson which is described by the $B\to D$ form factor. The hard subprocess is just given by the insertion of the colour-singlet four-fermion operator $O_\mathrm{singlet}=\bar c\gamma^\mu (1-\gamma_5)b \, \bar d \gamma_\mu (1-\gamma_5) u$. Therefore, it does not depend on the longitudinal momentum fraction $v$ of the two quarks that form the emitted $\pi^-$. Consequently, with $T^I_{ij}(v)$ being $v$-independent, the $v$-integral reduces to the normalization condition for the pion wave function. So (\ref{ffhl}) reproduces the result of naive factorization if we neglect gluon exchange. In the heavy-quark limit the decay amplitude scales as
\begin{equation}\label{BDpicount}
  A(\bar B_d\to D^+ \pi^-) \sim G_F m_b^2 F^{B\to D}(0) f_\pi \sim G_F m_b^2 \Lambda_\mathrm{QCD}
\end{equation}

\subsubsection*{Factorizable contributions}
\begin{figure}
  \begin{center}
    \psfig{figure=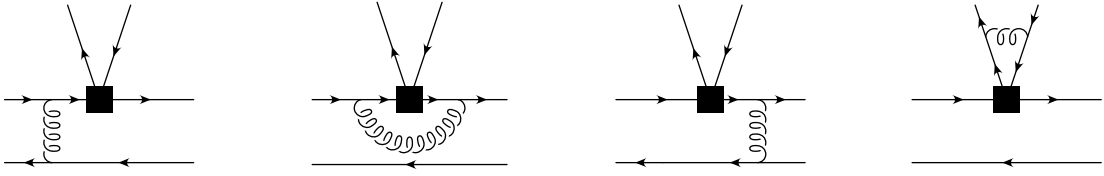}
  \end{center}
  \caption{Factorizable diagrams at ${\cal O}(\alpha_s)$ that are part of the form factor (first three diagrams) or renormalize the conserved light-quark $V-A$ current (last diagram).\label{fig:factNLOnocalc}}
\end{figure}

We now have to show that radiative corrections to Fig.~\ref{fig:factLO} are either suppressed by $\alpha_s$ or $\Lambda_\mathrm{QCD}/m_b$ or are already contained in the definition of the form factor or the decay constant of the pion. For example the first three diagrams in Fig.~\ref{fig:factNLOnocalc}
%%%%%%%%%%%%%%%%%%%%%%%%%%%%%%%%%%%%%%%%%%%%%%%%%%%%%
%%%%%%%%%%%%%%%%%%%%%%%%%%%%%%%%%%%%%%%%%%%%%%%%%%%%%
are part of the form factor and do not contribute to the hard-scattering kernels. In particular, the leading contributions from the region in which the gluon is soft (first and second diagram in Fig.~\ref{fig:factNLOnocalc}) are absorbed into the physical form factor. This is the one that appears in (\ref{ffll}) and (\ref{ffhl}) and is also the one directly measured in experiments. The fourth diagram in Fig.~\ref{fig:factNLOnocalc} simply renormalizes the conserved $(\bar ud)$ light-quark $V\!-\!A$ current and cancels the wave-function renormalization of the quarks in the emitted pion.

\subsubsection*{``Non-factorizable'' vertex corrections}
The diagrams containing gluon exchanges that do not belong to the form factor for the $B\to D$ transition or the pion decay constant are called ``non-factorizable''.

\begin{figure}
  \begin{center}
    \psfig{figure=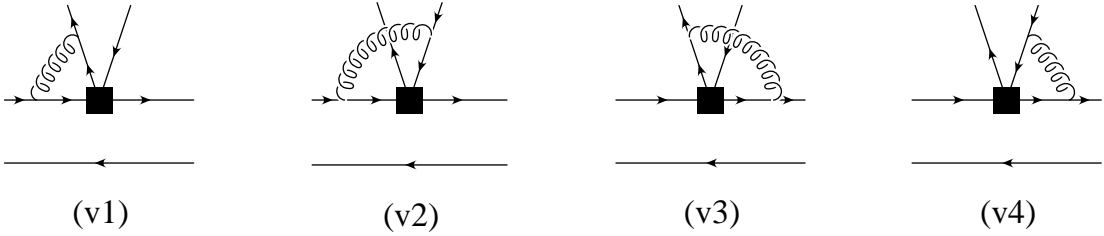}
  \end{center}
  \caption{``Non-factorizable'' vertex corrections.\label{fig:factNLOvertex}}
\end{figure}

The vertex corrections of Fig.~\ref{fig:factNLOvertex}
%%%%%%%%%%%%%%%%%%%%%%%%%%%%%%%%%%%%%%%%%%%%%%%%%%%%%
%%%%%%%%%%%%%%%%%%%%%%%%%%%%%%%%%%%%%%%%%%%%%%%%%%%%%
violate the naive factorization ansatz (\ref{nfact}). Nevertheless they are calculable and form an ${\cal O}(\alpha_s)$ correction to the hard-scattering kernel $T^I_{ij}(u)$. Although the diagrams in Fig.~\ref{fig:factNLOvertex} exhibit seperately a collinear and soft infrared divergence, we can show that these divergences cancel in the sum of all four diagrams. The cancellation of the soft divergences is a manifestation of Bjorken's colour-transparency argument and the collinear divergences cancel due to collinear Ward identities. So the sum of the four diagrams involves only hard-gluon exchange at leading power.

At ${\cal O}(\alpha_s)$ the only vertex correction in $\bar B_d \to D^+ \pi^-$ comes from the colour-octet operator $O_\mathrm{octet}=\bar c\gamma^\mu (1-\gamma_5) T^a b \, \bar d \gamma_\mu (1-\gamma_5) T^a u$. We choose the quark and antiquark momentum in the pion as
\begin{equation}\label{l1l2}
  l_1 =vk+l_\perp +\frac{\vec l_\perp^2}{4vE}n_- \qquad\quad l_2 =vk-l_\perp +\frac{\vec l_\perp^2}{4\bar vE}n_-
\end{equation}
with $k=E(1,0,0,1)$ the pion momentum, $E=p_b\cdot k/m_B$ the pion energy, and $n_-=(1,0,0,-1)$. We want to show that the transverse momentum $l_\perp$ can be neglected at leading power and that the contributions from the soft-gluon region and from gluons collinear to $k$ are power suppressed. As we have to keep $l_\perp$ for the moment we use (\ref{PLCDA}) before the $l_\perp$ integration in (\ref{phiPdef}):
\begin{equation}
  \langle \pi(k)|u(z)_\alpha \bar d(0)_\beta |0\rangle =\int dv \frac{d^2 l_\perp}{16\pi^3} \frac{1}{\sqrt{2N}} \psi^*(v,l_\perp) (\gamma_5\!\not\! k)_{\alpha\beta} e^{i l_1\cdot z}
\end{equation}
By this means the diagrams in Fig.~\ref{fig:factNLOvertex} give for the $O_\mathrm{octet}$ insertion
\begin{eqnarray}\label{o8NLO}
  \lefteqn{\langle D^+\pi^-|O_\mathrm{octet}|\bar B_d\rangle_{{\cal O}(\alpha_s)}=}\\
  &&=i g_s^2\frac{C_F}{2}\int \! \frac{d^4q}{(2\pi)^4}\langle D^+|\bar c A_{1}(q) b|\bar B_d\rangle\frac{1}{q^2}\int^1_0 \!du\frac{d^2l_\perp}{16\pi^3}\frac{\psi^*(v,l_\perp)}{\sqrt{2 N}}\,{\rm tr}[\gamma_5\!\not\! k A_2(l_1,l_2,q)] \nonumber
\end{eqnarray}
where
\begin{eqnarray}\label{a1q}
  A_{1}(q)&=&\frac{\gamma^\lambda(\not\! p_c-\not\! q+m_c)\Gamma_\mu}{2p_c\cdot q-q^2}-\frac{\Gamma_\mu(\not\! p_b+\not\! q+m_b)\gamma^\lambda}{2p_b\cdot q+q^2}\\ \label{a2q}
  A_2(l_1,l_2,q)&=&\frac{\Gamma^\mu(\not\! l_2+\not\! q)\gamma_\lambda}{2l_2\cdot q+q^2}-\frac{\gamma_\lambda(\not\! l_1+\not\! q)\Gamma^\mu}{2l_1\cdot q+q^2}
\end{eqnarray}
Here, $\Gamma_\mu=\gamma_\mu(1-\gamma_5)$, and $p_b$ and $p_c$ are the momenta of the $b$ and $c$ quark, respectively. We now show that the integral over $q$ does not contain infrared divergences at leading power in $\Lambda_\mathrm{QCD}/m_b$.

In the soft region all components of $q$ become small simultaneously, which we describe by the scaling $q\sim\lambda$. Power counting then shows that each diagram in Fig.~\ref{fig:factNLOvertex} is logarithmically divergent. But exactly the smallness of $q$ allows the following simplification of $A_2$:
\begin{equation}
  A_2(l_1,l_2,q)=\frac{\bar v k_\lambda}{\bar v k\cdot q}\Gamma^\mu -\frac{v k_\lambda}{v k\cdot q}\Gamma^\mu +{\cal O}(\lambda)={\cal O}(\lambda)
\end{equation}
where we used that a $\not\! k$ to the extreme left or right of an expression gives zero due to the on-shell condition for the external quark lines. So the soft contribution is of relative order $\Lambda_\mathrm{QCD}/m_b$ or smaller and hence suppressed relative to the hard contribution.

For $q$ collinear with the light-cone momentum $k$ of the pion we have the scaling
\begin{equation}
  q^+\sim\lambda^0,  \qquad  q_\perp\sim\lambda,  \qquad q^-\sim\lambda^2
\end{equation}
so that
\begin{equation}
  d^4 q \sim\lambda^4,  \qquad  k\cdot q \sim \lambda^2,  \qquad  q^2 \sim\lambda^2
\end{equation}
The divergence is again logarithmic. Substituting $q=\alpha k+\ldots$ we simplify $A_2$ as
\begin{equation}
  A_2(l_1,l_2,q)=\frac{2(\bar v+\alpha)k_\lambda\Gamma^\mu}{2 l_2\cdot q+q^2}-\frac{2(v+\alpha)k_\lambda\Gamma^\mu}{2 l_1\cdot q+q^2}\propto k_\lambda
\end{equation}
using again the on-shell relations. To leading power, $A_2$ is proportional to $k_\lambda$ in the collinear region. Therefore, we obtain with $k^2=0$
\begin{equation}
  k_\lambda A_1\sim \frac{\not\! k(\not\! p_c+m_c)\Gamma}{2\alpha p_c\cdot k}-\frac{\Gamma(\not\! p_b+m_b)\!\not\! k}{2\alpha p_b\cdot k}=0
\end{equation}
This completes the proof of the absence of infrared divergences at leading power in the hard-scattering kernel for $\bar B_d \to D^+ \pi^-$ to one-loop order. For the proof at two-loop order one can similarly analyze the Feynman integrands corresponding to the diagrams in momentum configurations that can give rise to singularities. This proof can be found in \cite{BBNS} for $B\to D\pi$. A simpler two-loop order proof will be performed in part III for the one-particle irreducible contributions in $B\to\gamma\gamma$.

As $|l_\perp|\ll E$ we can expand $A_2$ in $|l_\perp|/E$, which amounts to simply neglecting $l_\perp$ altogether. Then $l_1=vk$ and $l_2=\bar vk$, and we again can use the $l_\perp$ integrated pion wave function (\ref{phiPdef}) to simplify the matrix element of $O_\mathrm{octet}$ in (\ref{o8NLO}) to
\begin{eqnarray}
  \lefteqn{\langle D^+\pi^-|O_\mathrm{octet}|\bar B_d\rangle_{{\cal O}(\alpha_s)}=} \nonumber\\
  &&=-g_s^2\frac{C_F}{8N}\int\!\frac{d^4q}{(2\pi)^4} \langle D^+|\bar c A_1(q) b|\bar B_d\rangle \frac{1}{q^2} f_\pi \int_0^1\! dv \Phi_\pi(v) {\rm tr}[\gamma_5 \!\not\! k A_2(vk,\bar vk,q)]\\
  &&\propto F^{B\to D}(0) \int_0^1\! dv \,T_\mathrm{octet}(v,z)\Phi_\pi(v)
\end{eqnarray}
where $z=m_c/m_b$, $F^{B\to D}(0)$ is the form factor that parametrizes the $\langle D^+|c [\ldots]b| \bar B_d\rangle$ matrix element, and $T_\mathrm{octet}(v,z)$ is the hard-scattering kernel whose epxlicit expression can be found in \cite{BBNS}.

\subsubsection*{Penguin diagrams}
\begin{figure}
  \begin{center}
    \psfig{figure=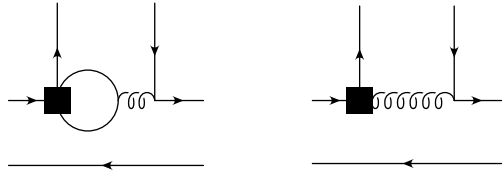}
  \end{center}
  \caption{Penguin diagram and chromomagnetic dipole operator contribution to $B\to M_1 M_2$.\label{fig:factpen}}
\end{figure}

In general, a non-leptonic $B$ decay can also get contributions from penguin diagrams and the chromomagnetic dipole operator $Q_8$ as depicted in Fig.~\ref{fig:factpen}
%%%%%%%%%%%%%%%%%%%%%%%%%%%%%%%%%%%%%%%%%%%%%%%%%%%%%
%%%%%%%%%%%%%%%%%%%%%%%%%%%%%%%%%%%%%%%%%%%%%%%%%%%%%
They would both give calculable ${\cal O}(\alpha_s)$ corrections to the hard-scattering kernel $T^I_{ij}(u)$. For $\bar B_d\to D^+ \pi^-$ however, neither of the two contributions exists because of the flavour structure of the final state.

\subsubsection*{Hard spectator interaction}
\begin{figure}
  \begin{center}
    \psfig{figure=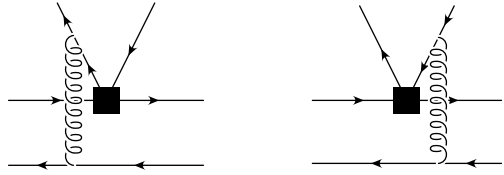}
  \end{center}
  \caption{``Non-factorizable'' spectator interactions.\label{fig:factNLOspect}}
\end{figure}

In (\ref{ffhl}) we claimed that there is no contribution from hard-spectator interaction to heavy-light final states at leading power in the heavy-quark expansion. Soft contributions from diagrams in Fig.~\ref{fig:factNLOspect}
%%%%%%%%%%%%%%%%%%%%%%%%%%%%%%%%%%%%%%%%%%%%%%%%%%%%%
%%%%%%%%%%%%%%%%%%%%%%%%%%%%%%%%%%%%%%%%%%%%%%%%%%%%%
cannot be absorbed into the $B\to D$ form factor or the pion decay constant. For factorization to hold true, no such soft contributions may appear at leading power. Although they are present in each of the two diagrams separately, to leading power they cancel in the sum over the two gluon attachments to the $\bar ud$ pair by the same colour-transparency argument that was applied to the ``non-factorizable'' vertex corrections. Power counting for the summed amplitude of the two diagrams shows explicitely the power suppression relative to (\ref{BDpicount}).

\subsubsection*{Annihilation topologies}
\begin{figure}
  \begin{center}
    \psfig{figure=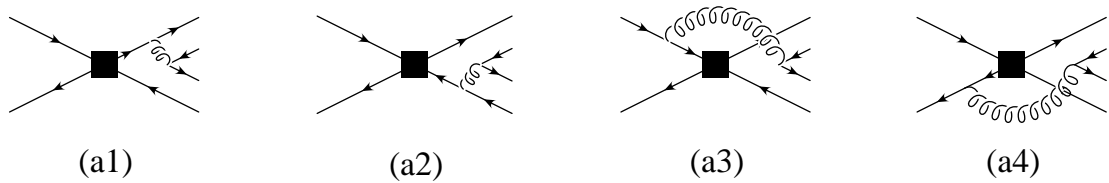}
  \end{center}
  \caption{Annihilation topologies.\label{fig:factann}}
\end{figure}

Finally, we will recall the power suppression of both hard and soft part of annihilation topologies as in Fig.~\ref{fig:factann}.
%%%%%%%%%%%%%%%%%%%%%%%%%%%%%%%%%%%%%%%%%%%%%%%%%%%%%
%%%%%%%%%%%%%%%%%%%%%%%%%%%%%%%%%%%%%%%%%%%%%%%%%%%%%
According to (\ref{phiBcount}) the light quark that goes into the $D$ meson must always be soft. Therefore, the virtuality of the gluon can be at most $m_b\Lambda_\mathrm{QCD}$. The potentially largest contribution can be achieved if both quarks the gluon splits into are soft. This configuration, however, is power suppressed via the pion wave function integral.

%=      Limitations of QCD factorization       =
\subsection{Limitations of QCD factorization}\label{subsec:limitfact}
The factorization formulas (\ref{ffll}) and (\ref{ffhl}) are exact in the heavy-quark limit $m_b \to \infty$. This is a theoretically very well defined limit. In the real world, on the other hand, $m_b$ is fixed to about $4-5\,\mathrm{GeV}$ and one may question the accuracy of the heavy-quark limit. The generic size of power corrections is ${\cal O}(\Lambda_\mathrm{QCD}/m_b)\approx {\cal O}(10\%)$. Yet, sometimes power corrections turn out to be numerically larger than suggested by this naive estimate.

Due to a conspiracy of small parameters, like Wilson coefficients, CKM elements, or colour suppression factors, the leading-power, factorizable term can be somehow suppressed, making power corrections disproportionately large. A related issue is that the strong-interaction phases appearing at ${\cal O}(\alpha_s)$ in QCD factorization might receive large corrections from soft phases of ${\cal O}(\Lambda_\mathrm{QCD}/m_b)$, because $\alpha_s$ and $\Lambda_\mathrm{QCD}/m_b$ are of comparable size.

Especially for $B$ decays into two two light mesons there sometimes is another enhancement of power-suppressed effects connected with the curious numerical fact that
\begin{equation}
  2\mu_\pi \equiv \frac{2m_\pi^2}{m_u+m_d}\approx 3\,\mathrm{GeV}
\end{equation}
is much larger than its naive scaling estimate $\Lambda_\mathrm{QCD}$. The $Q_6$ contribution to the diagram in Fig.~\ref{fig:factLO} for $\bar B_d \to \pi^+ \pi^-$ is an example for such a {\em chirally enhanced} power correction. Such numerically large corrections introduce a substantial uncertainty in some decay modes.

The last point we want to mention is the hybrid role of the charm quark. In our power counting we treated the charm quark as heavy, taking the heavy-quark limit for fixed $m_c/m_b$. However, in reality $m_c$ is not markedly large compared to $\Lambda_\mathrm{QCD}$. In particular, the power suppression of the hard-spectator term in (\ref{ffhl}) for $B$ decays into a heavy and a light meson is $\Lambda_\mathrm{QCD}/m_c$ only, if one employs the limit $m_b \gg m_c \gg \Lambda_\mathrm{QCD}$.

Although the issue of power corrections still deserves thorough investigation we believe that the QCD factorization approach is a big step forward in the treatment of exclusive $B$ decays. For the first time we really have a method based on solid theoretical grounds, namely the heavy-quark limit. Time will show if its phenomenological value is comparable. Proposals of how to estimate or even measure the up to now unknown power corrections are already on the market \cite{AK,DH} and more work on this subject is expected.

%=====================================================
%=      Other Approaches to Non-Leptonic Decays      =
%=====================================================

\section{Other Approaches to Non-Leptonic Decays}
\label{sec:otherNL}

We conclude this chapter with a mini review of attempts to treat hadronic matrix elements in non-leptonic $B$ decays. Some different formulations and generalizations of the naive factorization idea were presented already in section~\ref{sec:naive}. In these approaches a small set of phenomenological parameters is introduced in order to parametrize important non-factorizable effects. Yet, no attempt is made to calculate these parameters from first principles. Another class of approaches aims at a dynamical understanding of non-leptonic weak decays starting from QCD and making a controlled set of approximations. The third class, finally, is based on classifications in terms of flavour topologies or Wick contractions. We now will briefly discuss some representatives of the second and third class of approaches.

\subsection{Dynamical approaches}
\label{subsec:dynapp}
QCD is a beautiful and most probably even the correct theory to describe strong interaction effects in weak decays of hadrons. Still, the main drawback of QCD is that it is not yet solved in the nonperturbative domain. So, especially for the study of non-perturbative properties of QCD, many more or less well-defined approximations were invented to get insight into QCD dynamics at least in some limiting situations.

\subsubsection*{Large-N expansion}
QCD simplifies substantially in the limit of a large number $N$ of colour degrees of freedom \cite{1/N}. One can treat $SU(N)$ QCD as a theory of weakly interacting mesons, and an expansion in $1/N$ then amounts to a loop expansion in this meson theory \cite{BGR}. Most important for us is that the factorization of non-leptonic decay amplitudes becomes exact in the large-$N$ limit. Hence, an expansion in powers of $1/N$ provides a natural framework in which to discuss the structure of non-factorizable corrections. For kaon decays a combination of the $1/N$ expansion with chiral perturbation theory led to important insight into non-factorizable effects \cite{BBG}. For $B$ decays, however, chiral perturbation theory does not apply. Still, one can use the large-$N$ limit to show that non-factorizable corrections to class-I $B$ decays are generally suppressed by two powers of $1/N$, whereas for class-II decays they have the same $1/N$ scaling as the leading factorizable contribution.

\subsubsection*{Lattice gauge theory}
Although lattice field theory is the carrier of hope for many non-perturbative problems, no real results have been obtained so far in the evaluation of exclusive non-leptonic matrix elements. The Maiani-Testa No-go theorem \cite{MT} even forbade lattice QCD to calculate the matrix elements for non-leptonic two-body decays from first principles until Lellouch and L\"uscher found a way to possibly still do so \cite{LL}. Yet, there are still many technical difficulties with lattice QCD and non-leptonic $B$ decays, in particular concerning final-state interactions and the inclusion of penguin diagrams.

\subsubsection*{QCD sum rules}
Although intrinsically limited in their numerical accuracy, in many areas QCD sum rules have been established as a serious competitor to lattice gauge theory computations. In particular for the form factors at large recoil, QCD sum rules might be equal or even superior to lattice results for still some time. Some attempts were made to apply QCD sum rules to non-leptonic weak decays \cite{nonlepSR}. Non-factorizable effects were studied for $B\to J/\psi K_S$ \cite{KR} and $\bar B_d\to D^0 \pi^0$ \cite{IH}. Recently, Khodjamirian suggested to study power corrections for two-body $B$ decays in the framework of light-cone QCD sum rules \cite{AK}. An important conceptual limitation for QCD sum rules in non-leptonic decays is that it will not be possible to obtain a realistic description of final-state interactions if the projection on the external hadron state is performed using an ad hoc continuum subtraction.

\subsubsection*{Large-energy effective theories}
Dugan and Grinstein tried to formalize the concept of colour transparency by introducing a {\em large-energy effective theory} (LEET) to describe the soft interactions of gluons with a pair of fast-moving quarks inside a pion \cite{DGfact}. Yet, due to the presence of collinear singularities, the LEET is not the correct low-energy theory for non-leptonic $B$ decays. In the context of the LEET effective Lagrangian no light-cone distribution amplitudes can be introduced to absorb the factorizable collinear singularities. We refer to \cite{BF,CLOPR,BFPS} for further developments of ``LEET.'' 

\subsubsection*{PQCD}
The goal of perturbative QCD (pQCD) or other hard-scattering approaches is likewise the separation of soft and hard physics in the $B$-decay matrix element \cite{pQCD}. Similar to QCD factorization the decay amplitudes are expressed as a convolution of a hard-scattering amplitude and meson wave functions. Yet, there are fundamental differences in the details. In the pQCD approach the $B\to M$ form factors are assumed to be perturbatively calculable. It is postulated that soft contributions to the form factors are strongly suppressed by Sudakov effects. The form factor is then counted as being of ${\cal O}(\alpha_s)$ in perturbation theory, and thus the hierarchy of the various contributions to the decay amplitudes in pQCD is very different from that in QCD factorization. One consequence hereof is that naive factorization is not recovered in any limit.

Especially for the claim that Sudakov effects suppress long-distance contributions sufficiently for the form factors to be calculable reliably and precisely in perturbation theory, serious doubts were raised \cite{DGS}. Furthermore, the estimate of the strong-interaction phase within pQCD is both model dependent and numerically sensitive to effects that are poorly under control \cite{BBNS}.

\subsection{The diagrammatic way and flavour symmetry approaches}
\label{subsec:diagapp}

There exist several strategies to classify the various contributions to non-leptonic weak decay amplitudes in a convenient way, and then to use symmetry arguments to derive relations between different decay processes. We shall briefly discuss the diagrammatic approach, a parametrization introduced by Buras and Silvestrini, and a flavour $SU(3)$ decomposition of decay amplitudes.

\subsubsection*{The diagrammatic way}
\begin{figure}
  \begin{center}  
    \input{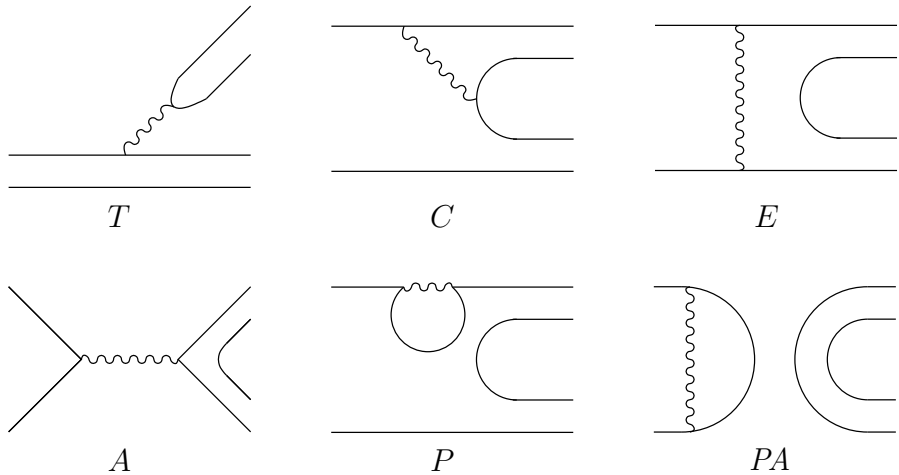}
  \end{center}
\caption{The six building blocks of the diagrammatic way: Tree emission of an external $W$ boson ($T$); internal colour-suppressed $W$ emission ($C$); Exchange of the $W$ boson between the quark lines of the initial state meson ($E$); annihilation of the $B$ meson and subsequent decay of the $W$ boson into the final state particles ($A$); penguin diagram with a ``horizontal'' $W$ exchange ($P$); penguin annihilation ($PA$).\label{fig:diag}}
\end{figure}

Especially for the description of charmless non-leptonic two-body decays of $B$ mesons the six diagrams in Fig.~\ref{fig:diag} were used \cite{diag}.
%%%%%%%%%%%%%%%%%%%%%%%%%%%%%%%%%%%%%%%%%%%%%%%%%%%%%%%%%%%%%%%%%%%
%%%%%%%%%%%%%%%%%%%%%%%%%%%%%%%%%%%%%%%%%%%%%%%%%%%%%%%%%%%%%%%%%%%
Because the amplitudes $A$, $PA$, and $E$ are suppressed with at least one factor of $1/\mbox{(mass of the decaying heavy meson)}$ compared with $T$, $C$, and $P$, they are often neglected altogether. The fact that no explicit conntection to the effective theory of weak interaction exists is also a disadvantage. Instead, Feynman diagrams with propagators in the full theory are used. But this describes the situation at large scales ${\cal O}(M_W)$ rather than at the much better suited mass scale of the decaying heavy meson.

\subsubsection*{The parametrization of Buras and Silvestrini}
\begin{figure}
  \begin{center}  
    \input{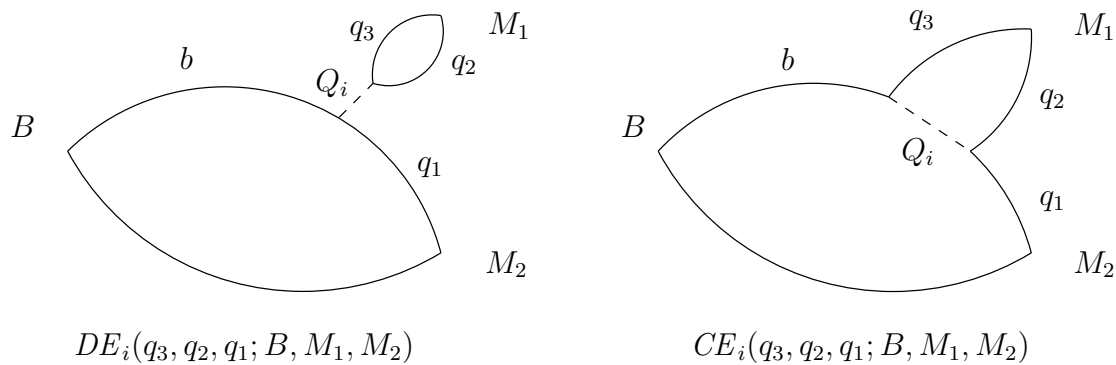}
  \end{center}
\caption{Disconnected and connected emission topology leading to the parameters ${\it DE}$ and ${\it CE}$, respectively. The dashed lines represent insertions of the operators $Q_i$.\label{fig:BSdiag}}
\end{figure}

The basis for the parametrization of Buras and Silvestrini (BS) are Wick contractions of the effective Hamiltonian ${\cal H}_\mathrm{eff}^{\Delta B=1}$ \cite{BS}. The different topologies arise from the different possibilities to connect the quark lines of the initial and final state mesons, taking into account the operator insertion. One obtains seven different topologies, which appear in a connected and disconnected version each. This leads to fourteen renormalization scale and scheme independent parameters, which allow to describe all amplitudes for non-leptonic two-body decays of $B$ mesons. An example for the simplest topology leading to the parameters ${\it DE}$ and ${\it CE}$ is given in Fig.~\ref{fig:BSdiag}.
%%%%%%%%%%%%%%%%%%%%%%%%%%%%%%%%%%%%%%%%%%%%%%%%%%%%%%%%%%%%%%%%%%%
%%%%%%%%%%%%%%%%%%%%%%%%%%%%%%%%%%%%%%%%%%%%%%%%%%%%%%%%%%%%%%%%%%%
The matrix elements of the operators $Q_i$ depend on the renormalization scale and scheme for the operators which has to be cancelled by a corresponding dependence of the Wilson coefficients. Usually, several operators are involved in this cancellation as the operators mix under renormalization. The fourteen effective parameters of Buras and Silvestrini are linear combinations of operator matrix elements and corresponding Wilson coefficients. Insertion of $Q_1$ and $Q_2$ into ${\it DE}$ and ${\it CE}$ topologies of Fig.~\ref{fig:BSdiag} yields the first two scale and scheme independent combinations of Wilson coefficients and operator matrix elements:
\begin{eqnarray}
  E_1&=& C_1\, {\it DE}_1 + C_2 \,{\it CE}_2 \nonumber \\
  E_2&=& C_1\, {\it CE}_1 + C_2 \,{\it DE}_2
\end{eqnarray}
which are generalizations of $a_1^{\rm eff}\langle Q_1 \rangle$ and $a_2^{\rm eff}\langle Q_2 \rangle$ in \cite{NS} and (\ref{a12eff}). Since all BS parameters are combinations of Wilson coefficients and special contractions of local operators, they are in principle calculable in QCD. Unfortunately, so far the BS parametrization is in particular because of the large number of parameters only of limited use for phenomenological analyses. In combination with a dynamical approach such as QCD factorization, however, one could obtain model-independent predictions for special decay channels or make statements about the relative importance of the various BS parameters.

\subsubsection*{$SU(3)$ flavour symmetry}
Compared with the typical scale of QCD $\Lambda_{\overline{MS}}^{(4)}\approx 325\, \mbox{MeV}$, the masses of the light quarks  $m_u\approx 5\, \mbox{MeV}$, $m_d\approx 10\, \mbox{MeV}$, and, less exactly,  $m_s(m_c)\approx 130\,\mbox{MeV}$ are small and approximately degenerate. Because the strong interaction is flavour blind, it therefore satisfies a flavour-$SU(3)$ symmetry, which is explicitely broken via the inexact mass degeneracy. This symmetry property was used both for $D$- and $B$-meson decays to determine for instance CKM elements from experimental data \cite{diagFSI,su3}.

Let us consider the decay $B\to P P$ as an example. We follow the conventions of \cite{GL}. The Hamilton operator ${\cal H}_\mathrm{eff}^{\Delta B=1}$ connects the $B$ mesons in the initial state $i$ with the two pseudoscalar mesons $P$ in the final state $f$. In view of the $SU(3)$-flavour symmetry the most general interaction Hamiltonian is composed of the representations $\mathbf R$ appearing in ${\mathbf f} \otimes \bar{\mathbf i}$. The $i$ and $f$ denote here both states and $SU(3)$ representations. A completely reduced $SU(3)$ amplitude can be denoted by $\langle f\|\mathbf{R}_I\|i \rangle$ where $I$ denotes the eigenvalue of the isospin Casimir operator $I^2$. Due to Bose statistics we only have to consider the completely symmetric representations for a $B$ decay into two light pseudoscalar octet mesons:
\begin{equation}
  (\mathbf8 \otimes \mathbf8)_S=\mathbf1 \oplus \mathbf8_S \oplus \mathbf{27}
\end{equation}
The physical amplitude for the decay of a particle of the $SU(3)$ representation  $\mathbf{R}_c$ with quantum numbers $\nu_c$ into two particles of the representation $\mathbf{R}_a$ and $\mathbf{R}_b$ with quantum numbers $\nu_a$ and $\nu_b$, respectively, can be decomposed into the reduced matrix elements as follows:
\begin{eqnarray}
  \lefteqn{A(i_{\nu_c}^{R_c} \to f_{\nu_a}^{R_a} f_{\nu_b}^{R_b})=} \\
  &&=(-1)^{(I_3+\frac{Y}{2}+\frac{T}{3})_{\overline{R}_c}} \sum_{R',\nu',R,\nu} \left(
   \begin{array}{ccc} R_a & R_b & R' \\ \nu_a & \nu_b & \nu' \end{array}
  \right) \left(
    \begin{array}{ccc} R' & \overline{R}_c & R \\ \nu' & -\nu_c & \nu\end{array}
  \right) \langle R' \| R_\nu \| R_c \rangle\nonumber
\end{eqnarray}
where the triality $T$ guarantees the reality of the phase factor. The quantities in round brackets are Clebsch-Gordan coefficients. They connect, via the Wigner-Eckart theorem, the irreducible tensor operators in-between two states, which are base vectors of irreducible representations. Because the Clebsch-Gordan matrices are orthogonal, the reduced $SU(3)$ amplitudes build an orthonormal basis. In a decomposition of amplitudes, ordered by particle content of the final state, and then by the number of units of strangeness changed, as e.g.
\begin{equation}
  \mathbf{u}_{B \to PP}^{\Delta S=0} = \mathcal{O}_{B \to PP}^{\Delta S=0} \mathbf{v}_{B \to PP}^{\Delta S=0}
\end{equation}
the matrices $\mathcal{O}_{B \to f}^{\Delta S}$ thus are orthogonal. This implies that the reduced $SU(3)$ amplitudes $\mathbf{v}_{B \to PP}^{\Delta S=0}$ are related with the decay amplitudes $\mathbf{u}_{B \to PP}^{\Delta S=0}$ via the transposed of $\mathcal{O}_{B \to PP}^{\Delta S=0}$. If there is a physical reason for a reduced $SU(3)$ amplitude to vanish, the corresponding combination of physical amplitudes also vanishes. For $\Delta S=0$ $B\to PP$ decays for example, no matrix elements with representations $\overline{\mathbf{24}}$ or $\overline{\mathbf{42}}$ can contribute in the limit of exact $SU(3)$-flavour symmetry. There are five such reduced $SU(3)$ amplitudes, leading to five amplitude relations, the simplest of them being the isospin relation
\begin{equation}
  -\sqrt{2} A(B_u^+ \to \pi^+ \pi^0) + A(B_d^0 \to\pi^+ \pi^-) + \sqrt{2} A(B_d^0 \to \pi^0 \pi^0)=0
\end{equation}
Similar relations can be derived for $D$ decays \cite{diplom}.

% =========================================================
% =     part II: B -> V gamma                             =
% =========================================================
\part{The Radiative Decays $B\to V\gamma$ at NLO in QCD}

% ===== Basic Formulas ====================================
\chapter{Basic Formulas for $B\to V\gamma$}
\label{ch:basicsBVgam}

The second part is devoted to the analysis of the exclusive radiative $B$-meson decays with a vector meson $V$ in the final state, i.e. in particular the decays $B\to K^* \gamma$ and $B\to\rho\gamma$. We present a factorization formula for the $B\to V\gamma$ matrix elements, which is based  on the heavy-quark limit $m_b\gg\Lambda_{QCD}$. This allows us to factorize perturbatively calculable contributions from nonperturbative form factors and universal light-cone distribution amplitudes. Our results are complete to next-to-leading order in QCD and to leading power in the heavy-quark limit and include the branching ratios and estimates of direct CP violation, isospin and U-spin breaking effects. In chapter \ref{ch:basicsBVgam} we present the structure of the QCD factorization formulas. These formulas are applied to $B\to K^* \gamma$ and $B\to\rho\gamma$ in chapter \ref{ch:phenBVgam}. The results given here have to a great extent already been presented in \cite{BBVgam}. We go more into detail in the work on hand, add a proof for the factorizability of annihilation contributions, and, above all, include the effects of QCD penguin operators.

%==================================
%=      What is the Problem?      =
%==================================

\section{What is the Problem?}
\label{sec:whatisproblem}

The radiative transitions $b\to s\gamma$ and $b\to d\gamma$ are among the most valuable probes of flavour physics. Proceeding at rates of order $G^2_F\alpha$, they are systematically enhanced over other loop-induced, non-radiative, rare decays, which are proportional to $G^2_F\alpha^2$. In fact, the Cabibbo-favoured $b\to s\gamma$ modes belong to the small number of rare decays that are experimentally accessible already at present. The experimental measurement of the inclusive branching fraction was quoted already in~(\ref{bsgamexp}). The branching ratios for the exclusive channels have been determined by CLEO \cite{CHEN} and more recently also by BELLE \cite{ABE} and {\sc BaBar} \cite{BaBarBKg}:
\begin{equation}\label{b0kgamex}
  B(B^0\to K^{*0}\gamma)^\mathrm{exp.}=\left\{\begin{array}{ll}
  (4.55\pm 0.70\pm 0.34)\cdot 10^{-5} & \cite{CHEN} \\
  (4.96\pm 0.67\pm 0.45)\cdot 10^{-5} & \cite{ABE} \\
  (4.23\pm 0.40\pm 0.22)\cdot 10^{-5} & \cite{BaBarBKg} \end{array}\right.
\end{equation}
\begin{equation}\label{bpkgamex}
  B(B^+\to K^{*+}\gamma)^\mathrm{exp.}=\left\{\begin{array}{ll}
  (3.76\pm 0.86\pm 0.28)\cdot 10^{-5} & \cite{CHEN} \\
  (3.89\pm 0.93\pm 0.41)\cdot 10^{-5} & \cite{ABE} \\
  (3.83\pm 0.62\pm 0.22)\cdot 10^{-5} & \cite{BaBarBKg} \end{array}\right.
\end{equation}
Although for the $B\to\rho\gamma$ decays so far only upper limits exist \cite{BaBarBrhog}, they are already very close to the Standard Model expectations:
\begin{equation}\label{brhogamexp}
  \begin{array}{rcll}
    B(B^0\to\rho^0\gamma) & < & 1.5\cdot 10^{-6} & \mbox{at 90\% C.L.}\\
    B(B^+\to\rho^+\gamma) & < & 2.8\cdot 10^{-6} & \mbox{at 90\%  C.L.}
  \end{array}
\end{equation}
On the theoretical side, the flavour-changing-neutral-current (FCNC) reactions $b\to s(d)\gamma$ are characterized by their high sensitivity to New Physics and by the particularly large impact of short-distance QCD corrections \citer{CFMRS,GOSN}.

As we already mentioned in section \ref{sec:EH}, the inclusive $b\to s\gamma$ mode can be calculated perturbatively, whereas the treatment of the exclusive channels is in general more complicated. Seen from the experimental point of view, however, the situation is the opposite way round. Measurements of inclusive rates are rather difficult to perform, as one has to sum over all final states containing e.g. a photon and a strange quark. Furthermore, a lower cutoff in the photon energy has to be imposed in order to get along with the background coming mainly from the nonleptonic charged current processes $b\to c/u\, q\bar q\, \gamma$. Theoretical input on the photon energy spectrum is then needed to translate the measured ``kinematic'' branching ratio into the total one. The exclusive decays, on the other hand, show a very clean experimental signature, which is particularly important for the difficult environment of hadron machines as the Fermilab Tevatron or the LHC at CERN. Once again the conservation law, that ``the product of theoretical difficulty and experimental effort needed to solve a problem is constant,'' seems to hold.

The main difficulty in calculations of exclusive hadronic decays is how to evaluate the matrix elements of the operators sandwiched between final $V\gamma$ and initial $B$ state. These matrix elements lump together all the nonperturbative long-distance dynamics below a scale of ${\cal O}(m_B)$. From Lorentz covariance and gauge invariance, the $B\to V\gamma$ transition matrix elements have the following general form
\begin{equation}\label{BVgamgen}
  \langle V(k,\eta) \gamma(q,\epsilon)| Q_i |B\rangle = -\frac{e}{2\pi^2} m_B F_V(0)\left[\varepsilon^{\mu\nu\lambda\rho}\epsilon_\mu\eta_\nu k_\lambda q_\rho +i(\epsilon\cdot\eta\, k\cdot q-\epsilon\cdot k\, \eta\cdot q)\right]
\end{equation}
This is just a convenient parametrization in terms of the form factor $F_V(q^2=0)$, which constitutes now the challenge to be determined. For $q^2\neq 0$, i.e. an off-shell photon, we have two form factors: one in front of each of the two terms in~(\ref{BVgamgen}), which are gauge invariant separately. The form factors are calculable in principle from lattice QCD \cite{UDQCD}. Since the lattice calculations provide results over a limited region of high $q^2$, i.e. small recoil,  extra assumptions are needed to scale down the form factors to the case of an onshell external photon with $q^2=0$. Via the light-cone sum-rule (LCSR) approach \cite{CZ,LCSR} a more direct access to the $q^2=0$ region is opened \cite{BB98,ABS}. But the theoretical accuracy of sum rule calculations is, due to the duality approximation made, irreducibly restricted to ${\cal O}(10\%)$. There are further prescriptions on the market how to include bound state effects in $B\to V\gamma$ \cite{GSW,AAW}. However, hadronic models are used which do not allow a clear separation of short- and long-distance dynamics and a clean distinction of model-dependent and model-independent features. In section \ref{sec:relcalc} we comment on these and related calculations in the context of QCD factorization.

%=======================================
%=      The Factorization Formula      =
%=======================================

\section{The Factorization Formula}
\label{sec:factBVgam}

In the heavy-quark limit, a systematic treatment of the hadronic matrix element is possible \cite{BBVgam,BFS,AP}. In this case, the following factorization formula is valid
\begin{equation}\label{fformBVgam}
  \langle V\gamma(\epsilon)|Q_i|\bar B\rangle = \left[ F^{B\to V}(0)\, T^I_{i} + \int^1_0 d\xi\, dv\, T^{II}_i(\xi,v)\, \Phi_B(\xi)\, \Phi_V(v)\right] \cdot\epsilon
\end{equation}
where $\epsilon$ is the photon polarization 4-vector. With the $B\to V$ transition form factor $F^{B\to V}$ and $\Phi_B$ and $\Phi_V$, the leading twist light-cone distribution amplitudes of the $B$ meson and the vector meson, respectively, we encounter universal, nonperturbative objects. They describe the long-distance dynamics of the matrix elements, which is factorized from the perturbative, short-distance interactions expressed in the hard-scattering kernels $T^I_{i}$ and $T^{II}_i$. The QCD factorization formula~(\ref{fformBVgam}) holds for all operators in~(\ref{q1def}--\ref{p8def}) up to corrections of relative order $\Lambda_{QCD}/m_b$. We include the calculable contributions of all operators unless they are both of subleading power and suppressed by $\alpha_s(\mu_b)$ with $\mu_b={\cal O}(m_b)$.

\begin{figure}[t]
   \epsfysize=4cm
   \epsfxsize=6cm
   \centerline{\epsffile{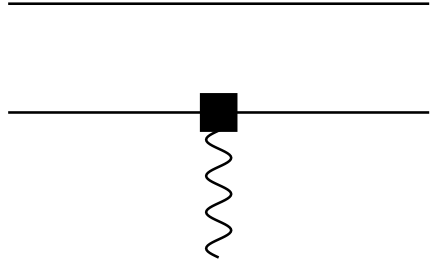}}
\caption{Contribution of the magnetic penguin operator $Q_7$ described by $B\to V$ form factors. All possible gluon exchanges between the quark lines are included in the form factors and have not been drawn explicitly. \label{fig:q7}}
\end{figure}

In the leading-logarithmic approximation and to leading power in the heavy-quark limit, $Q_7$ gives the only contribution to the amplitude of $\bar B\to V\gamma$, and the factorization formula~(\ref{fformBVgam}) is trivial. The matrix element is simply expressed in terms of the standard form factor, $T^I_{7}$ is a purely kinematical function, and the spectator term $T^{II}_7$ is absent. An illustration is given in Fig.~\ref{fig:q7}.
%%%%%%%%%%%%%%%%%%%%%%%%%%%%%%%%%%%%%%%%%%%%%%%%%%%%%%%%%%%%%%%%%%%
%%%%%%%%%%%%%%%%%%%%%%%%%%%%%%%%%%%%%%%%%%%%%%%%%%%%%%%%%%%%%%%%%%%
The matrix element reads
\begin{equation}\label{q7f1f2}
  \langle V(k,\eta) \gamma(q,\epsilon)|Q_7|\bar B\rangle = -\frac{e}{2\pi^2}m_b\, c_V F_V \left[\varepsilon^{\mu\nu\lambda\rho}\epsilon_\mu\eta_\nu k_\lambda q_\rho +i (\epsilon\cdot\eta\, k\cdot q-\epsilon\cdot k\, \eta\cdot q)\right]
\end{equation}
where $c_V=1$ for $V=K^*$, $\rho^-$ and $c_V=1/\sqrt{2}$ for $V=\rho^0$. Our phase conventions coincide with those of \cite{BB98,ABS}. In particular, we have $F_V>0$, $\varepsilon^{0123}=-1$, and the phases of $V$ (with flavour content $\bar qq'$) and $\bar B$ are such that
\begin{eqnarray}\label{fbfv}
  \langle V(k,\eta)|\bar q\sigma_{\mu\nu}q'|0\rangle &=& -i(\eta_\mu k_\nu-\eta_\nu k_\mu)f^\perp_V \\
  \langle 0|\bar u\gamma_\mu\gamma_5 b|\bar B(p)\rangle &=& +i f_B p_\mu
\end{eqnarray}
with positive $f_B$, $f^\perp_V$.

\begin{figure}[t]
   \epsfysize=4cm
   \epsfxsize=12cm
   \centerline{\epsffile{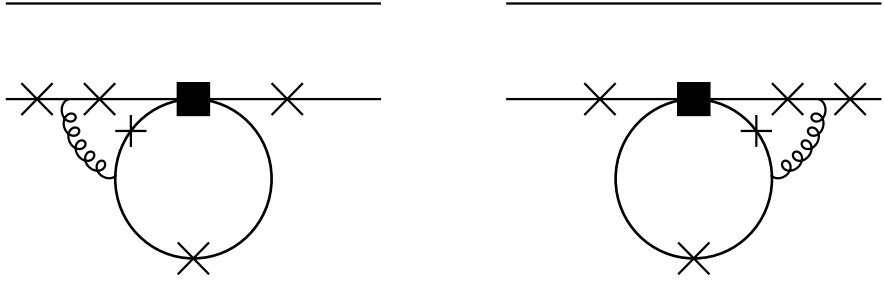}}
\caption{${\cal O}(\alpha_s)$ contribution to the hard-scattering kernels $T^I_{i}$ from four-quark operators $Q_{1\ldots 6}$. The crosses indicate the places where the emitted photon can be attached. Diagrams where the gluon is exchanged in the loop vanish for the same reason as the mixing of $Q_{1\ldots 4}$ into $Q_7$ vanishes at one-loop level: chirality conservation in the quark loop and QED gauge invariance.\label{fig:qit1}}
\end{figure}

\begin{figure}[t]
   \epsfysize=2.5cm
   \epsfxsize=12cm
   \centerline{\epsffile{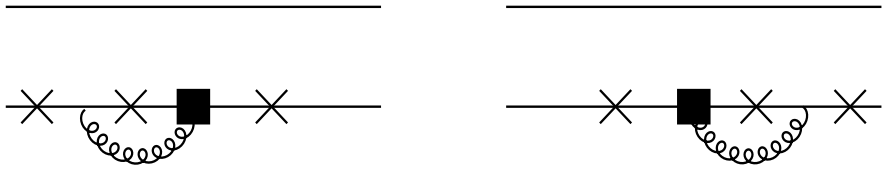}}
\caption{${\cal O}(\alpha_s)$ contribution to the hard-scattering kernels $T^I_{8}$ from chromomagnetic penguin operator $Q_8$.\label{fig:q8t1}}
\end{figure}

\begin{figure}[t]
   \epsfysize=4cm
   \epsfxsize=15cm
   \centerline{\epsffile{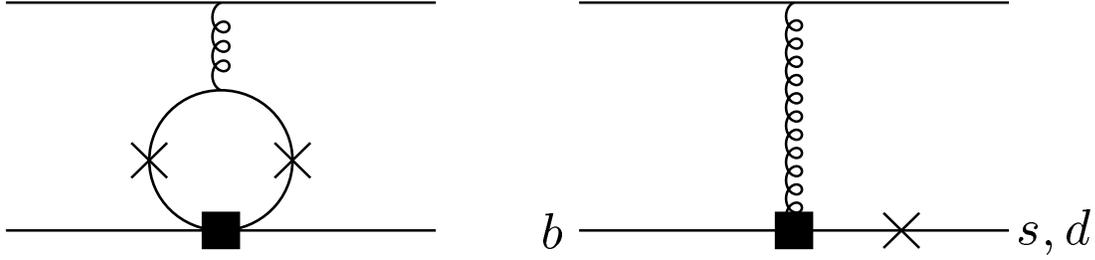}}
\caption{${\cal O}(\alpha_s)$ contribution to the hard-scattering kernels $T^{II}_i$ from four-quark operators $Q_i$ (left) and from $Q_8$ (right). \label{fig:qit2}}
\end{figure}

Only at ${\cal O}(\alpha_s)$ the matrix elements of the other seven operators start contributing. In this case the factorization formula becomes nontrivial. The diagrams for the ``type I'' or ``hard vertex'' contributions are shown in Fig.~\ref{fig:qit1} 
%%%%%%%%%%%%%%%%%%%%%%%%%%%%%%%%%%%%%%%%%%%%%%%%%%%%%%%%%%%%%%%%%%%
%%%%%%%%%%%%%%%%%%%%%%%%%%%%%%%%%%%%%%%%%%%%%%%%%%%%%%%%%%%%%%%%%%%
for $Q_{1\ldots 6}$ and in Fig.~\ref{fig:q8t1}
%%%%%%%%%%%%%%%%%%%%%%%%%%%%%%%%%%%%%%%%%%%%%%%%%%%%%%%%%%%%%%%%%%%
%%%%%%%%%%%%%%%%%%%%%%%%%%%%%%%%%%%%%%%%%%%%%%%%%%%%%%%%%%%%%%%%%%%
for $Q_8$. The calculation of these diagrams leads to infrared-finite functions of $m_q/m_b$, where $m_q$ is the mass of the internal quarks in the loop diagram. The ``type II'' or ``hard spectator'' contributions are those where the $B$-meson-spectator quark participates in the hard scattering. The non-vanishing contributions to $T_i^{II}$ are shown in Fig.~\ref{fig:qit2}. 
%%%%%%%%%%%%%%%%%%%%%%%%%%%%%%%%%%%%%%%%%%%%%%%%%%%%%%%%%%%%%%%%%%%
%%%%%%%%%%%%%%%%%%%%%%%%%%%%%%%%%%%%%%%%%%%%%%%%%%%%%%%%%%%%%%%%%%%
The hard gluon therein probes the momentum distribution of the quarks in the $B$ meson and the light vector meson. To evaluate these diagrams we need the light-cone distribution amplitudes (LCDA) of the participating mesons. In the case of the $B$ meson we have to leading power the formula given in~(\ref{BLCDA}). If we choose $n$ in~(\ref{BLCDA}) appropriately, we need besides the normalization conditions only the first negative moment of $\Phi_{B1}(\xi)$, which we parametrized by the quantity $\lambda_B$ in~(\ref{lambdef}). The vector meson in $\bar B\to V\gamma$ decays is transversly polarized to leading power. This can be described by $\Phi_\perp$, the leading-twist and leading-power distribution amplitude for a light vector mesons with transverse polarization, defined by the first line in equation~(\ref{VLCDA}):
\begin{equation}\label{phiperpdef}
  \langle V(k,\eta)|q'_i(z)_\alpha \bar q_j(0)_\beta |0\rangle = -\frac{f^\perp_V}{8N}\delta_{ij}\left[\not\!\eta,\not\!k\right]_{\alpha\beta} \int_0^1\! dv\, e^{i\bar vkx}\Phi_\perp(v)
\end{equation}

\begin{figure}
  \begin{center}  
    \input{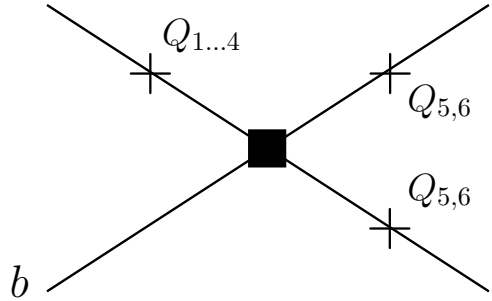}
  \end{center}
\caption{Annihilation contribution to the $\bar B\to V\gamma$ decay. The dominant mechanism for operators $Q_{1\ldots 4}$ is the radiation of the photon from the light quark in the $B$ meson, as shown. This amplitude is suppressed by one power of $\Lambda_{QCD}/m_b$, but it is still calculable in QCD factorization. Radiation of the photon from the remaining three quark lines is suppressed by $(\Lambda_{QCD}/m_b)^2$ for operators $Q_{1\ldots 4}$. For operators $Q_{5,6},$ however, radiation of the final state vector meson quarks is power suppressed diagrammatically but gives a contribution from the leading power projector.\label{fig:ann}}
\end{figure}

The diagrams in Figs.~\ref{fig:q7}--\ref{fig:qit2} are not yet the complete set of possible contributions to $\bar B\to V\gamma$ decays. One of the further possibilities is weak annihilation, depicted in Fig.~\ref{fig:ann}.
%%%%%%%%%%%%%%%%%%%%%%%%%%%%%%%%%%%%%%%%%%%%%%%%%%%%%%%%%%%%%%%%%%%
%%%%%%%%%%%%%%%%%%%%%%%%%%%%%%%%%%%%%%%%%%%%%%%%%%%%%%%%%%%%%%%%%%%
If we calculate this contribution for $Q_{1\ldots 4}$ using the leading-power projection onto the meson $V$ in~(\ref{phiperpdef}) we get a vanishing result because the trace over an odd number of Dirac matrices is zero. Yet, a non-vanishing result arises from the projector
\begin{equation}\label{fvmveta}
  \langle V(k,\eta)| \bar q\gamma_\nu q' |0\rangle = f_V m_V \eta_\nu
\end{equation}
which, however, is suppressed by one power of $\Lambda_{QCD}/m_b$ compared to~(\ref{fbfv}),~(\ref{phiperpdef}) for transverse polarization $\eta_\nu$ ($f_V$, $f^\perp_V$, $m_V\sim \Lambda_{QCD}$, $k\sim m_b$). For operators $Q_{5,6}$ one gets a contribution from the leading power projector when the photon is emitted from the final state vector meson, which is power suppressed diagrammatically. Hence, weak annihilation contributions are power suppressed. When the photon is not emitted as in Fig.~\ref{fig:ann}, the annihilation contributions are either vanishing or even stronger suppressed because the quark propagators are scaling as $1/m_b$ instead of $1/\Lambda_{QCD}$ and only the subleading projector appears..

Despite their power suppression, the dominant annihilation amplitudes can be computed within QCD factorization. For the explicit proof to ${\cal O}(\alpha_s)$ we refer to Appendix \ref{app:WAproof}. Physically, the annihilation contributions factorize because the colour-transparency argument applies to the emitted, highly energetic vector meson in the heavy-quark limit \cite{BBNS}. A similar observation was already made in \cite{GP}. Grinstein and Pirjol, however, claim that the ${\cal O}(\alpha_s)$ corrections to Fig.~\ref{fig:ann} would vanish identically in the chiral limit and to leading-twist, because the projector in~(\ref{phiperpdef}) gives zero when applied to a current with an odd number of Dirac matrices. But, by the same argument, the diagram in Fig.~\ref{fig:ann} would vanish even at leading order in $\alpha_s$, which is not the case. The proper treatment of the ${\cal O}(\alpha_s)$ correction for $Q_{1,2}$ should employ the subleading-power projections related to the wave functions $\Phi_V^\parallel$, $g_V^{\perp\,(v)}$, $g_V^{\perp\,(a)}$ \cite{BB96,BB98}, corresponding to the use of~(\ref{fvmveta}) at ${\cal O}(\alpha^0_s)$. 

Since weak annihilation is a power correction, we will content ourselves with the lowest order result (${\cal O}(\alpha^0_s)$) for our estimates below. In particular, we shall use the annihilation effects to estimate isospin-breaking corrections in $B\to V\gamma$ decays. The reason for including this class of power corrections is that they come with a numerical enhancement from the large Wilson coefficients $C_{1,2}$ ($C_1\approx 3|C_7|$) or are numerically enhanced in case of the $Q_{5,6}$ penguin operator contributions.

\begin{figure}[t]
   \epsfysize=4cm
   \epsfxsize=15cm
   \centerline{\epsffile{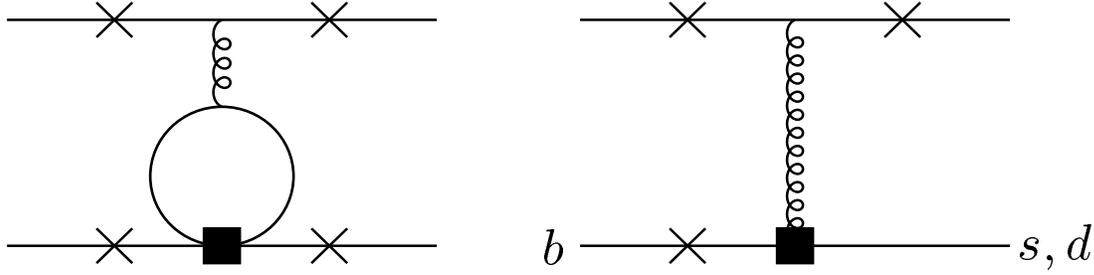}}
\caption{Other contributions that are power-suppressed in the heavy-quark limit. The penguin-annihilation diagrams (left) can be calculated in QCD factorization, but for the subleading-power $Q_8$ matrix elements (right) where the photon is emitted from the final state vector meson, factorization breaks down.\label{fig:subl}}
\end{figure}

Finally, Fig.~\ref{fig:subl} displays the remaining diagrams with a power suppression. 
%%%%%%%%%%%%%%%%%%%%%%%%%%%%%%%%%%%%%%%%%%%%%%%%%%%%%%%%%%%%%%%%%%%
%%%%%%%%%%%%%%%%%%%%%%%%%%%%%%%%%%%%%%%%%%%%%%%%%%%%%%%%%%%%%%%%%%%
The penguin-annihilation diagrams (left) are power suppressed in a way similar to the ordinary annihilation topologies. They are additionally suppressed by $\alpha_s$. The spectator diagrams from $Q_8$ (right) lead to amplitudes that are either manifestly power-suppressed (photon emitted from the quark line in the upper right) or that are superficially of leading power, but vanish when the leading-order projections are performed (photon emitted from either of the quarks forming the $B$ meson). Recall, however, that photon emission from the light-quark line from $Q_8$ is a leading-power effect (see Fig. \ref{fig:qit2}).

%=========================================
%=      The Hard Vertex Corrections      =
%=========================================
\section{The Hard Vertex Corrections}
\label{sec:typeI}

To ${\cal O}(\alpha_s)$ the amplitude for $\bar B\to V\gamma$ reads, with the short-hand notation $\langle Q_i\rangle \equiv \langle V\gamma| Q_i| \bar B\rangle$, for our standard or the CMM operator basis
\begin{eqnarray}
  \label{abkgamQ}
    A(\bar B\to V\gamma) &=& \frac{G_F}{\sqrt{2}} \sum_{p=u,c}\lambda_p^{(s)} \Bigg(C_1 \langle Q_1^{(p)}\rangle + C_2 \langle Q_2^{(2)}\rangle +\sum_{i=3}^8 C_i\langle Q_i\rangle\Bigg)\\
  \label{abKgamP}
    &=& \frac{G_F}{\sqrt{2}} \sum_{p=u,c}\lambda_p^{(s)} \Bigg(Z_1 \langle P_1^{(p)}\rangle + Z_2 \langle P_2^{(2)}\rangle +\sum_{i=3}^8 Z_i\langle P_i\rangle\Bigg)\nonumber
\end{eqnarray}
The leading-order matrix element $\langle P_7\rangle =\langle Q_7\rangle$ is given in~(\ref{q7f1f2}). At subleading order in $\alpha_s$ the matrix elements $\langle Q_{1\ldots 6,8}\rangle$ need to be computed from the diagrams in Figs. \ref{fig:qit1} -- \ref{fig:qit2}. We have to combine results of the matrix elements obtained in different operator bases: ours and the one of Chetyrkin, Misiak, and M\"unz (CMM). The final amplitude has to be independent of the choice of an operator basis, of course. Even partial results which have a physical meaning have to be independent of the chosen operator basis. This will be used subsequently. We will treat the type I contributions in the CMM~(\ref{p1def}--\ref{p8def}) and the type II contributions in our operator basis~(\ref{q1def}--\ref{q8def}).

Let us now have a closer look at the type I or hard vertex contributions of Figs.~\ref{fig:qit1} and \ref{fig:q8t1}. The results for these diagrams can be inferred from \citer{GHW,BCMU2}. In these papers the diagrams were computed to obtain the matrix elements for the inclusive mode $b\to s\gamma$ at next-to-leading order. In this context Figs. \ref{fig:qit1} and \ref{fig:q8t1} represented the virtual corrections to the inclusive matrix elements of $Q_{1,8}$ \cite{GHW} or $P_{1\ldots 6, 8}$ \cite{BCMU2}. In our case they determine the kernels $T^I_{1\ldots 6}$ and $T^I_{8}$. As required for the consistency of the factorization formula, these corrections must be dominated by hard scales $\sim m_b$ and hence must be infrared finite. This is indeed the case. Only recently, Buras, Czarnecki, Misiak, and Urban completed the calculation of the NLO matrix elements including the penguin operators \cite{BCMU2}. However, they calculated in the CMM basis. Nevertheless, we can re-interprete their results as the perturbative hard-scattering kernels for the exclusive process. This is because the NLO corrections to the matrix elements have a physical effect on the inclusive rate. For the exclusive decay the type I corrections are completely independent of the type II contributions, which appear at ${\cal O}(\alpha_s)$ for the first time. Therefore, we can use different operator bases for type I and type II corrections. The results from \cite{BCMU2} imply
\begin{equation}\label{q1me1}
  \langle P_{i}\rangle^I=\langle Q_7\rangle \frac{\alpha_s C_F}{4\pi} G_{i} \qquad i\neq 7
\end{equation}
where $C_F=(N^2-1)/(2N)$, with $N=3$ the number of colours, and
\begin{displaymath}
  \begin{array}{lcl}
    G_1(z_p) &=& \displaystyle \frac{52}{81}\ln\frac{\mu}{m_b}+ g_1(z_p) \\[4mm]
    G_2(z_p) &=& \displaystyle -\frac{104}{27}\ln\frac{\mu}{m_b}+ g_2(z_p) \\[4mm]
    G_3      &=& \displaystyle \frac{44}{27}\ln\frac{\mu}{m_b}+ g_3 \\[4mm]
    G_4(z_c) &=& \displaystyle \frac{38}{81}\ln\frac{\mu}{m_b}+ g_4(z_c) \\[4mm]
    G_5      &=& \displaystyle \frac{1568}{27}\ln\frac{\mu}{m_b}+ g_5 \\[4mm]
    G_6(z_c) &=& \displaystyle -\frac{1156}{81}\ln\frac{\mu}{m_b}+ g_6(z_c) \\[4mm]
    G_8      &=& \displaystyle \frac{8}{3}\ln\frac{\mu}{m_b} + g_8 \\[8mm]
    g_1(z) &=& \displaystyle \frac{833}{972}-\frac{1}{4}[a(z)+b(z)]+\frac{10i\pi}{81}\\[4mm]
    g_2(z) &=& \displaystyle -\frac{833}{162} +\frac{3}{2}[a(z) +b(z)]-\frac{20i\pi}{27}\\[4mm]
    g_3    &=& \displaystyle \frac{598}{81} +\frac{2\pi}{\sqrt{3}} +\frac{8}{3} X_b -\frac{3}{4}a(1) +\frac{3}{2}b(1) +\frac{14i\pi}{27}\\[4mm]
    g_4(z) &=& \displaystyle -\frac{761}{972} -\frac{\pi}{3\sqrt{3}} -\frac{4}{9} X_b +\frac{1}{8} a(1) +\frac{5}{4} b(z) -\frac{37i\pi}{81}\\[4mm]
    g_5    &=& \displaystyle \frac{14170}{81} +\frac{8\pi}{\sqrt{3}} +\frac{32}{3} X_b -12 a(1) +24 b(1) +\frac{224i\pi}{27}\\[4mm]
    g_6(z) &=& \displaystyle \frac{2855}{486} -\frac{4\pi}{3\sqrt{3}} -\frac{16}{9} X_b -\frac{5}{2} a(1) +11 b(1) +9 a(z) +15 b(z) -\frac{574i\pi}{81}\\[4mm]
    g_8    &=& \displaystyle \frac{11}{3} -\frac{2\pi^2}{9} +\frac{2i\pi}{3}
  \end{array}
\end{displaymath}
where
\begin{eqnarray}
  \label{Xb}
    X_b &=& \int_0^1 \!dx \int_0^1 \!dy \int_0^1 \!dv x y \ln[v+x(1-x)(1-v)(1-v+v y)]\approx -0.1684\\
  \label{as}
    a(z) &=& \frac{8}{9} \int_0^1 \!dx \int_0^1 \!dy \int_0^1 \!dv \left\{  
[2 - v + x y (2 v-3)] \ln [v z + x (1 - x) (1 - v) (1 - v + v y)] 
\right. \nonumber\\
    && \hspace{17mm} \left. 
+[1-v+xy(2v-1)] \ln [ z - i\varepsilon - x(1-x)yv] \right\} + \frac{43}{9} + \frac{4i\pi}{9}\\
  \label{bs}
  b(z) &=& \frac{4}{81}\ln z +\frac{16}{27}z^2 +\frac{224}{81}z -\frac{92}{243} +\frac{4i\pi}{81}\nonumber\\
  && +\frac{-48z^2-64z+4}{81}\sqrt{1\!-\!4z} f(z) -\frac{8}{9} z^2 \left( \frac{2}{3}z-1 \right) f(z)^2\\
  && - \frac{8}{9} \int_0^1 \!dx \int_0^1 \!dy\; \frac{\frac{1}{2} y^2 (y^2-1) x (1-x) + (2-y) u_1 \ln u_1 + (2y^2-2y-1) u_2 \ln u_2}{(1-y)^2}\nonumber
\end{eqnarray}
with $u_k = y^k x(1-x)+(1-y) z$ and
\begin{equation}
  f(z) = \theta(1\!-\!4z) \left( \ln \frac{1+\sqrt{1\!-\!4z}}{1-\sqrt{1\!-\!4z}} -i \pi \right) -2i \theta(4z\!-\!1) \arctan \frac{1}{\sqrt{4z\!-\!1}}
\end{equation}
The results for $a(1)$ and $b(1)$ read
\begin{eqnarray}
  a(1) &\simeq& 4.0859 + \frac{4i\pi}{9}\\
  b(1) &=& \frac{320}{81} - \frac{4 \pi}{3 \sqrt{3}} + \frac{632\pi^2}{1215} - \frac{8}{45} \left[ \frac{d^2 \ln \Gamma(x)}{dx^2} \right]_{x=\frac{1}{6}}+ \frac{4i\pi}{81} \simeq 0.0316 + \frac{4i\pi}{81}
\end{eqnarray}
For $z<0.3$ the functions $a(z)$ and $b(z)$ are very accurately given by their expansion in $z$:
\begin{displaymath}
  \begin{array}{lcl}
    a(z)  &=& \frac{16}{9} \left\{ \left[ \frac{5}{2} -\frac{\pi^2}{3} -3 \zeta(3) + \left( \frac{5}{2} - \frac{3\pi^2}{4} \right) \ln z + \frac{1}{4} \ln^2 z + \frac{1}{12} \ln^3 z \right] z \right. \\[2mm] 
    && \hspace{6mm} +\left[\frac{7}{4} +\frac{2\pi^2}{3} -\frac{\pi^2}{2} \ln z -\frac{1}{4} \ln^2 z +\frac{1}{12} \ln^3 z \right] z^2 \\[2mm]
    && \hspace{6mm} + \left[ -\frac{7}{6} -\frac{\pi^2}{4} + 2\ln z - \frac{3}{4}\ln^2 z \right] z^3  \\[2mm] 
    && \hspace{6mm} +\left[ \frac{457}{216} - \frac{5\pi^2}{18} -\frac{1}{72} \ln z -\frac{5}{6} \ln^2 z \right] z^4 \\[2mm]
    && \hspace{6mm} +\left[ \frac{35101}{8640} - \frac{35\pi^2}{72} -\frac{185}{144} \ln z -\frac{35}{24} \ln^2 z \right] z^5 \\[2mm] 
    && \hspace{6mm} + \left[ \frac{67801}{8000} - \frac{21\pi^2}{20} - \frac{3303}{800} \ln z - \frac{63}{20} \ln^2 z \right] z^6 \\[2mm]
    && \hspace{6mm}+ i \pi \left[ \left( 2 -\frac{\pi^2}{6} + \frac{1}{2} \ln z + \frac{1}{2} \ln^2 z \right) z + \left( \frac{1}{2} -\frac{\pi^2}{6} - \ln z + \frac{1}{2} \ln^2 z \right) z^2 \right. \\[2mm]
    && \hspace{14mm} \left. \left. + z^3 + \frac{5}{9} z^4 + \frac{49}{72} z^5  + \frac{231}{200} z^6 \right] \right\} \\[2mm]
    && + {\cal O}(z^7 \ln^2 z)\\[6mm]
    b(z) &=& \hspace{-2mm} -\frac{8}{9} \left\{ \left( -3 +\frac{\pi^2}{6} - \ln z \right) z - \frac{2\pi^2}{3} z^{3/2} \right. \\[2mm]
    && \hspace{6mm} + \left( \frac{1}{2} +\pi^2 - 2 \ln z - \frac{1}{2} \ln^2 z \right) z^2  \\[2mm] 
    && \hspace{6mm} + \left(-\frac{25}{12} -\frac{\pi^2}{9} - \frac{19}{18} \ln z + 2 \ln^2 z \right) z^3 \\[2mm]
    && \hspace{6mm} + \left[ -\frac{1376}{225} + \frac{137}{30} \ln z + 2 \ln^2 z + \frac{2\pi^2}{3} \right] z^4 \\[2mm] 
    && \hspace{6mm} + \left[ -\frac{131317}{11760} + \frac{887}{84} \ln z + 5 \ln^2 s + \frac{5\pi^2}{3} \right] z^5 \\[2mm]
    && \hspace{6mm} + \left[ -\frac{2807617}{97200} + \frac{16597}{540} \ln z + 14 \ln^2 z + \frac{14\pi^2}{3} \right] z^6 \\[2mm] 
    && \hspace{6mm} \left. + i \pi \left[ -z + (1 - 2 \ln z) z^2 + \left(-\frac{10}{9} + \frac{4}{3} \ln z \right) z^3 + z^4 + \frac{2}{3} z^5 + \frac{7}{9} z^6 \right] \right\} \\[2mm]
    && + {\cal O}(z^7 \ln^2 z)
\end{array}
\end{displaymath}
We denote with
\begin{equation}\label{zdef}
  z_q=\frac{m^2_q}{m^2_b}
\end{equation}
the squared ratio of the mass of the internal quark in the loop with respect to the $b$ quark mass.

%============================================
%=      The Hard Spectator Corrections      =
%============================================
\section{The Hard Spectator Corrections}
\label{sec:typeII}

\begin{figure}[t]
   \begin{center}
     \psfig{figure=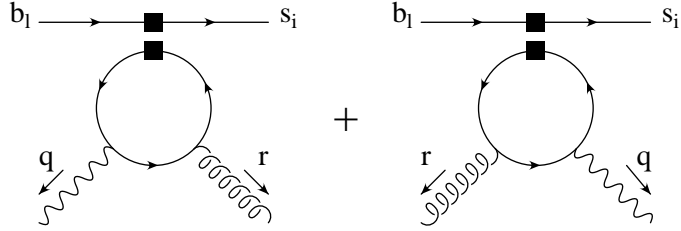}
   \end{center}
\caption{The building block for the $Q_{1\ldots 6}$ contributions. The indices denote the colour of the quarks. The other possible topology is a $\bar s q \times \bar q b$ instead of the $\bar s b \times \bar q q$ operator insertion leading to diagrams with a continuous fermion line. It is equivalent to inserting Fierz-transformed four-quark operators into the displayed topology.\label{fig:BB}}
\end{figure}

We now turn to the mechanism where the spectator participates in the hard scattering. To find the correction for $\langle Q_{1\ldots 6}\rangle$ we compute the left diagrams in Fig. \ref{fig:qit2} using the light-cone projectors in~(\ref{BLCDA}) and~(\ref{phiperpdef}). We first calculate the ``building blocks'' $B_{\alpha\beta}$ of Fig.~\ref{fig:BB}.
%%%%%%%%%%%%%%%%%%%%%%%%%%%%%%%%%%%%%%%%%%%%%%%%%%%%%%%%%%%%%%%%%%%
%%%%%%%%%%%%%%%%%%%%%%%%%%%%%%%%%%%%%%%%%%%%%%%%%%%%%%%%%%%%%%%%%%%
We obtain to leading power, with the photon on-shell and the gluon off-shell,
\begin{equation}\label{BBab}
  B^{(V-A)\times (V\mp A)}_{\alpha\beta} =\mp\frac{iQ_q g_s T^a_{il}}{8\pi^2}\varepsilon_{\rho\sigma\tau\delta}\gamma^\delta\gamma_5\left(g_\alpha^\rho g_\beta^\sigma \left(r^\tau-q^\tau\right) +\frac{r^\sigma q^\tau}{q\cdot r}\left(g_\alpha^\rho q_\beta -g_\beta^\rho r_\alpha\right)\right) (1-\gamma_5)\Delta i
\end{equation}
for the insertion of a $(V\!-\!A)\times (V\!-\!A)$ operator, like in $Q_{1\ldots 4}$, and a $(V\!-\!A)\times (V\!+\!A)$ structure, needed for $Q_{5,6}$, respectively. The relative minus sign of the building block for $(V\!-\!A)\times (V\!-\!A)$ and $(V\!-\!A)\times (V\!+\!A)$ insertions can be easily understood using Furry's theorem. Because a loop diagram with an odd number of vector currents vanishes, it is the sign of the axialvector current inserted in the loop that determines the overall sign. The integral occuring in~(\ref{BBab}) is
\begin{equation}
  \Delta i = 8 \,q\cdot r\int_0^1 \!dx\int_0^{\bar x}\!dy\frac{xy}{m_q^2-r^2 x\bar x-2r\cdot q\,xy-i\delta}
\end{equation}
The $1/\epsilon$ pole of the building block vanishes in the naive dimensional regularization scheme. The use of this scheme is a priori not consistent for the topologies in figure~\ref{fig:BB} because traces with $\gamma_5$ appear. Using another regularization scheme or calculating the corresponding topologies with a continuous fermion line will give a non-vanishing pole contribution. This, however, is mass independent and possible finite anomaly terms emerging from it would be cancelled via the GIM mechanism. Therefore the result is indeed finite and we can safely calculate in $D=4$. We checked that the building block $B^{(V-A)\times (V\mp A)}_{\alpha\beta}$ is the same when calculated with the Fierz-transformed operators inserted into the other possible topology of Fig.~\ref{fig:BB}. Apart from the fact that we set $D=4$ already, kept the leading power part only, and an overall factor of 4 due to the definition of the operator basis, our building block $B^{(V-A)\times(V-A)}_{\alpha\beta}$ is the same as $J_{\alpha\beta}$ in \cite{GHW}. The index $\alpha$ is contracted with the photon polarization vector $\epsilon^\alpha$ and $\beta$ is the gluon index.

For the QCD penguin operators we also have to consider the insertion of the Fierz-transformed operators if $q=s$ or $q=b$. One has the choice of inserting either both possible Fierz versions of a four-quark operator into one topology of a given diagram or one Fierz version into the two possible topologies. The result in the end is the same. For the topology in Fig.~\ref{fig:BB}, however, we can relate the $(V\!-\!A)\times (V\!-\!A)$ and $(V\!-\!A)\times (V\!+\!A)$ operators via Furry's theorem, which saves oneself the calculation of the diagrams with a continuous fermion line. We still performed both calculations to cross-check.

A Fierz transformation of a $(V\!-\!A)\times (V\!-\!A)$ operator again gives a $(V\!-\!A)\times (V\!-\!A)$ operator. Therefore, we can use $B^{(V-A)\times (V-A)}_{\alpha\beta}$ for insertions of $Q_3$ and $Q_4$. Yet, Fierz-transforming $Q_5=(\bar s_i b_l)_{V-A} \sum_q (\bar q_j q_k)_{V+A} \, \delta_{il} \delta_{kj}$ leads to
\begin{equation}\label{q5fierz}
  Q_5^\mathrm{Fierz}=-2 \sum_q (\bar s_i q_k)_{S+P} (\bar q_j b_l)_{S-P} \, \delta_{il} \delta_{kj}
\end{equation}
with the shorthand $(\bar q_1 q_2)_{S\pm P}:=\bar q_1 (1\pm \gamma_5)q_2$. An insertion of such an operator (without the sum over $q$) into the building block diagrams of Fig.~\ref{fig:BB} leads to
\begin{eqnarray}\label{BBSP}
  B^{(S+P)\times(S-P)}_{\alpha\beta} &=& -\frac{iQ_q g_s T^a_{jl}}{2\pi^2} m_q (1+\gamma_5)\Bigg[\left(\frac{q_\beta r_\alpha}{q\cdot r} -g_{\alpha\beta}\right) \Delta i\\
  && \qquad\qquad\qquad\qquad\qquad +\big(q\cdot r g_{\alpha\beta} -r_\alpha q_\beta -i\varepsilon (q,r,\gamma_\alpha,\gamma_\beta)\big) \Delta j\Bigg]\nonumber
\end{eqnarray}
with
\begin{equation}
  \Delta j = \int_0^1 \!dx\int_0^{\bar x}\!dy\frac{2}{m_q^2-r^2 x\bar x-2r\cdot q\,xy-i\delta}
\end{equation}
As the result in~(\ref{BBSP}) contains an overall $m_q$ to leading power only the case $q=b$ has to be considered wheras $q=s$ is of subleading power.

\begin{figure}[t]
   \epsfysize=10cm
   \epsfxsize=12cm
   \centerline{\epsffile{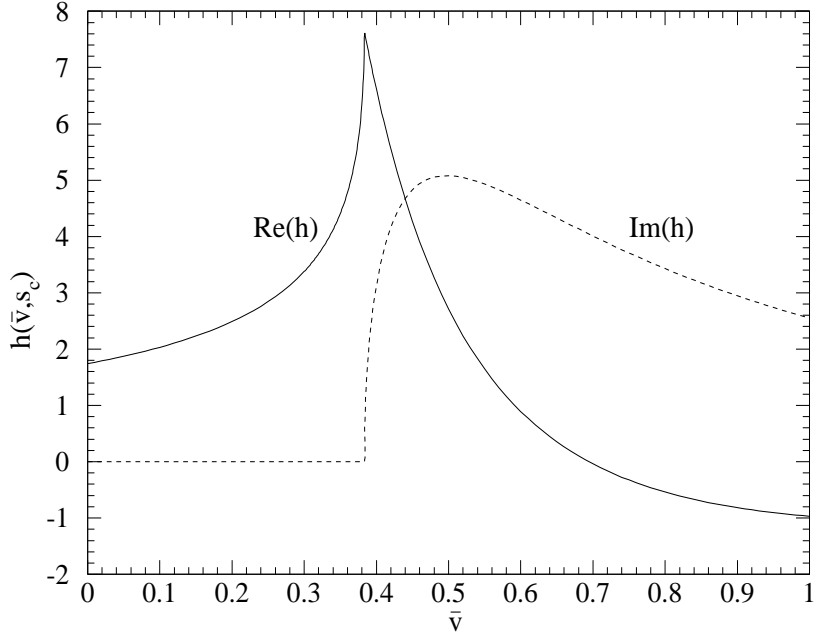}}
\caption{The hard-scattering kernel $h(\bar v,z_c)$, for a charm quark running in the loop, as a function of $\bar v$. \label{fig:hcvbar}}
\end{figure}

If we insert all $B_{\alpha\beta}$ into the left diagrams of Fig.~\ref{fig:qit2} we see that to leading power $r=\bar v k$, the momentum of the spectator quark in the vector meson, i.e. the former light spectator quark of the $B$ meson which was hit by the gluon and absorbed into the vector meson. We can express the result for the $\langle Q_i\rangle$ again in terms of the $Q_7$ matrix element and have
\begin{equation}\label{q1me2}
  \langle Q_i\rangle^{II}=\frac{\alpha_s(\mu_h) C_F}{4\pi} H^V_i(z_q) \langle Q_7\rangle \qquad i\neq 7
\end{equation}
with
\begin{eqnarray}\label{h1s}
  H^V_1(z_p)   &=& -\frac{2\pi^2}{3 N}\frac{f_B f^\perp_V}{F_V m_B^2} \int^1_0 d\xi\frac{\Phi_{B1}(\xi)}{\xi}\int^1_0 dv\, h(\bar v,z_p) \Phi^V_\perp(v)\\
  H^V_2      &=& 0\\
  H^V_3      &=& -\frac{1}{2}\left[ H^V_1(1) +H^V_1(0)\right]\\
  H^V_4(z_c) &=&  H^V_1(z_c)-\frac{1}{2}H^V_1(1)\\ \label{h5s}
  H^V_5      &=&  2 H^V_1(1)\\ \label{h6s}
  H^V_6(z_c) &=& -H^V_1(z_c)+\frac{1}{2}H^V_1(1) =-H^V_4(z_c)
\end{eqnarray}
where the factor $-1/2$ in $H_3^V$ is the ratio $Q_d/Q_u$ of down-type and up-type quark charges and we set the masses of the light quarks $u$, $d$, and $s$ to zero. To ${\cal O}(\alpha_s)$ the matrix element of $Q_2$ is zero because of its colour structure. $Q_5$ contributes to leading power to the hard spectator scattering only for $q=b$. In $H_{4,6}^V$ the contributions of an up, down, and strange quark in the loop cancel out if their masses are neglected. The first integral in~(\ref{h1s}) is just parametrized by $\lambda_B$ of eq.~(\ref{lambdef}) and the hard-scattering function $h(u,z)=\Delta i/u$ in the second integral is given by
\begin{equation}\label{hus}
  h(u,z)=\frac{4z}{u^2}\left[L_2\!\left(\frac{2}{1-\sqrt{\frac{u-4z+i\varepsilon}{u}}}\right)+ L_2\!\left(\frac{2}{1+\sqrt{\frac{u-4z+i\varepsilon}{u}}}\right)\right] -\frac{2}{u}
\end{equation}
$L_2$ is the dilogarithmic function
\begin{equation}\label{dilog}
  L_2(x)=-\int^x_0 dt\frac{\ln(1-t)}{t}
\end{equation}
The function $h(u,z)$ is real for $u\leq 4z$ and develops an imaginary part for $u > 4z$. At small values of $u$ it has the expansion
\begin{equation}\label{husexp}
  h(u,z)=\frac{1}{6z}+\frac{1}{45 z^2}u+ {\cal O}\left(\frac{u^2}{z^3}\right)
\end{equation}
It is also regular for $z\to 0$
\begin{equation}\label{hus0}
  h(u,0)=-\frac{2}{u}
\end{equation}
and real for $z\to 1$. Consequently only the charm quark running in the loop leads to an imaginary part. The function $h(\bar v,z_c)$ for this case is displayed in Fig.~\ref{fig:hcvbar}.
%%%%%%%%%%%%%%%%%%%%%%%%%%%%%%%%%%%%%%%%%%%%%%%%%%%%%%%%%%%%%%%%%%%
%%%%%%%%%%%%%%%%%%%%%%%%%%%%%%%%%%%%%%%%%%%%%%%%%%%%%%%%%%%%%%%%%%%

Inserting the building block $B^{(S+P)\times(S-P)}_{\alpha\beta}$ of~(\ref{BBSP}) into the diagrams in Fig.~\ref{fig:qit2} and performing the projections onto the $B$ meson and vector meson wave function gives to leading power a non-vanishing result only for the $\Delta i$ term in~(\ref{BBSP}). This insertion contributes due to the colour structure only to $H_5^V$ (and not to $H_6^V$), which in turn can be expressed via $H_1^V$ as done in~(\ref{h5s}).\medskip

The correction to $\langle Q_8\rangle$ from the hard spectator interaction comes from the second diagram in Fig. \ref{fig:qit2}. One finds 
\begin{equation}\label{q8me2}
  \langle Q_8\rangle^{II}=\langle Q_7\rangle \frac{\alpha_s(\mu_h) C_F}{4\pi} H^V_8
\end{equation}
where
\begin{equation}\label{h8}
  H^V_8=\frac{4\pi^2}{3 N}\frac{f_B f^\perp_V}{F_V m^2_B} \int^1_0 d\xi\frac{\Phi_{B1}(\xi)}{\xi} \int^1_0 dv\frac{\Phi^V_\perp(v)}{v}
\end{equation}
Here, the hard-scattering function is simply $1/v$ such that the second integral can be performed analytically, leading to the combination $3(1-\alpha^V_1 +\alpha^V_2 +\ldots)$ of Gegenbauer moments in~(\ref{phiperpgbp}).

Because the hard spectator corrections are not proportional to the form factor, we had to remove in $H^V_{1\ldots 6}(z)$ and $H_8^V$ the $F_V$ contained in $\langle Q_7 \rangle$. The gluons in Fig. \ref{fig:qit2} transfer a momentum of order $\mu_h \sim \sqrt{\Lambda_{QCD} m_b}$. Therefore, we set $\alpha_s=\alpha_s(\mu_h)$ in~(\ref{q1me2}) and~(\ref{q8me2}). For our numerical analysis we shall use  $\mu_h=\sqrt{\Lambda_h \mu}$ with $\Lambda_h=0.5$ GeV and $\mu={\cal O}(m_b)$.\medskip

Finally, we can combine the results obtained so far and write, summing the up- and charm-quark contribution
\begin{equation}\label{abkgam}
  A(\bar B\to V\gamma)=\frac{G_F}{\sqrt{2}} \sum_p \lambda_p^{(q)} a^p_7(V\gamma) \langle V\gamma|Q_7|\bar B\rangle
\end{equation}
where at NLO
\begin{eqnarray}
  a^p_7(V\gamma) &=& C_7 + \frac{\alpha_s(\mu) C_F}{4\pi} \left(\sum_{i=1,2} Z_i(\mu) G_i(z_p)+ \sum_{j=3\ldots 6,8} Z_j(\mu) G_j\right) \nonumber \\
  && \quad\,\,+\frac{\alpha_s(\mu_h) C_F}{4\pi} \left( C_1(\mu_h) H^V_1(z_p) +\sum_{j=3\ldots 6,8} C_j(\mu_h) H^V_j\right)\label{a7vgam}
\end{eqnarray}
Here, the NLO expression for $C_7$ has to be used while the leading order values are sufficient for the other Wilson coefficients. The explicit formulas for the Wilson coefficients can be found in \cite{CMM,BBL} and in Appendix~\ref{app:WC}.

The scale dependence of the matrix element of $Q_7$ is reflected in the running of the product of $b$-quark mass and form factor, which is explicitely given as
\begin{equation}\label{runmbff}
  \left( m_b \cdot F_V \right)[\mu] = \left( m_b \cdot F_V \right)[m_b]\left(1-\frac{\alpha_s(\mu)}{4\pi} 8C_F \ln\frac{\mu}{m_b}\right)
\end{equation}
This dependence on the scale $\mu$ has to be taken into account when the residual scale dependence of physical quantities is investigated.

%============================================
%=      The Annihilation Contribution       =
%============================================
\section{The Annihilation Contribution}
\label{sec:ann}

So far, the flavour of the light quark in the $B$ meson and therefore the $B$ meson charge were irrelevant. Yet, if we consider annihilation contributions we become sensitive to whether we are dealing with the decay of a $\bar B^0$ or $B^-$ meson and thus to isospin breaking effects. In \cite{BBVgam} we considered the annihilation contributions from operators $Q_{1,2}$ only. They are CKM suppressed for $B\to K^*\gamma$, but give a sizeable contribution to $B\to\rho\gamma$ decays. Kagan and Neubert found that in particular the operator $Q_6$ gives a large contribution to $B\to K^*\gamma$ because for it the CKM suppression is lifted \cite{KN}. So in the following we will calculate annihilation contributions from $Q_{1\ldots 6}$. On the other hand, we decided not to include the subleading power contributions in Fig.~\ref{fig:subl} because they are additionally suppressed by $\alpha_s$ or not calculable. We will further comment on this in the following section.

If the photon emission is from the light quark in the $B$ meson the annihilation amplitude contains
\begin{equation}\label{bVdef}
  b^V =\frac{2\pi^2}{F_V}\frac{f_B m_V f_V}{m_B m_b \lambda_B}
\end{equation}
whereas the $Q_{5,6}$ insertion with the photon emitted from one of the vector meson constituent quarks leads to
\begin{equation}\label{dVdef}
  d^V_{\stackrel{(-)}{v}} =-\frac{4 \pi^2}{F_V}\frac{f_B f_V^\perp}{m_B m_b} \int_0^1 \frac{dv}{\stackrel{(-)}{v}} \Phi_V^\perp(v)
\end{equation}
Recalling that $F_\rho\sim m^{-3/2}_b$, $f_B\sim m^{-1/2}_b$ in the heavy-quark limit, we note that $b^V,d^V\sim \Lambda_{QCD}/m_b$. This shows explicitly the power suppression of weak annihilation. The integral in (\ref{dVdef}) again can be performed and leads to $3(1 \mp\alpha^V_1 +\alpha^V_2 +\ldots)$ for $v$ and $\bar v$ in the denominator, respectively. The generic $Q_{5,6}$ annihilation matrix elements therefore have the opposite sign and an approximately twice as large absolute value than the ones from $Q_{1\ldots 4}$.

The annihilation amplitude for $B^-\to\rho^-\gamma$ for example is
\begin{equation}\label{annB0}
  A_{ann}(B^-\to\rho^-\gamma) =\frac{G_F}{\sqrt{2}} Q_u\left[ \lambda_u^{(d)} a_1 b^\rho +\!(\lambda_u^{(d)} \!+\!\lambda_c^{(d)})\!\left(a_4 b^\rho +\!\frac{Q_s}{Q_u} a_6 d_v^\rho +a_6 d^\rho_{\bar v}\right)\right] \langle \rho^-\gamma| Q_7 |B^-\rangle
\end{equation}
where the $a_i$ denote the following combinations of Wilson coefficients, evaluated in leading logarithmic approximation
\begin{eqnarray}
  a_{1,2} &=& C_{1,2}+\frac{1}{N} C_{2,1} \label{a12def}\\
  a_4     &=& C_4 +\frac{1}{N}C_3 \label{a4def}\\
  a_6     &=& C_6 +\frac{1}{N}C_5 \label{a6def}
\end{eqnarray}
The annihilation components are included in the decay amplitudes by substituting
\begin{eqnarray}
  \label{a7uann} a_7^u &\to& a_7^u +a_{ann}^u\\
  \label{a7cann} a_7^c &\to& a_7^c +a_{ann}^c
\end{eqnarray}
where
\begin{eqnarray}\label{aann}
      a_{ann}^u(\bar K^{*0}\gamma) & = & Q_d \left[ a_4 b^{K^*} +a_6(d_v^{K^*} +d^{K^*}_{\bar v})\right]\\
      a_{ann}^u(K^{*-}\gamma)      & = & Q_u \left[ a_1 b^{K^*} +a_4 b^{K^*} +Q_s/Q_u a_6 d_v^{K^*} +a_6 d^{K^*}_{\bar v}\right]\\[0.1cm]
      a_{ann}^u(\rho^0\gamma)      & = & Q_d \left[-a_2 b^\rho +a_4 b^\rho +a_6(d^\rho_v +d^\rho_{\bar v})\right]\\
      a_{ann}^u(\rho^-\gamma)      & = & Q_u \left[ a_1 b^\rho +a_4 b^\rho +Q_s/Q_u a_6 d_v^\rho +a_6 d^\rho_{\bar v}\right]\\[0.2cm]
      a_{ann}^c(\bar K^{*0}\gamma) & = & Q_d \left[ a_4 b^{K^*} +a_6(d_v^{K^*} +d^{K^*}_{\bar v})\right]\\
      a_{ann}^c(K^{*-}\gamma)      & = & Q_u \left[ a_4 b^{K^*} +Q_s/Q_u a_6 d_v^{K^*} +a_6 d^{K^*}_{\bar v}\right]\\[0.1cm]
      a_{ann}^c(\rho^0\gamma)      & = & Q_d \left[ a_4 b^\rho +a_6(d^\rho_v +d^\rho_{\bar v})\right]\\
      a_{ann}^c(\rho^-\gamma)      & = & Q_u \left[ a_4 b^\rho +Q_s/Q_u a_6 d_v^\rho +a_6 d^\rho_{\bar v}\right]
\end{eqnarray}
In $\bar B^0\to\rho^0\gamma$ the minus sign of $a_2$ comes from the relative sign between the up-quark and down-quark components of the $\rho^0$ wave function. It is the up quarks only that produce the $\rho^0$ in the $Q_{1,2}$ annihilation process while only the down quarks are relevant in $\langle Q_7\rangle$.

We include annihilation contributions in our numerical analysis unless otherwise stated.

%===============================
%=      Power Corrections      =
%===============================

\section{A Comment on Power Corrections}
\label{sec:power}

Here, we want to comment on the role of power corrections in $B\to V\gamma$ decays. Let us first repeat that the annihilation effect from operator $Q_{1}$ gives a numerically important power correction because it comes with a relative enhancement factor of $|C_1/C_7|\sim 3$. This leads to a $40\%$ correction in the up- and a $4\%$ correction in the total amplitude of the charged mode $B^-\to\rho^-\gamma$. Note that the parameter $Q_u b^V$, describing the generic effect of the $Q_{1\ldots 4}$ annihilation terms, is $18\%$, which is consistent with a $\Lambda_{QCD}/m_b$ correction of canonical size. The contribution $Q_u d^V$ from $Q_{5,6}$, however, can be as large as $40\%$ which is larger than naively expected. But it is accompanied by the very small Wilson coefficient combination $a_6\approx -0.03$.

The power-suppressed penguin annihilation diagrams of Fig.~\ref{fig:subl} can be calculated within QCD factorization, but they contain an $\alpha_s({\cal O}(m_b))$ and hence their effect is safely negligible. For the same reason we do not include QCD corrections to the annihilation diagrams in Fig.~\ref{fig:ann}.

The subleading power contributions for the matrix element of the chromomagnetic dipole operator $Q_8$ deserve some special inspection. For one of them factorization breaks down. Kagan and Neubert already calculated the diagram in Fig.~\ref{fig:subl} where the photon is emitted from the vector meson spectator \cite{KN}. This diagram is explicitely suppressed by power counting because both the gluon and quark propagator are hard. So it is sufficient to consider the projection onto the leading-power part of the vector meson wave function. We arrive at a result containing the integral
\begin{equation}\label{intphiperp}
  \alpha_s(\mu)\int_0^1\!dv\frac{\Phi_V^\perp(v)}{\bar v^2}
\end{equation}
which suffers from a logarithmic divergence as $v\to 1$ if $\Phi_V^\perp(v)$ vanishes only linearly at the endpoints. Exactly such infrared divergences signal the breakdown of factorization at subleading power. As this contribution is dominated by soft physics we cannot bother the not necessarily small $\alpha_s$ in~(\ref{intphiperp}) to simply neglect it. Although the endpoint singularity can be regulated by introducing a cutoff in the convolution integral \cite{KN} we refrain from including this contributions in our phenomenological analysis. The reason is that no explicit numerical enhancement, as for the annihilation contributions, is present and that there are other sub-leading power corrections which are not (yet?) calculable. These we discuss next.

At the same level of power counting we also have contributions from the photon emission from the other three possible quark lines in the $Q_8$ diagram. As already mentioned in section~\ref{sec:factBVgam}, diagrammatically these contributions are of leading power. Yet, the leading order projection gives a non-vanishing result only for the contribution in Fig.~\ref{fig:qit2}. But if we start to consider the contribution from the subleading power diagram with the projection onto the leading power wave function, we also have to treat the leading power diagrams with projections onto subleading power LCDAs. These are calculable within QCD factorization, but the $Q_8$ diagram of Fig.~\ref{fig:qit2} involves not only the first negative moment of $\Phi_{B1}$ but also that of $\Phi_{B2}$. This is unknown and would have to be parametrized by a further non-perturbative quantity $\lambda_{B2}$. But there are contributions which are even less known. Among these are the contributions from keeping the transverse momentum $l_\perp$ of the spectator quark inside the $B$ meson. This would require to modify the $B$-meson wave functions in~(\ref{BLCDA}). Yet, compared with the annihilation contributions, the power corrections from $Q_8$ are numerically absolutely insignificant. Therefore we will not include them in our numerical analysis in the following chapter.

In contrast to nonleptonic modes with pseudoscalar mesons in the final state, no chirally enhanced power corrections from the light-cone expansion of meson wave functions arise in our case. Finally, power corrections can also come from the loop effects with up- and  charm quarks whose leading-power contributions were computed in~(\ref{q1me2}). These power corrections correspond to the region of integration where the gluon becomes soft, that is $\bar v={\cal O}(\Lambda_{QCD}/m_b)$. Their contribution is nonperturbative and cannot be calculated in the hard-scattering formalism. Nevertheless, the expression in~(\ref{q1me2}) can be used to read off the scaling behaviour of these power corrections in the heavy-quark limit. For the charm loop the kernel approaches a constant $\sim m^2_b/m^2_c$ in the endpoint region as shown in~(\ref{husexp}). Taking into account the linear endpoint suppression of the wave function $\Phi_V^\perp$, the integral in~(\ref{q1me2}) over the region $\bar v\sim \Lambda_{QCD}/m_b$ therefore contributes a term of order $(\Lambda_{QCD}/m_b)^2\times (m_b/m_c)^2 = (\Lambda_{QCD}/m_c)^2$. It is interesting that we thus recover the power behaviour of soft contributions in the charm sector first pointed out in \cite{Voloshin}. This was discussed for the inclusive decay $b\to s\gamma$ in \citer{Voloshin,BIR} and for the exclusive mode $B\to K^*\gamma$ in \cite{KRSW}. Numerically, this correction is with $\sim 3\%$ in the decay rate very small. A similar consideration applies to the up-quark sector. In this case the endpoint behaviour of the kernel is singular~(\ref{hus0}), which now leads to a linear power suppression of the form $\Lambda_{QCD}/m_b$. This coincides with the scaling behaviour derived in \cite{BIR} in the context of the inclusive process.

In view of these remarks, we see no indication for large uncalculable power corrections. Further studies on this issue are definitely desireable, but would go beyond the scope of this work.

%=============================
%=      The Observables      =
%=============================

\section{The Observables}
\label{sec:obs}

In this section we present the formulas for the observables of interest in the phenomenology of $B\to V\gamma$ decays. These are the branching ratio, CP and isospin-breaking asymmetries, and U-spin breaking effects.

From the amplitude in~(\ref{abkgam}) the branching ratio is obtained as
\begin{equation}\label{brbvgam}
  B(\bar B\to V\gamma)=\tau_B\frac{G^2_F \alpha \,m^3_B m^2_b}{32\pi^4}\left(1-\frac{m^2_V}{m^2_B}\right)^{\!3} \Big|\sum_p \lambda_p^{(q)}\, a^p_7(V\gamma)\Big|^2 c_V^2 |F_V|^2
\end{equation}
where $q=s$ for $V=K^*$ and $q=d$ for $V=\rho$.

The rate for the CP-conjugated mode $B\to V\gamma$ is obtained by replacing $\lambda_p^{(q)} \to \lambda_p^{(q)*}$. We may then consider the CP asymmetry
\begin{equation}\label{acpbrgdef}
  {\cal A}_{CP}(V\gamma)=\frac{\Gamma(B\to\rho\gamma)-\Gamma(\bar B\to\rho\gamma)}{\Gamma(B\to\rho\gamma)+\Gamma(\bar B\to\rho\gamma)}
\end{equation}
with $\Gamma$ denoting the rate. A non-vanishing CP asymmetry appears at ${\cal O}(\alpha_s)$ only. Expanding ${\cal A}_{CP}$ in $\alpha_s$ we obtain
\begin{eqnarray}\label{acpbrgamsimp}
  {\cal A}_{CP}(V\gamma) &\approx& \frac{2\,{\rm Im}\lambda^{(q)*}_u\lambda^{(q)}_c}{|\lambda^{(q)}_t|^2}\, \frac{{\rm Im}a^{u*}_7 a^c_7}{|C_7|^2} \\
  &=&\frac{\alpha_s C_F}{4\pi}\, \frac{1}{|C_7|}\left[\sum_{i=1,2}Z_i\,\mathrm{Im}\left(G_i(z_c)-G_i(0)\right) +C_1\,\mathrm{Im}\left(H_1^V(z_c)-H_1^V(0)\right)\right] \nonumber\\ \label{acpsimp}
  && \qquad \left\{\begin{array}{cll} 2\bar\eta \lambda^2 & \mbox{for} & V=K^*\\[0.2cm] \displaystyle-\frac{2\bar{\eta}}{(1-\bar\rho)^2+\bar{\eta}^2} & \mbox{for} & V=\rho \end{array}\right.
\end{eqnarray}
where we neglected annihilation contributions and used only one $\alpha_s$. When expanding in $\alpha_s$ we in particular considered the denominator in (\ref{acpbrgdef}) to ${\cal O}(\alpha_s^0)$ only because the numerator starts with ${\cal O}(\alpha_s)$. In this approximation the effects from penguin operators cancel out completely in~(\ref{acpsimp}). We see that the CP asymmetry in $B\to K^*\gamma$ decays is suppressed to the one in $B\to\rho\gamma$ via the factor $-|\lambda_t^{(d)}/\lambda_t^{(s)}|^2=-\lambda^2\left[(1-\bar\rho^2)^2+\bar\eta^2\right]\approx -0.04$ and is thus very small.

Further interesting observables are the isospin-breaking quantities
\begin{eqnarray}\label{iso+0}
  \Delta^V_{+0} &=& \frac{c_V^2\Gamma(B^+\to V^+\gamma)}{\Gamma(B^0\to V^0\gamma)}-1\\ \label{iso-0}
  \Delta^V_{-0} &=& \frac{c_V^2\Gamma(B^-\to V^-\gamma)}{\Gamma(\bar{B}^0\to V^0\gamma)}-1\\ \label{isobreak}
  \Delta(V\gamma) &=& \frac{\Delta^V_{+0}+\Delta^V_{-0}}{2}
\end{eqnarray}
Within our approximations, isospin breaking is generated by weak annihilation. Isospin breaking was already discussed in \cite{AHL}, partially including NLO corrections. It was considered in more detail for $B\to K^*\gamma$ in \cite{KN}.

Another application of our results concerns an estimate of U-spin breaking effects in $B\to V\gamma$ decays. U-spin symmetry, the symmetry of strong interactions under exchange of $d$ and $s$ quarks, has been advocated as a tool to control hadronic uncertainties in tests of the Standard Model \cite{RF,GRG}. Defining
\begin{eqnarray}
  \Delta B(B\to K^*\gamma) &=& B(B^+\to K^{*+}\gamma) - B(B^-\to K^{*-}\gamma) \label{dbk} \\
  \Delta B(B\to \rho\gamma) &=& B(B^+\to \rho^{+}\gamma) - B(B^-\to \rho^{-}\gamma) \label{dbrho}
\end{eqnarray}
the quantity
\begin{equation}\label{dbb}
  \Delta B(B\to K^*\gamma)+\Delta B(B\to \rho\gamma) \equiv 0
\end{equation}
in the limit of U-spin symmetry and within the Standard Model. This has been discussed in \cite{GRG,JMS} and was considered in more detail in \cite{HM}.\medskip

Because all the contributions to $B\to V\gamma$ are proportional to the matrix element of $Q_7$ in~(\ref{q7f1f2}), the polarization of the final state photon is left-handed for a $\bar B$ and right-handed for a $B$ decay. Small corrections of ${\cal O}(m_s/m_b)$ come from the $m_s(1-\gamma_5)$ contribution in the operator $Q_7$ which we neglected. In many extensions of the Standard Model the $(1-\gamma_5)$ part in $Q_7$ can be considerably enhanced due to a chirality flip along a heavy fermion line in the electroweak loop process. If the New Physics contribution appears not already at tree level, the only effect is a new local $Q_7'$ operator with $\bar s\sigma^{\mu\nu} (1-\gamma_5)b\,F_{\mu\nu}$ structure and a corresponding Wilson coefficient $C_7'$. Such an operator would not affect our result for $a_7^p$ in (\ref{a7vgam}) but simply give an additional contribution $C_7'\langle Q_7'\rangle$ to (\ref{abkgam}). Here the Dirac structure of $\langle Q_7'\rangle$ differs from the one given in (\ref{q7f1f2}) by a relative minus sign in front of the $i (\epsilon\cdot\eta\, k\cdot q-\epsilon\cdot k\, \eta\cdot q)$ term.

 A real photon can convert inside the detector via the Bethe-Heitler process \cite{BH} into an electron-positron pair. Any significant measurement of an angular dependence of this conversion rate will be an indication for new physics. Yet, the experimental efficiency for this measurement is only at the few percent level \cite{GrP}.

% ===== Phenomenology B -> V gamma =======================
\chapter{Phenomenology of $B\to V\gamma$}
\label{ch:phenBVgam}

Here we want to evaluate the results of the previous chapter numerically for the $B\to K^*\gamma$ and $B\to\rho\gamma$ decays. Our choice of input parameters is summarized in table \ref{tab:input}. The default choice for the CKM angle $\gamma$ is $58^\circ$.
%%%%%%%%%%%%%%%%%%%%%%%%%%%%%%%%%%%%%%%%%%%%%%%%%%%%
%%%%%%%% table of input parameters   %%%%%%%%%%%%%%%
%%%%%%%%%%%%%%%%%%%%%%%%%%%%%%%%%%%%%%%%%%%%%%%%%%%%
\begin{table}
\renewcommand{\arraystretch}{1.2}
\begin{center}
\begin{tabular*}{140mm}{@{\extracolsep\fill}|c|c|c|c|c|c|}
\hline\hline
\multicolumn{6}{|c|}{CKM parameters and coupling constants}\\
\hline
$V_{us}$ & $V_{cb}$ & $\left|V_{ub}/V_{cb}\right|$ & 
$\Lambda_{\overline{MS}}^{(5)}$ & $\alpha$ & $G_F$\\
\hline
0.22 & 0.041 & $0.085 \pm 0.025$ & $(225 \pm 25)$ MeV & 1/137 & $1.166 \times 10^{-5} 
\mbox{GeV}^{-2}$\\
\hline
\end{tabular*}

\begin{tabular*}{140mm}{@{\extracolsep\fill}|c|c|c|c|c|}
\hline
\multicolumn{5}{|c|}{Parameters related to the $B$ mesons}\\
\hline
$m_B$ & $f_B$ \cite{SMR} & $\lambda_B$ & $\tau_{B^+}$ & $\tau_{B^0}$\\
\hline
5.28 GeV & (200$\pm$30) MeV & $(350 \pm 150)$ MeV & 1.65 ps & 1.55 ps\\
\hline
\end{tabular*}

\begin{tabular*}{140mm}{@{\extracolsep\fill}|c|c|c|c|c|c|}
\hline
\multicolumn{6}{|c|}{Parameters related to the $K^*$ meson \cite{BB98}}\\
\hline
$F_{K^*}$ & $f_{K^*}^\perp$ & $m_{K^*}$ & $\alpha^{K^*}_1$ & 
$\alpha^{K^*}_2$ & $f_{K^*}$ \\
\hline
$0.38 \pm 0.06$ & 185 MeV & 894 MeV & 0.2 & 0.04 & 230 MeV\\
\hline\hline
\multicolumn{6}{|c|}{Parameters related to the $\rho$ meson \cite{BB98}}\\
\hline
$F_\rho$ & $f_\rho^\perp$ & $m_\rho$ & $\alpha^\rho_1$  & $\alpha^\rho_2$ & 
$f_\rho$\\
\hline
$0.29 \pm 0.04$ & 160 MeV & 770 MeV & 0 & 0.2 & 200 MeV\\
\hline
\end{tabular*}

\begin{tabular*}{140mm}{@{\extracolsep\fill}|c|c|c|c|}
\hline
\multicolumn{4}{|c|}{Quark and W-boson masses}\\
\hline
$m_b(m_b)$ & $m_c(m_b)$ & $m_{t,\mbox{pole}}$ & $M_W$\\
\hline
$(4.2 \pm 0.2)$ GeV & $(1.3 \pm 0.2)$ GeV & 174 GeV & 80.4 GeV\\
\hline\hline
\end{tabular*}
\end{center}
\caption[]{Summary of input parameters.\label{tab:input}}
\end{table}
%%%%%%%%%%%%%%%%%%%%%%%%%%%%%%%%%%%%%%%%%%%%%%%%%%%%%%%%%%%%%%%%%

%===============================
%=      Numerical Results      =
%===============================

\section{Numerical results}
\label{sec:numBVgam}

We begin with the numerical result for the NLO QCD coefficients $a_7^p(K^*\gamma)$ as a typical example. For central values of all parameters, at $\mu=m_b$, and displaying separately the size of the various correction terms without annihilation effects, we find
\begin{eqnarray}
 \label{numa7cK}
  a_7^c(K^* \gamma) &=& \begin{array}[t]{cccc}-0.3221 & +0.0113 & -0.0796-0.0134i & -0.0166-0.0123i\\ C_7^{LO} & \Delta C_7^{NLO} & T^I\mbox{-contribution} & T^{II}\mbox{-contribution}\\ \end{array} \nonumber\\
  &=& -0.4070 -0.0257i.\\[0.2cm]
 \label{numa7uK}
  a_7^u(K^* \gamma) &=& \begin{array}[t]{cccc}-0.3221 & +0.0113 & -0.1397-0.0682i & +0.0338-0.0002i\\ C_7^{LO} & \Delta C_7^{NLO} & T^I\mbox{-contribution} & T^{II}\mbox{-contribution}\\ \end{array} \nonumber\\
  &=& -0.4167 -0.0684i.
\end{eqnarray}
We note a sizable enhancement of the leading order value, dominated by the $T^I$-type correction. This feature was already observed in the context of the inclusive case in \cite{GHW}. A complex phase is generated at NLO where the hard-vertex corrections $T^I$ and the hard-spectator interactions $T^{II}$ yield comparable effects for the charm amplitude. Although the hard-spectator interaction $H^V_1(0)$ for a massless up quark running in the loop is real, a tiny imaginary part is generated from the QCD penguin operators $Q_{4,6}$ also in the up amplitude. The numerical results for the $a_7^p(\rho\gamma)$ are very similar. Then, with the two equations above and (\ref{acpbrgamsimp}), we can immediately estimate the CP asymmetries ${\cal A}_{CP}^\mathrm{eq.(\ref{acpbrgamsimp})}(K^*\gamma)\approx -0.5\%$ and ${\cal A}_{CP}^\mathrm{eq.(\ref{acpbrgamsimp})}(\rho\gamma)\approx +13\%$ with annihilation contributions neglected. The strict expansion in $\alpha_s$ with the neglect of annihilation contributions and NLO corrections to the denominator of the CP asymmetry overestimates its numerical value.

For $B\to K^*\gamma$ the effect from the annihilation contributions to the charm-quark part of the amplitude is numerically small because only the penguin operators with tiny Wilson coefficients can contribute:
\begin{eqnarray}\label{acann0num}
  a_{ann}^c(\bar K^{*0}\gamma) &=& -\frac{1}{3}\Big[\begin{array}[t]{ccc} -0.0063 & +0.0138 & +0.0203\\ a_4 b^{K^*} & a_6 d_v^{K^*} & a_6 d_{\bar v}^{K^*}\end{array}\Big] =-0.0093\\ \label{acann1num}
  a_{ann}^c(K^{*-}\gamma) &=& +\frac{2}{3}\Big[\begin{array}[t]{ccc} -0.0063 & -0.0069 & +0.0203\\ a_4 b^{K^*} & Q_s/Q_u a_6 d_v^{K^*} & a_6 d_{\bar v}^{K^*}\end{array}\Big]= +0.0047
\end{eqnarray}
But as they come with opposite sign, they still lead to a $3.4\%$ difference on the amplitude level which results in a $-7.5\%$ isospin asymmetry. They also suffice to make the expectation for the branching ratio $B(\bar B^0\to \bar K^{*0}\gamma)=7.40\cdot 10^{-5}$ larger than that of $B(B^-\to K^{*-}\gamma)=7.25\cdot 10^{-5}$ although $\tau_{B^0}<\tau_{B^+}$. The operators $Q_{1,2}$ with large Wilson coefficients contribute to annihilation only in the up-quark sector. We have $a_1 b^{K^*}=0.2781$ which gives $a_{ann}^u(K^{*-}\gamma)=+0.1901$. This represents $45\%$ of the absolute value of $a_7^u(K^*\gamma)$. Yet, the strong CKM suppression $|\lambda_u^{(s)}/\lambda_c^{(s)}|\approx 0.02$ puts this large correction for $B\to K^*\gamma$ into perspective. If we consider the isospin breaking due to $Q_{1,2}$ only, we get $\Delta(K^*\gamma)=-0.9\%$. For $B\to\rho\gamma$ the numerical values for the $a^p_\mathrm{ann}$ are similar, but as both sectors of the effective Hamiltonian have the same order of magnitude the impact of annihilation contributions is larger.\medskip

\begin{figure}
\includegraphics[width=8cm]{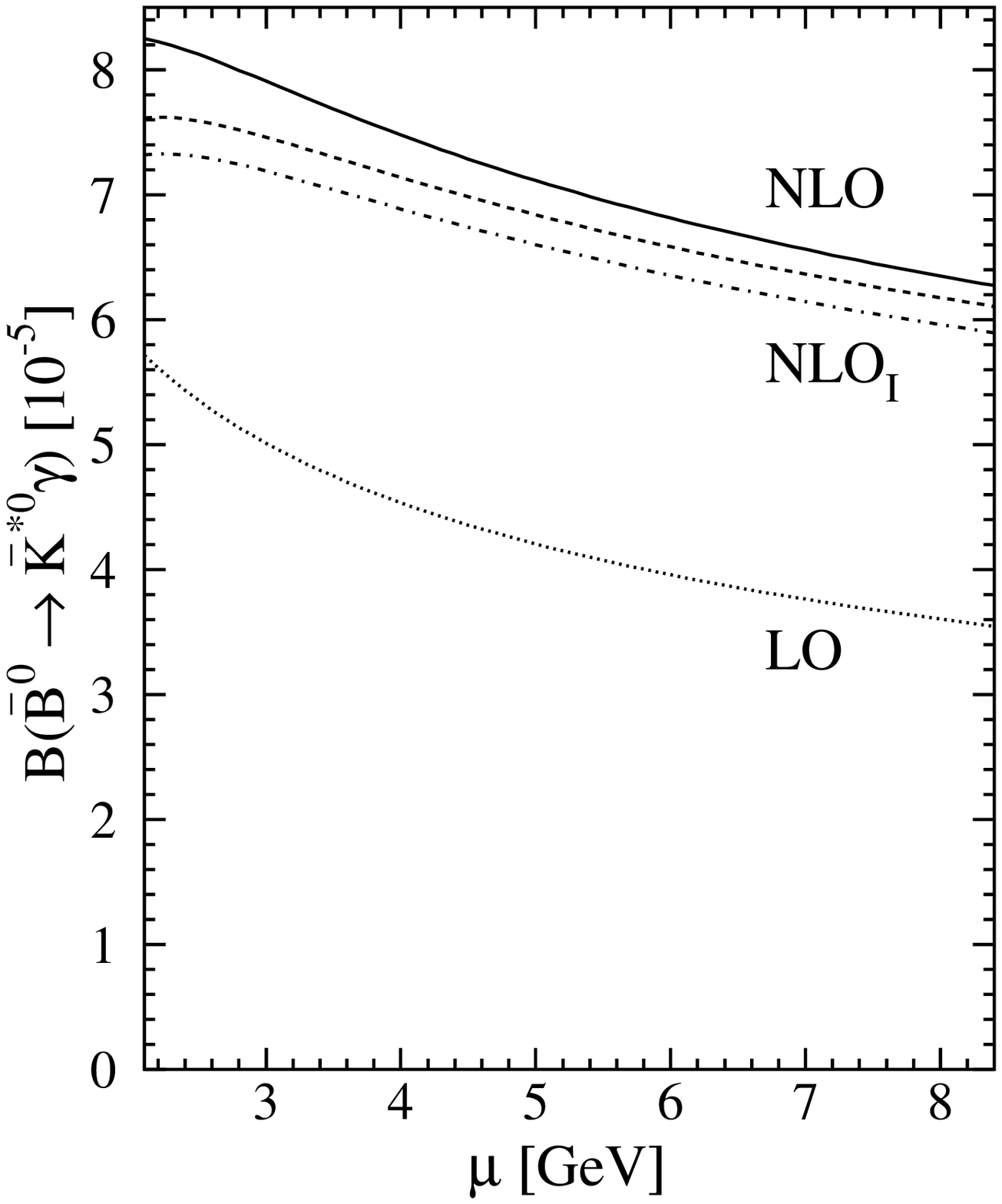}\includegraphics[width=8cm]{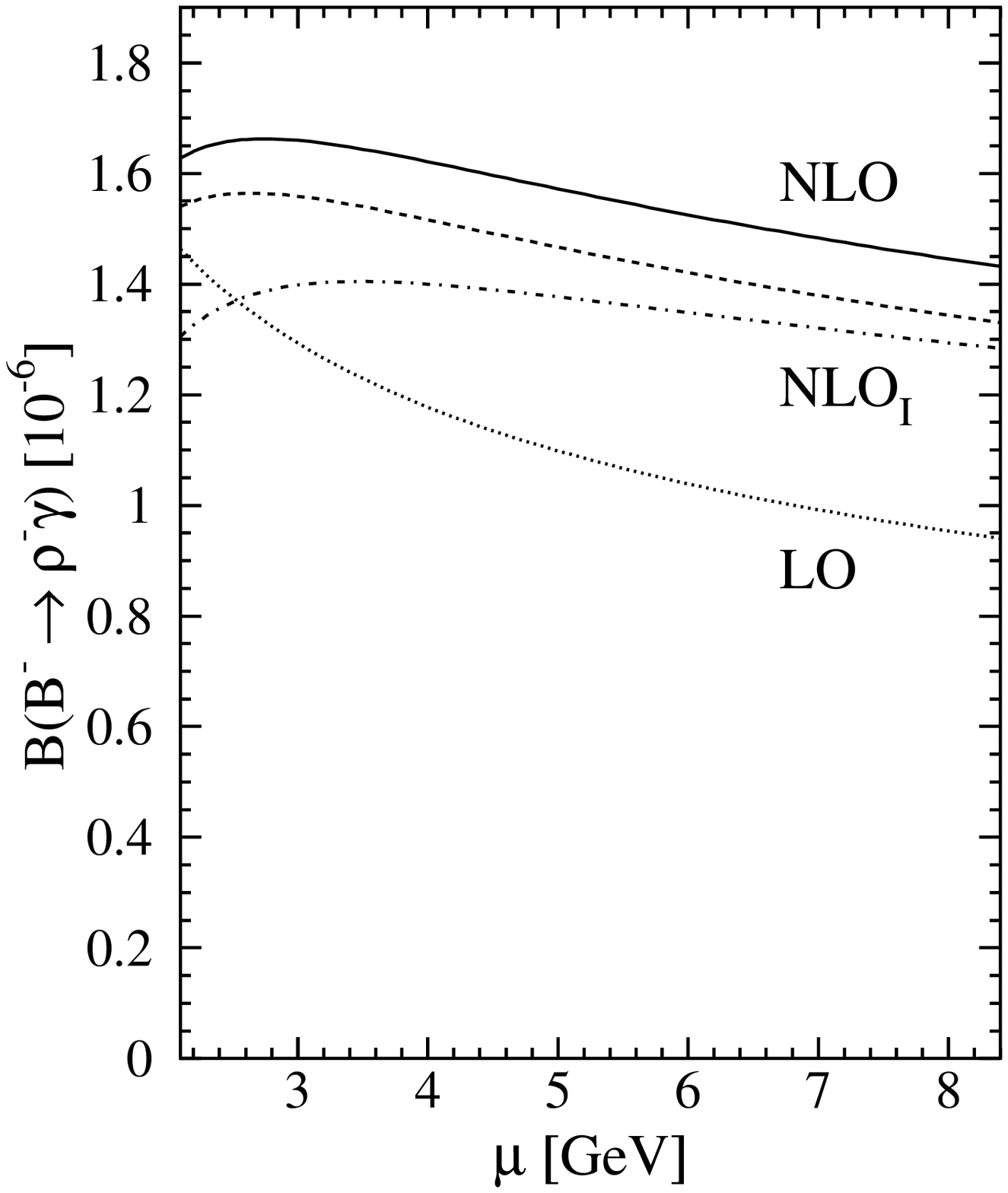}
\caption{Dependence of the branching fractions $B(\bar{B}^0 \to \bar{K}^{*0} \gamma)$ and $B(B^- \to \rho^- \gamma)$ on the renormalization scale $\mu$. The dotted line shows the LO result, the dash-dotted line the NLO result including type-I corrections only, and the solid line shows the complete NLO result for central values of all input parameters. For these three results the effect of annihilation contributions was included. The dashed line, finally, shows the NLO result without annihilation contributions.}
\label{fig:bkrhomu}
\end{figure}

We now turn to analyze the implications of the numerics at amplitude level for our observables of interest. The net enhancement of $a_7$ at NLO leads to a corresponding enhancement of the branching ratios, for fixed value of the form factor. This is illustrated in Fig. \ref{fig:bkrhomu} where we show the residual scale dependence for $B(\bar{B}\to \bar{K}^{*0}\gamma)$ and $B(B^-\to\rho^-\gamma)$ at leading and next-to-leading order.
%%%%%%%%%%%%%%%%%%%%%%%%%%%%%%%%%%%%%%%%%%%%%%%%%%%%%%%%%%%%%%%%%%%
%%%%%%%%%%%%%%%%%%%%%%%%%%%%%%%%%%%%%%%%%%%%%%%%%%%%%%%%%%%%%%%%%%%
As expected, the inclusion of the hard-vertex corrections $T^I$ reduces the scale dependence coming from the Wilson coefficients. The scale dependence of the complete NLO result, however, is ``deteriorated'' because the type-II contributions appear at ${\cal O}(\alpha_s)$ for the first time and therefore introduce a completely new scale dependence which would be cancelled at NNLO only. The effect of annihilation contributions on the branching ratios is small for our default choice of the CKM angle $\gamma =58^\circ$

The sensitivity of the $\bar{B}\to \bar{K}^{*0}\gamma$ and $B^-\to\rho^-\gamma$ branching ratios, and of the CP asymmetry ${\cal A}_{CP}(\rho^\pm\gamma)$ to variations in the relevant input parameters is summarized in table \ref{tab:output}.
%%%%%%%%%%%%%%%%%%%%%%%%%%%%%%%%%%%%%%%%%%%%%%%%%%%%%%%%%%%%%%%%%%%
%%%%%%%%          Table of output parameters               %%%%%%%%
%%%%%%%%%%%%%%%%%%%%%%%%%%%%%%%%%%%%%%%%%%%%%%%%%%%%%%%%%%%%%%%%%%%
\begin{table}
\renewcommand{\arraystretch}{1.2}
\begin{center}
\begin{tabular}{|l|c|c|c|}
\hline\hline
  & $B(\bar{B}^0 \to \bar{K}^{*0} \gamma)\, [10^{-5}]$ &  $B(B^- \to \rho^- \gamma)\, [10^{-6}]$ & ${\cal A}_{CP}(\rho^\pm\gamma)\, [\%]$\\
\hline\hline
central                         & 7.40           & 1.61          & 10.11\\
\hline\hline
$F_{K^*}$                       & +2.35/$-$2.03  &   --          &  --  \\
\hline
$F_{\rho}$                      &   --           & +0.42/$-$0.37 & +0.08/$-$0.06\\
\hline
$m_b/2 <\mu < 2m_b$             & +0.85/$-$1.13  & +0.05/$-$0.18 & +5.65/$-$2.51\\
\hline
$m_b$                           & +0.57/$-$0.55  & +0.12/$-$0.12 & +0.16/$-$0.14\\
\hline
$\lambda_B$                     & +0.40/$-$0.15  & +0.39/$-$0.13 & +0.31/$-$0.14\\
\hline
$m_c$                           & +0.38/$-$0.44  & +0.11/$-$0.11 & +1.21/$-$1.29\\
\hline
$\Lambda_{\overline{MS}}^{(5)}$ & +0.22/$-$0.23  & +0.04/$-$0.04 & +0.17/$-$0.19\\
\hline
$\left|V_{ub}/V_{cb}\right|$    & +0.05/$-$0.05  & +0.15/$-$0.13 & +3.54/$-$3.33\\
\hline
$\gamma=(58\pm 24)^\circ$       & +0.08/$-$0.11  & +0.34/$-$0.22 & +0.12/$-$2.20\\
\hline
$f_B$                           & +0.14/$-$0.14  & +0.06/$-$0.06 & +0.10/$-$0.05\\
\hline
$m_b\leftrightarrow m_B$        & +0.24/$-$0.66  & +0.11/$-$0.14 & +0.00/$-$0.81\\
\hline\hline
\end{tabular}
\end{center}
\caption[]{Predictions for branching ratios and CP asymmetries with the errors from the individual input uncertainties.\label{tab:output}}
\end{table}
%%%%%%%%%%%%%%%%%%%%%%%%%%%%%%%%%%%%%%%%%%%%%%%%%%%%%%%%%%%%%%%%%%%
Numerical differences compared to the results presented in table 2 of \cite{BBVgam} are due to a change in the input value for the $B$ meson decay constant from $f_B=180\,\mathrm{MeV}$ to $f_B=200\,\mathrm{MeV}$ ($+0.8\%$ for the $\bar{K}^{*0}$ and $+2.5\%$ for the $\rho^-$ mode branching ratio) and the inclusion of QCD penguin operator contributions ($+3.6\%$ and $-2.4\%$, respectively). The uncertainty of the branching fractions is currently dominated by the form factors $F_{K^*}$, $F_\rho$. They are taken from QCD sum rules \cite{BB98} with an error of $15\%$ and enter the branching ratio quadratically. This situation, however, can be systematically improved. In particular, our approach allows for a consistent perturbative matching of the form factor to the short-distance part of the amplitude. For $B(B^-\to\rho^-\gamma)$ a considerable effect also comes from the variation of the hadronic parameter $\lambda_B$ and the CKM angle $\gamma$. The vector meson decay constants $f_V^{(\perp)}$ and the Gegenbauer moments $\alpha^V_i$ are not known very precisely either, but they appear always in combination with $F_V$ and/or $\lambda_B$, which are both generously varied. The distinction between $m_b$ and $m_B$ in $G_i$, $H_i^V$, $b^V$, and $d^V$ is, strictly speaking, a subleading power issue. An estimate of this effect is given in the last row of our output tables.

Adding the uncertainties for the branching ratios in table~\ref{tab:output} in quadrature leads to an error estimate of ${\cal O}(35\%)$ and ${\cal O}(40\%)$ for the $B\to K^*\gamma$ and $B\to\rho\gamma$ mode, respectively. Taking this sizable uncertainties into account, the results for $B\to K^*\gamma$ are compatible with the experimental measurements quoted in~(\ref{b0kgamex}) and (\ref{bpkgamex}), even though the central theoretical values appear to be somewhat high. According to our predictions we expect the $B\to\rho\gamma$ decays to be seen soon at the $B$ factories as their experimental upper limits on the branching ratios quoted in (\ref{brhogamexp}) are only a factor of two above our results (we get $B(\bar B^0\to \rho^0\gamma)=0.76\cdot 10^{-6}$).

If these measurements of $B(B\to\rho\gamma)$ are available one can also consider the ratio of $B\to K^*\gamma$ and $B\to\rho\gamma$ branching ratios where for example $m_b^2$ of (\ref{brbvgam}) cancels. Furthermore, the ratio of form factors is better known because it are mostly $SU(3)$ breaking effects that make it differ from 1. The ratio of the rates can also be employed to get constraints on the $\bar\rho-\bar\eta$ plane of the CKM unitarity triangle.\medskip

The direct CP asymmetry, which is substantial for the $\rho\gamma$ modes, is much less dependent on the form factors because the $|F_V|^2$ cancels. Here, the largest theoretical uncertainty comes from the scale dependence. This is to be expected because the direct CP asymmetry is proportional to the perturbative strong phase difference, which arises at ${\cal O}(\alpha_s)$. If one wants to reduce the renormalization scale dependence for the CP asymmetry one could determine its NLL result. This would amount to calculating the absorptive parts of the diagrams in figures \ref{fig:qit1}--\ref{fig:qit2} with one additional gluon. But as the influence of power corrections on the CP asymmetry might be large as well, we see at present no urgent motivation to embark upon such a three-loop calculation.

\begin{figure}
%   \vspace{-3cm}
%   \epsfysize=10cm
   \epsfxsize=12cm
   \centerline{\epsffile{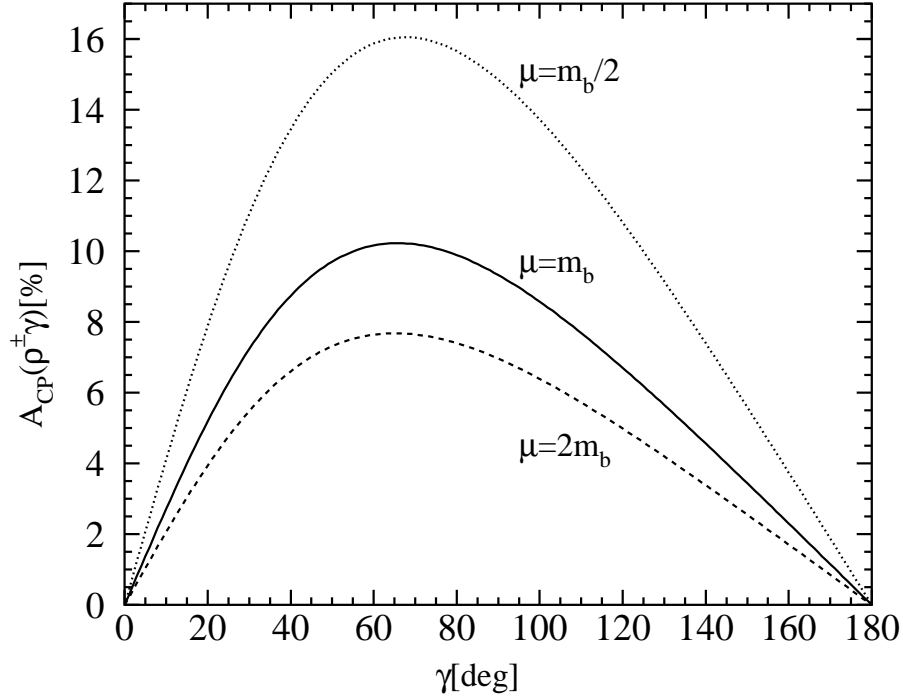}}
%   \vspace*{-15.5cm}
\caption{The CP asymmetry ${\cal A}_{CP}(\rho^\pm\gamma)$ as a function of the CKM angle $\gamma$ for three values of the renormalization scale $\mu=m_b/2$, $m_b$ and $2 m_b$.\label{fig:cpasymrho}}
\end{figure}

\begin{figure}
   \epsfxsize=10cm
   \centerline{\epsffile{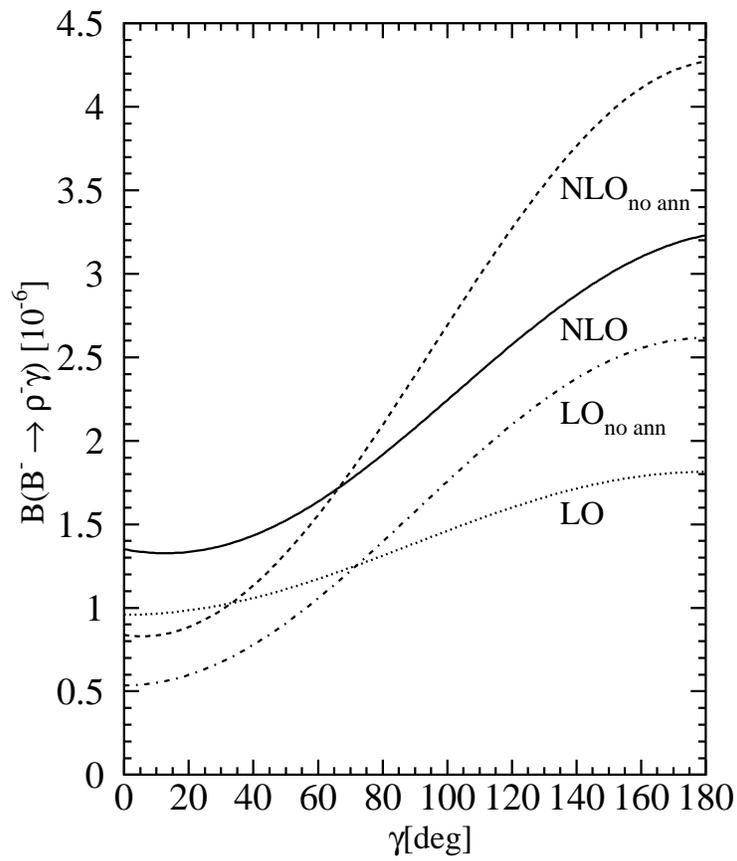}}
\caption{The branching fraction $B(B^- \to \rho^- \gamma)$ as a function of the CKM angle $\gamma$ at leading and next-to-leading order with and without annihilation contributions.\label{fig:brhogamma}}
\end{figure}

The decay $B\to\rho\gamma$ also depends sensitively on fundamental CKM parameters, such as $\left|V_{ub}/V_{cb}\right|$ and $\gamma$, and can thus in principle serve to constrain the latter quantities once measurements become available. This is further illustrated in Figs.~\ref{fig:cpasymrho} and \ref{fig:brhogamma} where the dependence on $\gamma$ is shown for ${\cal A}_{CP}(\rho^\pm\gamma)$ and $B(B^-\to\rho^-\gamma)$, respectively.
%%%%%%%%%%%%%%%%%%%%%%%%%%%%%%%%%%%%%%%%%%%%%%%%%%%%%%%%%%%%%%%%%%%
%%%%%%%%%%%%%%%%%%%%%%%%%%%%%%%%%%%%%%%%%%%%%%%%%%%%%%%%%%%%%%%%%%%
%%%%%%%%%%%%%%%%%%%%%%%%%%%%%%%%%%%%%%%%%%%%%%%%%%%%%%%%%%%%%%%%%%%
%%%%%%%%%%%%%%%%%%%%%%%%%%%%%%%%%%%%%%%%%%%%%%%%%%%%%%%%%%%%%%%%%%%
Including or omitting penguin operators has practically no effect on the CP asymmetries as was mentioned already after equation (\ref{acpsimp}). The ${\cal A}_{CP}(\rho^\pm\gamma)$ gets smaller by about $20\%$ if annihilation contributions are omitted whereas ${\cal A}_{CP}(\rho^0\gamma)=8.25\%$ (for the CKM angle $\gamma=58^\circ$) is practically not affected over the whole range of $\gamma$. The maximal value of the CP asymmetry in Fig.~\ref{fig:cpasymrho} is obtained for $\gamma\approx 65^\circ$, near the value of $\gamma$ prefered by standard unitarity triangle fits. For $B\to K^*\gamma$ the CP asymmetry is negative and smaller by a factor $|\lambda_t^{(s)}/\lambda_t^{(d)}|\approx 25$. We get typically ${\cal A}_{CP}(K^{*0}\gamma)=-0.3\%$.\medskip

Let us now comment on Fig.~\ref{fig:brhogamma}. The $\gamma$ dependence of the leading and next-to-leading order $B^-\to\rho^-\gamma$ branching ratios can be understood when looking at the amplitudes $a_7^u(\rho\gamma)=-0.4154-0.0685i$ and $a^u_\mathrm{ann}(\rho^-\gamma)=0.1884$, which are accompanied by $\lambda_u^{(d)}=|V_{ud}||V_{ub}|e^{-i\gamma}$. The leading order value for $a_7^u$ without annihilation effects is simply $C_7^{LO}=-0.3221$. Its absolute value gets reduced when the positive $a^u_\mathrm{ann}(\rho^-\gamma)$ is added and hereby also the dependence on $\gamma$ gets reduced. The same story is repeated at next-to-leading order. Among the $a^u$ amplitudes the NLO $a_7^u$ has the largest absolute value and thus the largest $\gamma$ dependence, which is again reduced by the inclusion of the annihilation contribution. For our default choice of $58^\circ$ for $\gamma$ we see that the effect of annihilation contributions on the branching ratio is small. Although for this $\gamma$ the real part of the total amplitude is increased when annihilation is included, this is compensated by the imaginary part getting smaller.\medskip

\begin{figure}
\includegraphics[width=8cm]{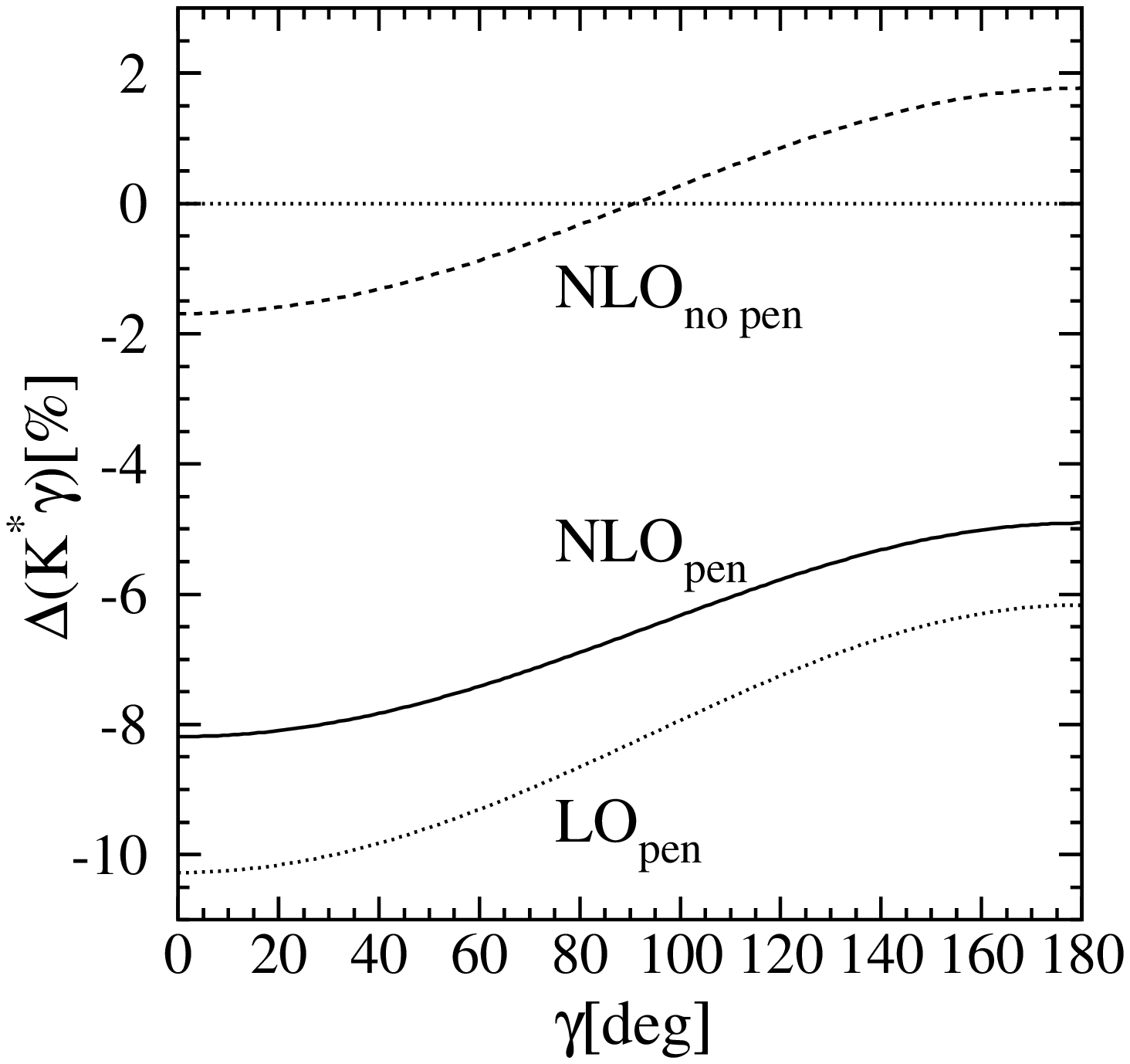}\includegraphics[width=8cm]{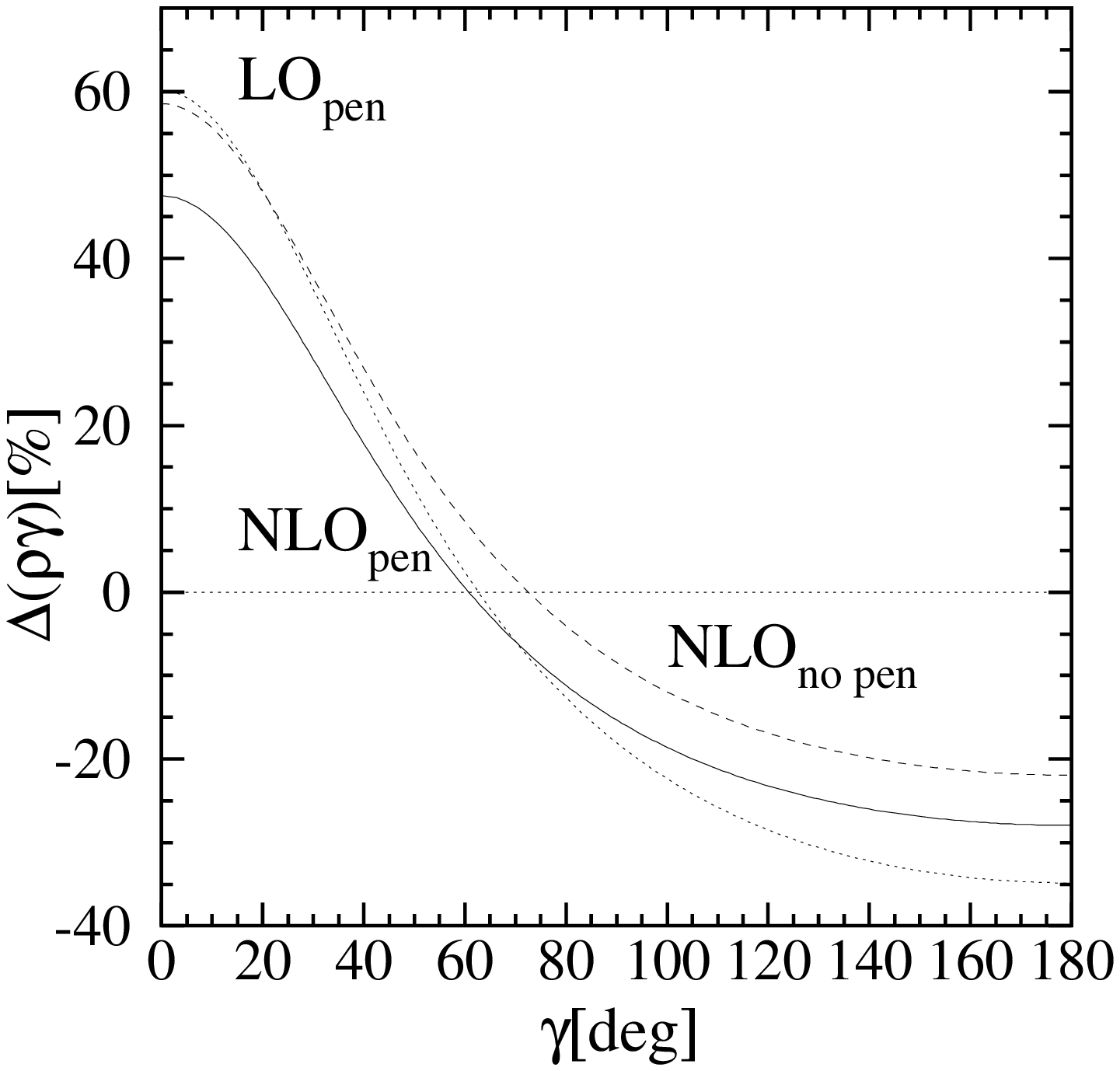}
\caption{The isospin-breaking asymmetries $\Delta(K^*\gamma)$ and $\Delta(\rho\gamma)$ as a function of the CKM angle $\gamma$ at leading and next-to-leading order with and without the inclusion of QCD penguin operator effects.\label{fig:isodeltaKrho}}
\end{figure}

Figure~\ref{fig:isodeltaKrho}
%%%%%%%%%%%%%%%%%%%%%%%%%%%%%%%%%%%%%%%%%%%%%%%%%%%%%%%%%%%%%%%%%%%
%%%%%%%%%%%%%%%%%%%%%%%%%%%%%%%%%%%%%%%%%%%%%%%%%%%%%%%%%%%%%%%%%%%
finally displays the isospin breaking quantities $\Delta (K^*\gamma)$ and $\Delta(\rho\gamma)$ as a function of the CKM angle $\gamma$. Within our approximations isospin breaking is generated by weak annihilation. For comparison with our earlier work \cite{BBVgam} the dashed curves give the result without QCD penguin operator effects. As Kagan and Neubert \cite{KN} already realised, this has large impact on the prediction for the isospin asymmetry in $B\to K^*\gamma$. We comment briefly on the numerical results of Kagan and Neubert in the next section. The QCD penguin operators give a, albeit small, contribution to the not CKM suppressed charm amplitude in contrast to $Q^p_{1,2}$, which can contribute to the up amplitude only. Looking at equations (\ref{acann0num}) and (\ref{acann1num}) we see that the dominant effect is indeed coming from $Q_6$.

In table~\ref{tab:outputiso}
%%%%%%%%%%%%%%%%%%%%%%%%%%%%%%%%%%%%%%%%%%%%%%%%%%%%%%%%%%%%%%%%%%%%%
%%%%%%%%          Table of isospindelta output               %%%%%%%%
%%%%%%%%%%%%%%%%%%%%%%%%%%%%%%%%%%%%%%%%%%%%%%%%%%%%%%%%%%%%%%%%%%%%%
\begin{table}
\renewcommand{\arraystretch}{1.2}
\begin{center}
\begin{tabular}{|l||c|c||c|c|}
\hline\hline
  & $\Delta (K^*\gamma)\, [\%]$ &  $\Delta_{\mbox{\scriptsize no pen}}(K^*\gamma)\, [\%]$ & $\Delta (\rho\gamma)\, [\%]$  & $\Delta_{\mbox{\scriptsize no pen}}(\rho^*\gamma)\, [\%]$\\
\hline\hline
central                         & $-$7.46        & $-$0.92          & 2.02            & 10.04\\
\hline\hline
$F_{K^*}$                       & +0.94/$-$1.26  & +0.12/$-$0.16    &   --            & --\\
\hline
$F_{\rho}$                      &   --           &  --              & +1.23/$-$0.79   & +2.54/$-$1.77\\
\hline
$\mu$                           & +2.69/$-$5.29  & +0.05/$-$0.13    & +2.41/$-$4.37   & +1.89/$-$0.81\\
\hline
$m_b$                           & +0.46/$-$0.52  & +0.04/$-$0.04    & +0.01/$-$0.02   & +0.60/$-$0.54\\
\hline
$\lambda_B$                     & +1.71/$-$0.70  & +0.27/$-$0.63    & +15.56/$-$4.92  & +13.65/$-$4.13\\
\hline
$m_c$                           & +0.20/$-$0.24  & +0.03/$-$0.03    & +0.19/$-$0.22   & +0.03/$-$0.04\\
\hline
$\Lambda_{\overline{MS}}^{(5)}$ & +0.14/$-$0.13  & +0.01/$-$0.01    & +0.31/$-$0.28   & +0.14/$-$0.12\\
\hline
$\left|V_{ub}/V_{cb}\right|$    & +0.26/$-$0.26  & +0.27/$-$0.27    & +0.03/$-$1.62   & +0.03/$-$1.76\\
\hline
$\gamma=(58\pm 24)^\circ$       & +0.62/$-$0.46  & +0.66/$-$0.49    & +21.79/$-$14.10 & +23.25/$-$15.04\\
\hline
$f_B$                           & +1.04/$-$1.01  & +0.13/$-$0.13    & +1.15/$-$0.96   & +2.37/$-$2.17\\
\hline
$m_b\leftrightarrow m_B$        & +2.08/$-$1.72  & +0.17/$-$0.22    & +2.34/$-$0.29   & +4.18/$-$2.45\\
\hline\hline
\end{tabular}
\end{center}
\caption[]{Predictions for the isospin-breaking asymmetries $\Delta (K^*\gamma)$ and $\Delta (\rho\gamma)$, including and not including QCD-penguin-operator contributions, with the errors from the individual input uncertainties.\label{tab:outputiso}}
\end{table}
%%%%%%%%%%%%%%%%%%%%%%%%%%%%%%%%%%%%%%%%%%%%%%%%%%%%%%%%%%%%%%%%%%%%%
we summarize the sensitivity of $\Delta (K^*\gamma)$ and $\Delta(\rho\gamma)$ with and without the effects from QCD penguin operators to variations of the relevant input parameters. Especially the strong $\gamma$ dependence of $\Delta (\rho\gamma)$ around $\gamma=58^\circ$ is striking. It is also rather insensitive to the inclusion of penguin-operator and even next-to-leading order corrections (see Fig.~\ref{fig:isodeltaKrho}). Therefore the isospin asymmetry in the $\rho\gamma$ channel, which is far easier to measure than the CP asymmetry, could give valuable direct information on the CKM angle $\gamma$. Yet, one has to keep in mind that the errors from uncertainties in other input parameters, in particular $\lambda_B$, are sizeable as well.

Averaging the branching ratio measurements in (\ref{b0kgamex}) and (\ref{bpkgamex}) gives $B(B^0\to K^{*0}\gamma)=(4.44\pm 0.35)\cdot 10^{-5}$ and $B(B^+\to K^{*+}\gamma)=(3.82\pm 0.47)\cdot 10^{-5}$. This implies, when corrected for the $B$-meson lifetime differences, an isospin violation of $\Delta (K^*\gamma)^\mathrm{exp.}=(-19.2\pm 11.8)\%$. It is not significantly different from zero and consistent with the value $\Delta (K^*\gamma)=(-7.46^{+4.15}_{-5.89})\%$ from table~\ref{tab:outputiso} when adding all uncertainties in quadrature. Our corresponding prediction for the $\rho\gamma$ mode is $\Delta (\rho\gamma)=(2.02^{+27.01}_{-15.69})\%$.\medskip

The last application of our results we want to discuss is an estimate of U-spin breaking effects in $B\to V\gamma$ decays. Using our expressions and central values for all parameters we find for the differences of the charge conjugated branching ratios defined in (\ref{dbk}) and (\ref{dbrho})
\begin{eqnarray}
  \Delta B(B\to K^*\gamma) &=& -7.3\cdot 10^{-7}\\
  \Delta B(B\to\rho\gamma) &=& +4.3\cdot 10^{-7}
\end{eqnarray}
where we have chosen the CKM angle $\gamma=\pi/2$ to maximise the effects. The two quantities indeed have opposite signs, but their sum only partly cancels, leaving a U-spin breaking remainder of $-3.0\cdot 10^{-7}$ (for $\gamma=58^\circ$ we get $-2.6\cdot 10^{-7}$). This effect is almost entirely due to the difference $(F_{K^*}-F_\rho)$. For form-factor values different from those in table~\ref{tab:input} the U-spin breaking effect would approximately be rescaled proportional to $(F_{K^*}-F_\rho)$. For our choice the U-spin breaking remainder is of the same order of magnitude as the individual asymmetries. This example quantifies the limitations of the relation (\ref{dbb}) as a Standard Model test.

%==================================
%=      Related Calculations      =
%==================================

\section{Related Calculations}
\label{sec:relcalc}

Finally, we want to comment on related work done for $B\to V\gamma$. Beneke, Feldmann, and Seidel treat exclusive $B\to V l^+ l^-$ and $B\to V\gamma$ decays also in QCD factorization \cite{BFS}. They use the CMM operator basis (\ref{p1def}--\ref{p8def}) supplemented with the semileptonic operators $P_9$ and $P_{10}$
\begin{eqnarray}
  P_9 &=& \frac{\alpha}{8\pi} (\bar l l)_V (\bar s b)_{V-A}\\
  P_{10} &=& \frac{\alpha}{8\pi} (\bar l l)_A (\bar s b)_{V-A}
\end{eqnarray}
Another technical difference is that they use the formalism of \cite{BF} with our physical form factor $F_{K^*}$ replaced by its HQET analog $\xi_\perp^{(K^*)}$. In this form factor the hard corrections from diagrams as in Fig.~\ref{fig:factNLOnocalc} are taken explicitely into account. The analytic result of \cite{BFS} for the $B\to K^*\gamma$ branching ratio perfectly agrees with ours. The numerical central value is $8\%$ larger because a somewhat larger value for the $b$ quark mass was used. The interpretation is slightly different as well. Although within the error bars still consistent with the experimental averages, Beneke, Feldmann, and Seidel get a little worried by the fact that the prediction for the central value of the branching ratio is nevertheless nearly twice as large as the experimental central value. So they consider the possibility that the form factors at $q^2 =0$ are substantially different from what is obtained with QCD sum rules or in quark models. Fitting their result for the branching ratio to the experimental value results in $F_{K^*}=0.27\pm0.04$, which is almost $30\%$ smaller than the value we gave in table~\ref{tab:input}.

The paper of Kagan and Neubert \cite{KN} was mentioned already several times. Herein the leading isospin-breaking contributions to the $B\to K^* \gamma$ decay amplitudes are calculated. In particular a large effect from $Q_6$ is obtained. This brings their theoretical prediction in good agreement with the current experimental value. Yet, we have to keep in mind that there are still large error bars. The definition of the isospin-violating ratio
\begin{equation}
  \Delta^\mathrm{KN}_{0-}=\frac{\Gamma (\bar B^0 \to \bar K^{*0}\gamma) -\Gamma (B^- \to K^{*-}\gamma)}{\Gamma (\bar B^0 \to \bar K^{*0}\gamma) +\Gamma (B^- \to K^{*-}\gamma)}
\end{equation}
Kagan and Neubert use differs by approximately a factor $-1/2$ from our $\Delta_{-0}^{K^*}$ in~(\ref{iso-0}). In addition to the isospin-sensitive annihilation diagrams in Fig.~\ref{fig:ann} they also calculated those penguin-annihilation and $Q_8$ hard-spectator contributions of Fig.~\ref{fig:subl} where the photon is emitted from one of the upper light quark lines. As explained in section \ref{sec:power} we included hereof only the annihilation contributions because they are numerically enhanced, calculable, and not additionally suppressed by $\alpha_s({\cal O}(m_b))$. With this choice we get $\Delta^\mathrm{KN}_{0-}=(3.8^{+3.2}_{-2.2})\%$ for our input parameters. The main difference to the Kagan-Neubert value of $(8.0^{+2.1}_{-3.2})\%$ comes from other values of the form factor and $B$ meson decay constant used and from the different contributions included.

In \cite{GSW,AAW} the basic mechanisms for the $B\to V\gamma$ amplitudes at next-to-leading order were already discussed. However, a simple model was used to take into account the bound-state effects. Additionally, some heuristic methods had to be used to regulate the infrared divergences. These emerge from the quark self-energy diagrams and the $b\to s\gamma$ vertex correction and are part of the form factor in our QCD factorization approach. The use of hadronic models does not allow a clear separation of short- and long-distance dynamics or a clean distinction of model-dependent and model-independent features.

Another related work is that of Ali and Parkhomenko where also $B\to V\gamma$ is considered \cite{AP}. The first version of this work differed in some points from our results, but there is full agreement after revision of the paper.

Last but not least we comment on a perturbative QCD study of the $B\to K^* \gamma$ decay by Li and Lin \cite{LiLin}. They calculate most of the relevant diagrams, in particular the spectator contribution from $Q_1$. Yet, in our language, they include type-II corrections only, but thereof also power-suppressed contributions. The numerically much smaller result for the branching ratio is used to determine the $B$ meson wave function from a fit to the experimental data.

% =========================================================
% =     part III: B -> gamma gamma                        =
% =========================================================
\part{The Radiative Decays $B\to \gamma\gamma$}

% ===== Basic Formulas ====================================
\chapter{The Basic Formulas for $B\to\gamma\gamma$}
\label{ch:basicBgamgam}

In the third part we discuss the double radiative $B$-meson decays $B_s\to\gamma\gamma$ and $B_d\to\gamma\gamma$ in QCD factorization. We systematically discuss the various contributions to these exclusive processes in the heavy-quark limit $m_b\gg\Lambda_\mathrm{QCD}$. The factorization formula for the hadronic matrix elements of local operators in the weak Hamiltonian again allows us to separate perturbatively calculable hard scattering kernels from the nonperturbative $B$-meson light-cone distribution amplitude. With power counting in $\Lambda_{QCD}/m_b$ we identify leading and subleading contributions to $B\to\gamma\gamma$. Only one diagram contributes at leading power, but an important class of subleading contributions can also be calculated. The inclusion of these corrections is used to estimate CP asymmetries in $B\to\gamma\gamma$ and one-particle irreducible two-photon emission from light-quark loops in $B$ and $D\to\gamma\gamma$. These corrections represent the quark-level analogue of so-called long-distance contributions to these decays. In this chapter we give the basic formulas to leading logarithmic accuracy and discuss the leading-power and subleading-power contributions separately. Numerical results are presented and discussed in chapter \ref{ch:phenBgamgam}. Compared to our work in~\cite{BBgamgam} we again include the QCD penguin operator contributions.

%========================
%=      Motivation      =
%========================
\section{Motivation}
\label{sec:motivation}

For the double radiative decays $B\to\gamma\gamma$ experimentally so far only upper limits on the branching fractions exist:
\begin{equation}\label{brbsdggex}
  \begin{array}{rclll}
    B(B^0_s\to\gamma\gamma) & < & 1.48\cdot 10^{-4} & \mbox{at 90\% C.L.} & \cite{PDG}\\
    B(B^0_d\to\gamma\gamma) & < & 1.7\cdot 10^{-6} & \mbox{at 90\%  C.L.} & \cite{AUB}
  \end{array}
\end{equation}
The Standard Model expectations are roughly two orders of magnitude below these upper limits. $B\to\gamma\gamma$ decays have a rather clean experimental signature but are much better suited to be searched for at $e^+\,e^-$ $B$ factories because at hadron colliders the combinatorial background is probably too large \cite{IA}. Furthermore these decays are of interest because they could provide useful tests of QCD dynamics in $B$ decays. The $B\to\gamma\gamma$ modes realize the exceptional situation of nontrivial QCD dynamics related to the decaying $B$ in conjunction with a completely nonhadronic final state and simple two-body kinematics. In principle they also probe the CKM parameters $V_{ts}$ and $V_{td}$ and could allow us to study CP violating effects as the two-photon system can be in a CP-even or CP-odd state.

The theoretical treatments of $B_s\to\gamma\gamma$ performed so far \citer{LLY,EKP} all had to employ hadronic models to describe the $B_s$ meson bound state dynamics. A clear separation of short- and long-distance dynamics and a distinction of model-dependent and model-independent features were therefore not possible. This concerns especially the dynamics of the light-quark constituent inside the $B$, but also contributions from intermediate $J/\psi$, $\eta_c$, $\phi$, or $D_s^{(*)}$ meson states, which have been discussed as sources of potentially important long-distance effects.

%============================
%=      Basic Formulas      =
%============================
\section{Basic Formulas}
\label{sec:basicBgamgam}

The effective Hamiltonian for $b\to s\gamma\gamma$ is identical to the one for $b\to s\gamma$ transitions. In addition to the six four-quark operators $Q_{1\ldots 6}$ in (\ref{q1def}--\ref{q6def}) or $P_{1\ldots 6}$ in (\ref{p1def}--\ref{p6def}) there are ten more gauge-invariant dimension-six operators with the right flavour quantum numbers to contribute to a $b\to s$ transition. These are two-quark operators containing combinations of the covariant derivative $D_\mu=\partial_\mu +ie Q_f A_\mu + i g T^a A^a_\mu$, the photonic and gluonic field strength tensors, and their duals. Although these operators contain up to three photon fields, Grinstein, Springer and Wise showed that the equations of motion \cite{HP}
\begin{equation}\label{eom}
  i \,\slash\hspace{-0.7em}D q=m_q q \quad\mbox{and}\quad D_\mu F^{\mu\nu}=e Q_q \bar q \gamma^\nu q
\end{equation}
can be used to reduce them to the four-quark operator basis and the magnetic penguin operators $Q_7$ and $Q_8$ \cite{GSWbsg}. Up to corrections of order $1/M_W^2$ from higher-dimensional operators the effective Hamiltonian for $b\to s\gamma\gamma$ thus is just the one of (\ref{heff}) for $b\to s\gamma$. We use the standard operator basis (\ref{q1def}--\ref{q8def}). The most important operators are the magnetic penguin operator $Q_7$ and the four-quark operators $Q^p_{1,2}$. The effective Hamiltonian for $b\to d\gamma\gamma$ is obtained from (\ref{heff}--\ref{q8def}) by the replacement $s\to d$.

The amplitude for the $B\to\gamma\gamma$ decay has the general structure
\begin{eqnarray}\label{Bgggen}
  \lefteqn{{\cal A}\left(\bar B\to\gamma(k_1,\epsilon_1)\gamma(k_2,\epsilon_2)\right)\equiv \frac{G_F}{\sqrt{2}} \frac{\alpha}{3\pi} f_B \frac{1}{2}\langle\gamma\gamma|A_+ F_{\mu\nu} F^{\mu\nu}-i\, A_- F_{\mu\nu} \tilde F^{\mu\nu}|0\rangle} \nonumber\\
  &&= \frac{G_F}{\sqrt{2}} \frac{\alpha}{3\pi} f_B \left[ A_+\left(2k_1\cdot\epsilon_2 \, k_2\cdot\epsilon_1 -m_B^2 \epsilon_1\cdot\epsilon_2\right) -2i\,A_-\varepsilon(k_1,k_2,\epsilon_1,\epsilon_2)\right]
\end{eqnarray}
Here $F^{\mu\nu}$ and $\tilde F^{\mu\nu}$ are the photon field strength tensor and its dual, where
\begin{equation}\label{fdual}
  \tilde F^{\mu\nu}=\frac{1}{2}\varepsilon^{\mu\nu\lambda\rho}F_{\lambda_\rho}
\end{equation}
with $\varepsilon^{0123}=-1$ again. The decay rate is then given by
\begin{equation}\label{gbgg}
  \Gamma(\bar B\to\gamma\gamma)=\frac{G^2_F m^3_B f^2_B \alpha^2}{288\pi^3}\left(|A_+|^2 + |A_-|^2\right)
\end{equation}

In the heavy-quark limit we propose a factorization formula for the hadronic matrix elements of the operators in the effective Hamiltonian (\ref{heff}):
\begin{equation}\label{fformBgamgam}
  \langle \gamma(\epsilon_1)\gamma(\epsilon_2)|Q_i|\bar B\rangle = \int^1_0 d\xi\, T^{\mu\nu}_i(\xi)\, \Phi_{B1}(\xi)\epsilon_{1\mu} \epsilon_{2\nu}
\end{equation}
where the $\epsilon_i$ are the polarization 4-vectors of the photons and $\Phi_{B1}$ is the leading twist light-cone distribution amplitude of the $B$ meson defined in (\ref{BLCDA}). If the light-like vector $n$ is chosen appropriately parallel to one of the 4-momenta of the photons only the first negative moment of $\Phi_{B1}(\xi)$ appears, which we parametrize again via $\lambda_B={\cal O}(\Lambda_{QCD})$ as in (\ref{lambdef}).

Because there are no hadrons in the final state only one type of hard-scattering kernel $T$ (type II or hard-spectator contribution) enters the factorization formula. The QCD factorization formula (\ref{fformBgamgam}) holds up to corrections of relative order $\Lambda_{QCD}/m_b$. The form of (\ref{fformBgamgam}) with a simple convolution over the light-cone variable $\xi$ is appropriate for the lowest (leading logarithmic) order in $\alpha_s$, which we will use in the present analysis. A generalization to include transverse-momentum variables is likely to be necessary at higher orders in QCD \cite{BBNS,BF,BFPS,KPY}.

We conclude this section with a brief discussion of CP violation in $B\to\gamma\gamma$. The subscripts $\pm$ on $A_\pm$, defined in (\ref{Bgggen}) for $\bar B\to\gamma\gamma$, denote the CP properties of the corresponding two-photon final states, which are eigenstates of CP: $A_{CP=+1}$ is proportional to the parallel spin polarization $\vec\epsilon_1\cdot\vec\epsilon_2$ and $A_{CP=-1}$ is proportional to the perpendicular spin polarization $\vec\epsilon_1\times\vec\epsilon_2$ of the photons. In addition to $A_\pm$ we introduce the CP conjugated amplitudes $\bar A_\pm$ for the decay $B\to\gamma\gamma$ (decaying $\bar b$ anti-quark). Then the deviation of the ratios
\begin{equation}\label{rCPdef}
  r^\pm_{CP}=\frac{|A_\pm|^2-|\bar A_\pm|^2}{|A_\pm|^2+|\bar A_\pm|^2}
\end{equation}
from zero is a measure of direct CP violation. We shall focus here on this direct effect, which is specific for the $B\to\gamma\gamma$ decay, and will not consider CP violation originating in $B$--$\bar B$ mixing.

Due to the unitarity of the CKM matrix we can parametrize
\begin{eqnarray}\label{apar}
  A_\pm &=& \lambda_u a_u^\pm e^{i\alpha_u^\pm} +\lambda_c a_c^\pm e^{i\alpha_c^\pm}\\ \label{abarpar}
  \bar A_\pm &=& \lambda_u^* a_u^\pm e^{i\alpha_u^\pm} +\lambda_c^* a_c^\pm e^{i\alpha_c^\pm}
\end{eqnarray}
where $a_{u,c}^\pm$ are real hadronic matrix elements of weak transition operators and the $\alpha_{u,c}^\pm$ are their CP-conserving phases. We then have
\begin{equation}\label{rCPpar}
  r^\pm_{CP}=\frac{2 \,\mathrm{Im}(\lambda_u \lambda_c^*)a_u^\pm a_c^\pm \sin(\alpha_c^\pm-\alpha_u^\pm)}{|\lambda_u|^2 a_u^{\pm 2} +|\lambda_c|^2 a_c^{\pm 2} +2 \,\mathrm{Re}(\lambda_u\lambda_c^*)a_u^\pm a_c^\pm \cos(\alpha_c^\pm -\alpha_u^\pm)}
\end{equation}
The weak phase differences read
\begin{equation}
  \mathrm{Im}(\lambda_u^{(q)} \lambda_c^{(q)*})=
    \mp|\lambda_u^{(q)}||\lambda_c^{(q)}|\sin\gamma
\end{equation}
with the negative sign for $q=s$ and the positive sign for $q=d$. The strong phase differences $\sin(\alpha_c^\pm -\alpha_u^\pm)$ arise from final state interaction (FSI) effects, generated for instance via the Bander-Silverman-Soni (BSS) mechanism \cite{BSS}.

%===============================================
%=      B -> gamma gamma at leading power      =
%===============================================
\section{$B\to\gamma\gamma$ at leading power}
\label{sec:Bgamgamldgpow}

\begin{figure}[t]
\begin{center}
\psfig{figure=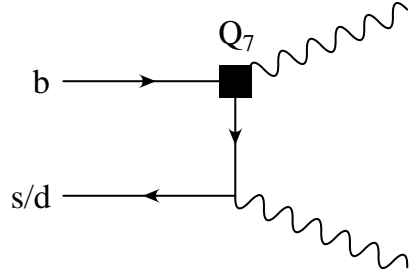}
\end{center}
\caption{The leading power contribution to the $B\to\gamma\gamma$ amplitude given by the magnetic penguin operator $Q_7$. The diagram with interchanged photons is not shown. Radiation from the $b$-quark line (Fig. \ref{fig:1PRslp}) is power suppressed.\label{fig:1PRlp}}
\end{figure}

To leading power in $\Lambda_{QCD}/m_b$ and in the leading logarithmic approximation of QCD only one diagram contributes to the amplitude for $B\to\gamma\gamma$. It is the one-particle reducible (1PR) diagram with the electromagnetic penguin operator $Q_7$, where the second photon is emitted from the $s$-quark line. An illustration is given in Fig.~\ref{fig:1PRlp}.
%%%%%%%%%%%%%%%%%%%%%%%%%%%%%%%%%%%%%%%%%%%%%%%%%%%%%%%%%%%%%%%%%%%
%%%%%%%%%%%%%%%%%%%%%%%%%%%%%%%%%%%%%%%%%%%%%%%%%%%%%%%%%%%%%%%%%%%
Evaluating the factorization formula (\ref{fformBgamgam}) for this diagram leads to the matrix element
\begin{eqnarray}\label{Q7lp}
  \lefteqn{\langle\gamma(k_1,\epsilon_1)\gamma(k_2,\epsilon_2)|Q_7|\bar B\rangle =}\\
  && i\frac{f_B \alpha}{\pi} Q_s \int^1_0 d\xi\, \frac{\Phi_{B1}(\xi)}{\xi}\left(2k_1\cdot\epsilon_2 k_2\cdot\epsilon_1-m_B^2\epsilon_1\cdot\epsilon_2 -2i\,\varepsilon(k_1,k_2,\epsilon_1,\epsilon_2)\right)\nonumber
\end{eqnarray}
Together with the effective Hamiltonian (\ref{heff}) and the definition of $A_\pm$ in (\ref{Bgggen}) this implies
\begin{equation}\label{apmlp}
  A_\pm=\lambda^{(q)}_t C_7 \frac{m_{B}}{\lambda_{B}}
\end{equation}
where CKM unitarity, $\lambda^{(q)}_u+\lambda^{(q)}_c=-\lambda^{(q)}_t$, has been used.

After summing over the photon polarizations the branching fraction to leading power becomes
\begin{equation}\label{brlp}
  B(B_q\to\gamma\gamma)=\tau_{B_q} \frac{G^2_F m^5_{B_q}}{192\pi^3}|\lambda^{(q)}_t|^2 C^2_7 \frac{4\alpha^2 f^2_{B_q}}{3\lambda^2_{B_q}}
\end{equation}
where $q=d,s$ for the decay of a $B^0_d$ or a $B^0_s$ meson, respectively.

In the present approximation the strong-interaction matrix elements multiplying $\lambda^{(q)}_u$ and $\lambda^{(q)}_c$ are identical and have no relative phase. The CP-violating quantities $r^\pm_{CP}$ defined in (\ref{rCPdef}) therefore vanish for the strictly leading-power result.

%================================================================
%=      Power-suppressed contributions to B -> gamma gamma      =
%================================================================
\section{Power-suppressed contributions to $B\to\gamma\gamma$}
\label{sec:Bgamgamsubldgpow}

\begin{figure}[t]
\begin{center}
\psfig{figure=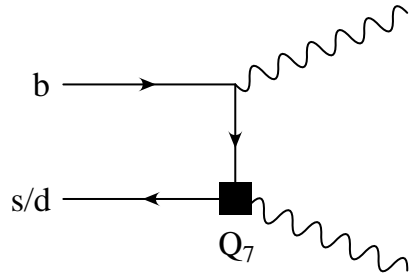}
\end{center}
\caption{The subleading power 1PR diagram of the magnetic penguin operator $Q_7$. The diagram with interchanged photons is not shown.\label{fig:1PRslp}}
\end{figure}

\begin{figure}[t]
\begin{center}
\psfig{figure=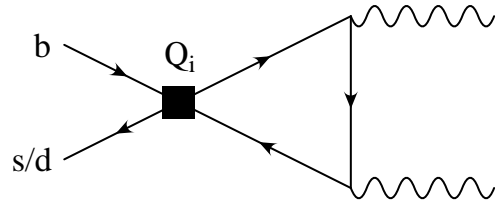}
\end{center}
\caption{The subleading power 1PI diagram. Here the four-quark operators $Q_i$ can contribute. The diagram with interchanged photons is not shown.\label{fig:1PI}}
\end{figure}

Subleading contributions come from the 1PR diagram, where the second photon is emitted from the $b$ quark line (Fig. \ref{fig:1PRslp}), and from the one-particle irreducible diagram (1PI) (see Fig. \ref{fig:1PI}).
%%%%%%%%%%%%%%%%%%%%%%%%%%%%%%%%%%%%%%%%%%%%%%%%%%%%%%%%%%%%%%%%%%%
%%%%%%%%%%%%%%%%%%%%%%%%%%%%%%%%%%%%%%%%%%%%%%%%%%%%%%%%%%%%%%%%%%%
%%%%%%%%%%%%%%%%%%%%%%%%%%%%%%%%%%%%%%%%%%%%%%%%%%%%%%%%%%%%%%%%%%%
%%%%%%%%%%%%%%%%%%%%%%%%%%%%%%%%%%%%%%%%%%%%%%%%%%%%%%%%%%%%%%%%%%%

To estimate the subleading 1PR contribution, we simply evaluate the graph in Fig.~\ref{fig:1PRslp} using the $B$-meson projector in (\ref{BLCDA}). The result equals $\lambda_B/m_B$ times the leading-power expression in (\ref{Q7lp}), clearly showing the power suppression of this mechanism. Other corrections of the same order can arise from higher-twist terms in the $B$-meson wave function. We shall neglect all those subleading terms for the matrix element of $Q_7$, keeping in mind that they could naturally contribute relative corrections of order $\sim 10\%$. We stress, however, that subleading effects in $\langle Q_7\rangle$ contribute equally to the up- and charm-quark components of the amplitude (see (\ref{heff})). Therefore they do not give rise, in particular, to relative FSI phases between these sectors, which would affect direct CP violation. The same is true for perturbative QCD corrections to $\langle Q_7\rangle$.

We next consider more closely the diagrams in Fig. \ref{fig:1PI}. These 1PI contributions, which come from the matrix elements of four-quark operators in (\ref{heff}), are of special interest for two reasons. First, they are the basic effects responsible for a difference between the up- and charm-quark sectors of the amplitude, including rescattering phases. Second, they represent the parton-level processes that are dual to $B\to\gamma\gamma$ amplitudes from $Q_i$ with hadronic intermediate states ($J/\Psi$, $D^{(*)}$, etc.), which are commonly considered as generic long-distance contributions. Investigating the amplitude in Fig. \ref{fig:1PI} could therefore shed light on this class of effects in $B\to\gamma\gamma$.

\begin{figure}[t]
\begin{center}
\psfig{figure=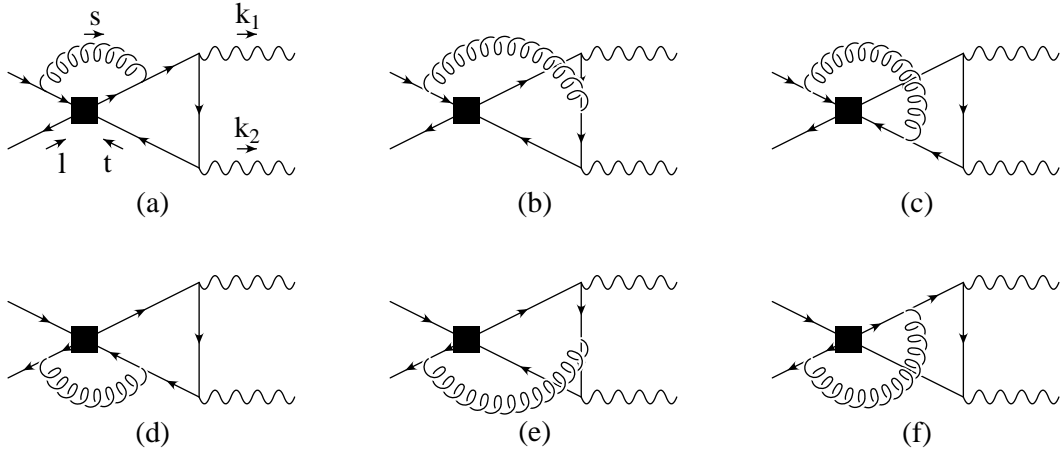}
\caption{The gluon corrections to the 1PI diagram. Quark labels are as in Fig.~\ref{fig:1PI} and the momentum assignment is everywhere as in diagram (a). The diagrams with interchanged photons are not shown.}
\label{fig:1PINLO}
\end{center}
\end{figure}

It shall now be argued that the 1PI contributions are calculable using QCD factorization. To this end we show at ${\cal O}(\alpha_s)$ that there are no collinear or soft infrared divergences at leading power in $1/m_b$. By leading power we mean here the lowest nonvanishing order in the power expansion, where the entire 1PI contribution starts only at subleading power with respect to the dominant mechanism in Fig. \ref{fig:1PRlp}. This suppression will be apparent from the explicit expressions given in (\ref{1PIQ1}--\ref{1PIQ6}) below. The relevant diagrams at ${\cal O}(\alpha_s)$ are shown in Fig.~\ref{fig:1PINLO}.
%%%%%%%%%%%%%%%%%%%%%%%%%%%%%%%%%%%%%%%%%%%%%%%%%%%%%%%%%%%%%%%%%%%
%%%%%%%%%%%%%%%%%%%%%%%%%%%%%%%%%%%%%%%%%%%%%%%%%%%%%%%%%%%%%%%%%%%
Following the approach  explained in the second reference of \cite{BBNS} for the case of $B\to D\pi$ decays, we examine the potentially IR-singular regions of the momentum-space integrals in Fig.~\ref{fig:1PINLO}. We demonstrate the absence of divergences or their mutual cancellation among different diagrams for the hard-scattering part of the amplitude at ${\cal O}(\alpha_s)$. As a consequence the matrix elements of four-quark operators can be written in factorized form as in (\ref{fformBgamgam}), where in the present case the kernel $T^{\mu\nu}_i$ is independent of $\xi$:
\begin{equation}\label{fform12}
  \langle \gamma(\epsilon_1)\gamma(\epsilon_2)|Q_{1\ldots 6}|\bar B\rangle = f_B T^{\mu\nu}_{1\ldots 6}\, \epsilon_{1\mu} \epsilon_{2\nu}
\end{equation}

Power counting shows that all diagrams in Fig.~\ref{fig:1PINLO} are IR finite in the soft-soft and hard-soft region as well as when the gluon is collinear with one of the photon momenta and the quark in the loop is soft or supersoft. But for a soft or supersoft gluon emitted from the $B$ meson spectator quark in addition to a collinear quark in the loop we encounter logarithmic IR-divergences. To leading power, however, setting the loop momentum proportional to the photon momentum, the diagrams vanish individually. The last region of phase space we have to consider is where both the gluon momentum $s$ and the momentum of the quark in the loop $t$ are collinear with one of the photon momenta, for example
\begin{equation}
  s=\alpha k_2 +s_\perp \qquad\qquad t=\beta k_2 +t_\perp
\end{equation}
where $\alpha,\beta\sim 1$ and $s_\perp, t_\perp\sim \Lambda_{QCD} \ll m_b$. This again leads to a logarithmic IR divergence. The leading $t=\beta k_2$ contribution vanishes again as in the (super)soft-collinear case. When the gluon is attached to the light spectator quark of the $B$ meson and one of the two collinear quarks in the loop (Fig.~\ref{fig:1PINLO}\hspace{0.5ex}d,e), however, this contribution is power enhanced by $m_b/\Lambda_{QCD}$ compared to the hard diagram without any gluon (Fig.~\ref{fig:1PI}). But for factorization to hold, we need that the logarithmic divergences have coefficients that are power suppressed. 
We therefore consider the next order in $\Lambda_{QCD}/m_b$ by keeping the transverse momentum of $t$ and analyze these terms in more detail. It turns out that the contributions from the loop where the gluon hits one of the collinear quarks are both proportional to $k_2^\alpha$. The factor $k_2^\alpha$ is contracted with the open index of the gluon emission from one of the $B$ meson constituent quarks. Substituting $s=\alpha k_2$ and using $p_b$ and $l$ for the momenta of the $b$ and the spectator quark in the $B$ meson, respectively, we obtain for this term
\begin{eqnarray}
  \lefteqn{\bar s \left[\frac{\Gamma (\not\! p_b -\alpha \!\!\not\! k_2 +m_b)\gamma_\alpha}{(p_b-\alpha k_2)^2-m_b^2} +\frac{\gamma_\alpha(-\not\! l+\alpha \!\!\not\! k_2)\Gamma}{(-l+\alpha k_2)^2}\right]b \, k_2^\alpha}\nonumber \\
  && =\bar s \left[\frac{\Gamma(\not\! p_b -\alpha \!\!\not\! k_2+m_b)(\not\!p_b-m_b)}{\alpha(p_b-\alpha k_2)^2-m_b^2} +\frac{\not\! l(-\not\! l+\alpha \!\!\not\! k_2)\Gamma}{(-l+\alpha k_2)^2}\right]b\nonumber \\
  && =0
\end{eqnarray}
Here we employed the equations of motion for the constituent quarks of the $B$ meson. It follows that the collinear-collinear divergences are also of sub-leading power. This completes our proof that at ${\cal O}(\alpha_s)$ for the 1PI contribution there are no collinear or soft infrared divergences at leading power in $1/m_b$ in the hard-scattering kernel. It supports our claim that these diagrams are calculable using QCD factorization. As no assumption on the Dirac structure $\Gamma$ of the inserted operator had to be made, this demonstration applies to all four-quark operators $Q_{1\ldots 6}$ in any possible Fierz ordering.\medskip

For the evaluation of the 1PI diagram in Fig.~\ref{fig:1PI} we can use the building blocks $B^{(V-A)\times (V\mp A)}_{\alpha\beta}$ and $B^{(S+P)\times(S-P)}_{\alpha\beta}$ of~(\ref{BBab}) and (\ref{BBSP}). We simply have to substitute $g_s T^a \to Q_q e ,\; q\to k_1 ,\; r\to k_2$, and can use the on-shell relations $k_i^2=0 ,\; \epsilon_i\cdot k_i=0$ for both photons. Having regard to the possible insertions of the Fierz-transformed QCD penguin operators, we find herewith the matrix elements
\begin{eqnarray}\label{1PIQ1}
  \langle Q^p_1\rangle &=& -\frac{f_{B_q}\alpha}{\pi} Q_u^2 g(z_p) \varepsilon(k_1,k_2,\epsilon_1,\epsilon_2)\\ \label{1PIQ2}
  \langle Q^p_2\rangle &=& N \langle Q^p_1\rangle\\ \label{1PIQ3}
\langle Q_3\rangle &=& -\frac{f_{B_q}\alpha}{\pi} \varepsilon(k_1,k_2,\epsilon_1,\epsilon_2)\Bigg[ N \sum_q Q_q^2 g(z_q) +Q_d^2 (g(0)+g(1))\Bigg]\\ \label{1PIQ4}
  \langle Q_4\rangle  &=& -\frac{f_{B_q}\alpha}{\pi} \varepsilon(k_1,k_2,\epsilon_1,\epsilon_2)\Bigg[ \sum_q Q_q^2 g(z_q) +N Q_d^2 (g(0)+g(1))\Bigg]\\ \label{1PIQ5}
  \langle Q_5\rangle &=& \frac{f_{B_q}\alpha}{\pi}\Bigg( \varepsilon(k_1,k_2,\epsilon_1,\epsilon_2) \Bigg[ N \sum_q Q_q^2 g(z_q) +Q_b^2 (g(1)+2)\Bigg] \\
  &&\qquad\quad -i Q_b^2\left(1-\frac{3}{2} g(1)\right)\Bigg[ 2 k_1\cdot \epsilon_2 k_2\cdot \epsilon_1 -m_{B_q}^2\epsilon_1\cdot \epsilon_2\Bigg]\Bigg) \nonumber\\ \label{1PIQ6}
  \langle Q_6\rangle &=& \frac{f_{B_q}\alpha}{\pi}\Bigg( \varepsilon(k_1,k_2,\epsilon_1,\epsilon_2) \Bigg[ \sum_q Q_q^2 g(z_q) +N Q_b^2 (g(1)+2)\Bigg] \\
  &&\qquad\quad -i N Q_b^2\left(1-\frac{3}{2} g(1)\right)\Bigg[ 2 k_1\cdot \epsilon_2 k_2\cdot \epsilon_1 -m_{B_q}^2\epsilon_1\cdot \epsilon_2\Bigg]\Bigg) \nonumber
\end{eqnarray}
with
\begin{equation}\label{gofz}
  g(z)=-2+4z\left[L_2\left(\frac{2}{1-\sqrt{1-4z+i\epsilon}}\right)+L_2\left(\frac{2}{1+\sqrt{1-4z+i\epsilon}}\right)\right]
\end{equation}
Here the dilogarithmic function was defined already in (\ref{dilog}), $z_q$ is again $m_q^2/m_b^2$, and $p=u,c$. We note that the function $g(z)$ is related to the kernel $h(u,z)$ defined in (\ref{hus}) in the context of $B\to V\gamma$ by $g(z)=h(1,z)$. The function $g(z)$ is regular for $z\to 0$, $g(0)=-2$, and has an imaginary part growing almost linearly with $z$ to sizeable values for $z=z_c$ and vanishing again for $z>1/4$. The imaginary part is due to the rescattering process $B\to q\bar q\to\gamma\gamma$ with an on-shell light-quark pair $q\bar q$ in the intermediate state. Since the process $B\to q\bar q$ is forbidden by helicity and angular momentum conservation for a massless quark $q$, the function $g(z)$ is real at $z=0$ (that is for $q=u,d,s$). The helicity suppression of the phase will be absent at higher order in $\alpha_s$. 

Including the effect of 1PI diagrams in the decay amplitudes, the quantities $A_\pm$ from (\ref{apmlp}) become
\begin{eqnarray}
  A^p_+ &=& -C_7 \frac{m_{B}}{\lambda_{B}} +\left( C_5 +N C_6\right) \left[ \frac{1}{2}g(1)-\frac{1}{3}\right] \label{apslp} \\
  A^p_- &=& -C_7 \frac{m_{B}}{\lambda_{B}}-\frac{2}{3} \left( C_1 + N C_2\right) g(z_p) -\left( C_3 -C_5\right) \left[ 2 g(z_c) +\frac{5}{6} g(1)\right] \nonumber\\
        && \qquad\quad\;\;\; -\left(C_4 -C_6\right)\left[ \frac{2}{3} g(z_c) +\frac{7}{6} g(1)\right]+\frac{20}{3}C_3 +4C_4 -\frac{16}{3}C_5 \label{amslp} 
\end{eqnarray}
where we have decomposed $A_\pm$ into
\begin{equation}\label{apmap}
A_\pm=\sum_{p=u,c} \lambda^{(q)}_p A^p_\pm
\end{equation}
neglected the light quark masses and used $g(0)=-2$. We see that the dominant effect is a modification of $A_-$ via the current-current operators $Q_1$ and $Q_2$. The imaginary part of the 1PI loop diagram with a charm quark running in the loop leads to a relative CP-conserving phase between the $\lambda_u$ and the $\lambda_c$ contribution. This gives a nonvanishing $r^-_{CP}$ while $r^+_{CP}$ is still zero because $A^u_+=A^c_+$. The effect of QCD penguin operators is small because of their tiny Wilson coefficients.

%===============================================================
%=      Long-distance contributions to B/D -> gamma gamma      =
%===============================================================
\section{Long-distance contributions to $B$ and $D\to\gamma\gamma$}
\label{sec:BDgamgamLD}

Long-distance contributions were calculated for both $B\to\gamma\gamma$ \cite{HI,HI2,CE,LZZ} and $D\to\gamma\gamma$ \cite{FSZ,BGHP}. Different mechanisms were considered:
\begin{itemize}
  \item $B$ and $D\to V\gamma$ followed by a $V\to\gamma$ conversion with the conversion factor supplied by the vector meson dominance (VMD) model. This gives long-distance contributions to the operator $Q_7$. Counting powers of $m_b$ in the explicit result for the $B_s\to\phi\gamma\to\gamma\gamma$ amplitude in \cite{HI} shows that it is suppressed with $\Lambda_{QCD}/m_b$ compared to the leading power result from Fig.~\ref{fig:1PRlp}. Yet, it is just one among the subleading power terms for the matrix element of $Q_7$. Another comes for example from the diagram in Fig.~\ref{fig:1PRslp}.

  We can apply a similar power counting to the $D\to\gamma\gamma$ VMD amplitude of \cite{BGHP}. In the limit of a heavy charm quark it is likewise power suppressed with $\Lambda_{QCD}/m_c$ compared to the short-distance amplitude. The limit $m_c\gg\Lambda_{QCD}$ is certainly questionable, but nevertheless some properties of charmed hadrons have been analyzed successfully in this limit \cite{HQS}. Again, the VMD contributions are just one out of several possible power corrections to the $Q_7$ matrix element.

  \item Single-particle unitarity contributions where the $D^0$ mixes with a spinless intermediate particle, which decays into a photon pair. Such a process corresponds to our 1PI contributions in Fig.~\ref{fig:1PI}. Relative to this contribution the amplitudes in \cite{BGHP} are at most of the same order in powers of the inverse ``heavy'' charm quark mass. 

  \item Two-particle unitarity contributions are other hadron level counterparts of our 1PI diagrams. Triangle loops with intermediate $D_s^{(*)}$ \cite{CE,LZZ} and $K^{(*)}$ \cite{BGHP} mesons were calculated for $B_s\to\gamma\gamma$ and $D\to\gamma\gamma$, respectively. As we will see in the next chapter, the effect of the 1PI contributions in $B\to\gamma\gamma$ on the branching ratio is very small. The authors of \cite{LZZ} also get a not too large effect, reestimating the result of \cite{CE}. Yet, one has to keep in mind that the effects from intermediate $D$ and $D^*$ mesons are only two of many possible hadronic intermediate states.

With obvious replacements we can use our 1PI results in (\ref{1PIQ1}--\ref{1PIQ6}) to estimate the quark-level analogue of the two-particle unitarity contributions to $D\to\gamma\gamma$ calculated in \cite{BGHP}. For the current-current operators we get for the 1PI amplitude
\begin{eqnarray}
  A_{\mathrm{1PI}}^{D\to\gamma\gamma} &=& \frac{G_F}{\sqrt{2}}\frac{\alpha}{\pi}f_D Q_d^2 \left( C_1 +N C_2\right)|V_{cs}||V_{us}|\left[g\!\left(\frac{m_s^2}{m_c^2}\right) -g(0)\right] i \, \varepsilon\!\left(\epsilon_1,\epsilon_2,k_1,k_2\right)\nonumber\\[0.1cm]
  &=& -2.4\cdot 10^{-11}\mathrm{GeV}^{-1}\left[ -0.2 +0.5i\pm 0.1i\right] i \, \varepsilon\!\left(\epsilon_1,\epsilon_2,k_1,k_2\right) \nonumber\\[0.1cm]
|A_{\mathrm{1PI}}^{D\to\gamma\gamma}|  &\approx& 1.3 \cdot 10^{-11}\mathrm{GeV}^{-1} \varepsilon\!\left(\epsilon_1,\epsilon_2,k_1,k_2\right) \label{ADgg1PI}
\end{eqnarray}
where we set $f_D=200\,\mathrm{MeV}$ and evaluated the Wilson coefficients at $m_c$. The contributions of a strange quark and a down quark in the loop are accompanied by CKM elements which are practically equal in absolute value but opposite in sign. This leads to a strong GIM cancellation in the square bracket. If we neglected the strange quark mass, the 1PI contribution would vanish altogether. This is a crucial difference of the $c\to u\gamma$ transition at one loop compared to the $b\to s\gamma$ transition and disturbs the hierarchy of higher order QCD terms. Leading logarithmic QCD corrections are known to enhance the amplitude by more than an order of magnitude. Including the two-loop QCD contributions increases the amplitude further by two orders of magnitude \cite{GHMW}. Therefore we would like to compare equation~(\ref{ADgg1PI}) with the $c\to u\gamma$ amplitude from the dominant two-loop diagrams $|A|=0.0047$ of \cite{GHMW}. Evaluating the factorization formula then leads to the following ``short distance'' amplitude:
\begin{eqnarray}\label{ADggSD}
  A_{\mathrm{SD}}^{D\to\gamma\gamma} &=& -\frac{G_F}{\sqrt{2}} \frac{\alpha}{\pi} f_D Q_u A \frac{m_D}{\lambda_D} \left[ 2i \, \varepsilon\!\left(\epsilon_1,\epsilon_2,k_1,k_2\right) -2k_1\!\cdot\! \epsilon_2 \, k_2\!\cdot\! \epsilon_1 +m_D^2\epsilon_1\!\cdot\!\epsilon_2\right] \nonumber\\[0.1cm]
  |A_{\mathrm{SD},-}^{D\to\gamma\gamma}| &\approx&12.8\cdot 10^{-11}\mathrm{GeV}^{-1}\varepsilon\!\left(\epsilon_1,\epsilon_2,k_1,k_2\right)
\end{eqnarray}
We see that the 1PI amplitude is smaller than the CP-odd part of the short distance amplitude. This should be understood as an estimate only because we relied on the heavy quark limit for the charm quark.

Burdman, Golowich, Hewett, and Pakvasa on the other hand find a large effect on the $D\to\gamma\gamma$ branching ratio from $K^+K^-$ intermediate states \cite{BGHP}. Their amplitude, however, is formally suppressed with three powers relative to our 1PI amplitude. In view of our above estimates the large effect could be overrated. Only one intermediate state was taken into account so that important contributions in the duality sum might be missing. We also expect cancellations due to the GIM mechanism as explicitely seen in our estimate above.

\item The chain process $B_s\to\phi\psi\to\phi\gamma\to\gamma\gamma$ was estimated to give very small effects on the $B\to\gamma\gamma$ decay amplitudes \cite{HI2}. It is multiply power suppressed compared to the leading power 1PR contribution.
\end{itemize}

% ===== Phenomenology =====================================
\chapter{The Phenomenology of $B\to\gamma\gamma$}
\label{ch:phenBgamgam}
Here we present numerical results for the expressions derived in the previous chapter. Additionally to the input parameters given in table~\ref{tab:input} we need the parameters related to the $B_s$ meson. These we summarize in table~\ref{tab:inputBs}.
%%%%%%%%%%%%%%%%%%%%%%%%%%%%%%%%%%%%%%%%%%%%%%%%%%%%
%%%%%%%% table of input parameters   %%%%%%%%%%%%%%%
%%%%%%%%%%%%%%%%%%%%%%%%%%%%%%%%%%%%%%%%%%%%%%%%%%%%
\begin{table}[tpb]
\renewcommand{\arraystretch}{1.2}
\begin{center}
\begin{tabular*}{140mm}{@{\extracolsep\fill}|c|c|c|c|}
\hline\hline
\multicolumn{4}{|c|}{Parameters related to the $B_s$ meson}\\
\hline
$m_{B_s}$ & $f_{B_s}$ \cite{SMR}& $\tau_{B_s}$ & $\lambda_{B_s}$\\
\hline
5.37 GeV & (230 $\pm$ 30) MeV & 1.49 ps & $(350 \pm 150)$ MeV\\
\hline\hline
\end{tabular*}
\end{center}
\caption[]{The input parameters for the $B_s$ meson.\label{tab:inputBs}}
\end{table}
%%%%%%%%%%%%%%%%%%%%%%%%%%%%%%%%%%%%%%%%%%%%%%%%%%%%
We will work to leading logarithmic order in QCD and use the running mass $\overline{m}_t(m_t)=166\,\mathrm{GeV}$ for the top quark as input for the Wilson coefficient $C_7$. For definiteness we employ the two-loop form of the running coupling $\alpha_s(\mu)$ as quoted in (\ref{runalpha}), which corresponds to $\alpha_s(M_Z)=0.118$ for $\Lambda^{(5)}_{\overline{MS}}=225\,{\rm MeV}$. With central values of all input parameters, at $\mu=m_b$, and using a nominal value of the CKM angle $\gamma=58^\circ$ we find for the branching ratios to leading logarithmic accuracy, evaluating (\ref{gbgg}), (\ref{apslp}--\ref{apmap}):
\begin{eqnarray}
  \label{brbsnum}B(\bar B_s\to\gamma\gamma) &=& 1.22\cdot 10^{-6}\\
  \label{brbdnum}B(\bar B_d\to\gamma\gamma) &=& 3.10\cdot 10^{-8}
\end{eqnarray}
Using the leading power amplitudes (\ref{apmlp}) without the 1PI contributions we get
\begin{eqnarray}
  \label{brbsldgpownum}B_\mathrm{ldg.pow.}(\bar B_s\to\gamma\gamma) &=& 1.17\cdot 10^{-6}\\
  \label{brbdldgpownum}B_\mathrm{ldg.pow.}(\bar B_d\to\gamma\gamma) &=& 3.12\cdot 10^{-8}
\end{eqnarray}
To display the size of power corrections from 1PI diagrams on amplitude level we compare $A^p_+$ and $A^p_-$ in (\ref{apslp}) and (\ref{amslp}) with the leading power result $A^p_{\pm,\mathrm{ldg.pow.}} =-C_7 m_{B_s}/\lambda_{B_s}$ for $B_s\to\gamma\gamma$
\begin{equation}\label{amucnum}
  \begin{array}{lclllcl}  
    A^{u/c}_+ &=& 4.872            &                             & +0.045                            &=& 4.917\\
    A^u_- &=& 4.872            & +0.419                      & -0.061-0.033\,i                   &=& 5.230-0.033\,i\\
    A^c_- &=& 4.872            & +0.203-0.534\,i             & -0.061-0.033\,i                   &=& 5.014-0.567\,i\\[0.2cm]
          & & \mbox{ldg. pow.} & Q_{1,2}\mbox{ contribution} & Q_{3\ldots 6}\mbox{ contribution} & &
  \end{array}
\end{equation}
where we showed separately the leading power $Q_7$ contribution and the effects from current-current and QCD-penguin operators to the 1PI contribution. For $B_d\to\gamma\gamma$ nominally only $A^p_{\pm,\mathrm{ldg.pow.}} =-C_7 m_{B_s}/\lambda_{B_s}=4.79$ is changed. The calculable power corrections thus are of the expected canonical size of ${\cal O}(10\%)$. The precise value of these power-suppressed effects depends strongly on the renormalization scale $\mu$. Typically, the relative size of the 1PI terms diminishes when the scale is lowered, both due to an increase in $C_7$ as well as the simultaneous decrease of the combination $a_2$. The effect of QCD penguin operators to the 1PI diagrams is small.

The main errors in (\ref{brbsnum}--\ref{amucnum}) come from the variation of the nonperturbative input parameters $\lambda_B$ and the decay constants $f_{B_d}$ and $f_{B_s}$ which are all poorly known and enter the branching ratios quadratically. The residual scale dependence is sizeable as well because we calculated at leading logarithmic accuracy only. The variation of the branching ratio with the scale would be less severe if next-to-leading QCD corrections were known. We expect the NLL corrections to increase the branching ratio as was the case for both the inclusive \cite{CMM} and exclusive \cite{BBVgam,BFS} $b\to s\gamma$ decays.

The effect of next-to-leading logarithmic corrections for the branching ratio of $b\to s\gamma$ can be reproduced by choosing a low renormalization scale ($\mu\approx m_b/2$) in the leading logarithmic expressions. If such a low scale were also relevant for $B\to\gamma\gamma$ the branching ratios would read
\begin{eqnarray}
  B(\bar B_s\to\gamma\gamma)_{\mu=m_b/2} &=& 1.52\cdot 10^{-6}\\
  B(\bar B_d\to\gamma\gamma)_{\mu=m_b/2} &=& 4.06\cdot 10^{-8}
\end{eqnarray}
However, we prefer to use the nominal $\mu=m_b$ and to quote the standard scale ambiguity with $m_b/2 <\mu < 2 m_b$ as a theoretical uncertainty.

For comparison we also show the results for the case where short-distance QCD effects are neglected altogether. This amounts to taking $\mu=M_W$ in the leading-order Wilson coefficients and leads to
\begin{eqnarray}
  B(\bar B_s\to\gamma\gamma)_{\mu=M_W} &=& 0.60\cdot 10^{-6}\\
  B(\bar B_d\to\gamma\gamma)_{\mu=M_W} &=& 1.22\cdot 10^{-8}
\end{eqnarray}
Clearly, the leading logarithmic QCD corrections lead to a substantial enhancement of the branching ratios (\ref{brbsnum}), (\ref{brbdnum}), similar to the case of $b\to s\gamma$.\medskip

The numerical evaluation of (\ref{rCPpar}) gives for central values of the input parameters and at $\mu=m_b$ the following predictions for the CP-violating ratios $r^-_{CP}$:
\begin{eqnarray}
  \label{rcpbsnum}r^-_{CP}(B_s) &=&  0.35\%\\
  \label{rcpbdnum}r^-_{CP}(B_d) &=& -9.66\%
\end{eqnarray}
The effect is negligible for $B_s\to\gamma\gamma$ as expected. For $B_d\to\gamma\gamma$ CP violation occurs at the level of about $10\%$. It is interesting to note that in $B\to\gamma\gamma$ a non-vanishing CP asymmetry appears already at ${\cal O}(\alpha_s^0)$. This is possible via a modified BSS mechanism: Instead of the usual QCD penguin, already the electroweak loop in Fig.~\ref{fig:1PI} itself leads to a CP-conserving phase. This phase difference defined in (\ref{rCPpar}) is nonvanishing for the CP-odd amplitude only: $\alpha_c^- -\alpha_u^- =-6.1^\circ$. Including the effects of QCD penguin operators does not lead to a CP-conserving phase difference $\alpha^+_c -\alpha^+_u$ and therefore $r^+_{CP}=0$.

\begin{figure}[t]
   \epsfxsize=12cm
   \centerline{\epsffile{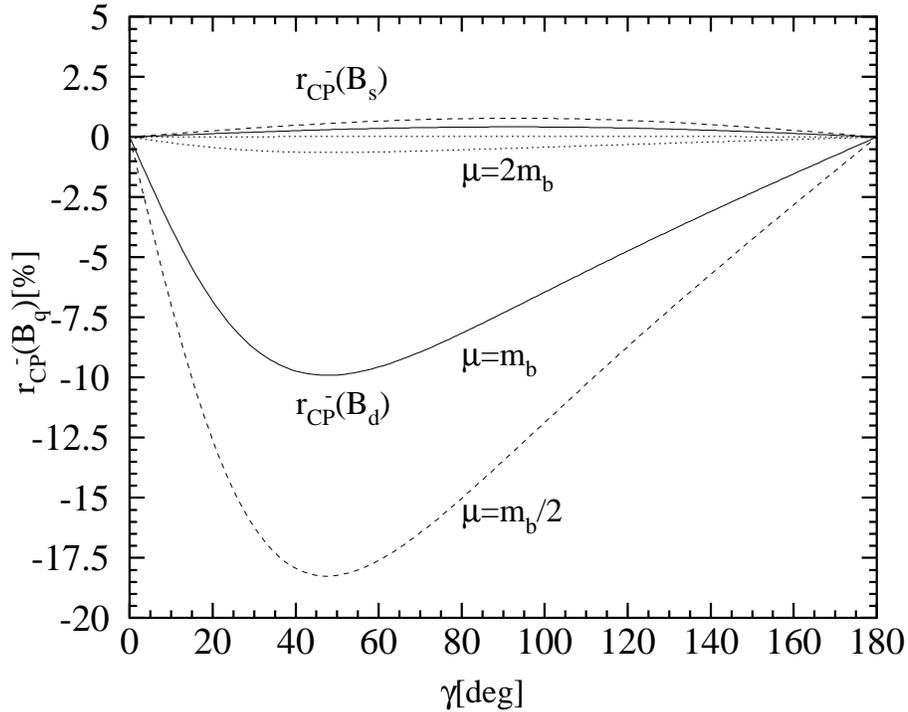}}
\caption{The CP-violating ratios $r^-_{CP}$ for decays of neutral $B_d$ 
  and $B_s$ mesons to two photons as a function of the CKM angle $\gamma$, 
  each for three values of the renormalization scale $\mu=m_b/2$, $m_b$ 
  and $2 m_b$.\label{fig:rCPgam}}
\end{figure}

The scale dependence of the CP asymmetry is rather strong, as can be seen in Fig.~\ref{fig:rCPgam}.
%%%%%%%%%%%%%%%%%%%%%%%%%%%%%%%%%%%%%%%%%%%%%%%%%%%%%%%%%%%%%%%%%%%
%%%%%%%%%%%%%%%%%%%%%%%%%%%%%%%%%%%%%%%%%%%%%%%%%%%%%%%%%%%%%%%%%%%
The ratio $r_{CP}$ also depends sensitively on fundamental CKM parameters such as the CKM angle $\gamma$ (Fig.~\ref{fig:rCPgam}). The extremal value of $r_{CP}^-(B_d)$ is obtained for $\gamma=50^\circ$. The sensitivity of the branching ratios and CP asymmetries to variations in the relevant input parameters is summarized in table \ref{tab:outputBgg}.
%%%%%%%%%%%%%%%%%%%%%%%%%%%%%%%%%%%%%%%%%%%
%%%%%%%% table of output    %%%%%%%%%%%%%%%
%%%%%%%%%%%%%%%%%%%%%%%%%%%%%%%%%%%%%%%%%%%
\begin{table}[tpb]
\renewcommand{\arraystretch}{1.1}
\begin{center}
\begin{tabular}{|l|c|c|c|c|}
\hline\hline
                          & $B(\bar B_d \to \gamma\gamma)$ & $B(\bar B_s\to \gamma\gamma)$ & $r^-_{CP}(B_d)$ & $r^-_{CP}(B_s)$\\
                          & $[10^{-8}]$                    & $[10^{-6}]$                   & $[\%]$          & $[\%]$\\
\hline\hline
central                   & 3.09                           & 1.22                          & $-$9.66         & 0.35
\\
\hline\hline
$\lambda_B$               & +6.40/$-$1.58                  & 2.44/$-$0.61                  &  +4.13/$-$4.02  & +0.15/$-$0.15\\
\hline
$f_B$                     & +1.00/$-$0.86                  & +0.34/$-$0.30                 & --              & --\\
\hline
$\mu$                     & +0.97/$-$0.66                  & +0.31/$-$0.21                 & +9.03/$-$8.13   & +0.30/$-$0.33\\
\hline
$m_c$                     & +0.10/$-$0.12                  & +0.03/$-$0.04                 & +1.79/$-$1.58   & +0.06/$-$0.06\\
\hline
$\gamma=(58\pm 24)^\circ$ & +1.28/$-$0.95                  & +0.01/$-$0.02                 & +1.66/$-$0.25   & +0.06/$-$0.12\\
\hline\hline
\end{tabular}
\end{center}
\caption[]{Predictions for branching ratios and CP asymmetries with the errors from the individual input uncertainties.\label{tab:outputBgg}}
\end{table}
%%%%%%%%%%%%%%%%%%%%%%%%%%%%%%%%%%%%%%%%%%%

Our predictions for the central values of the branching ratios are roughly two orders of magnitude below the current experimental bounds. The recent upper limit for $B(B_d\to\gamma\gamma)$ from {\sc BaBar} in (\ref{brbsdggex}) improved the previous limit from the L3 collaboration \cite{L3} by a factor of 20. With more integrated luminosity accumulated the upper bound will further be reduced. Yet, even with the planned total integrated luminosity of $500\,\mathrm{fb}^{-1}$ for each {\sc BaBar} and BELLE, the number of $B_d\to\gamma\gamma$ decays one expects to observe is only a single one. The high-luminosity option Super{\sc BaBar} proposes to collect a total integrated luminosity of $10\,\mathrm{ab}^{-1}$. Therefore the number of observed events for $B_d\to\gamma\gamma$ would be two dozens \cite{SuperBaBar}. As long as the $e^+ e^-$ $B$ factories run at the $\Upsilon (4s)$ resonance they produce no $B_s$ mesons and therefore cannot profit from the larger branching fraction to observe $B_s\to\gamma\gamma$ decays. At hadron machines as the Fermilab Tevatron or the LHC at CERN all sorts of $b$-flavoured hadrons are produced. CDF, for instance, expects to observe at the present run IIa at Tevatron in $2\,\mathrm{fb}^{-1}$ of data 28 million $b\to J/\psi X$ decays \cite{BTeV}. The number of $B_s$ mesons will therefore be roughly 7 million. It will consequently be very hard to distinguish $B_s\to\gamma\gamma$ events from the immense combinatorial background events.

As already mentioned, the dominant uncertainty for our prediction of the branching ratios comes from the variation of the hadronic parameter $\lambda_B$ (factor 2). It governs the uncertainty of the leading power amplitude. So far, our value for $\lambda_B$ quoted in table~\ref{tab:input} is only an educated guess with the uncertainty taken appropriately large. As $\lambda_B$ enters the leading power branching ratio quadratically a large uncertainty in this input parameter is translated into an even larger uncertainty of the branching ratio output. It would be highly desirable to get a better understanding of $\lambda_B$. We hope that this problem can be attacked soon with the help of QCD sum rule or lattice QCD techniques. As $\lambda_B$ parametrizes the first negative moment of the $B$ meson wave function it is a universal quantity and therefore independent of the precise decay under consideration. An extraction of it therefore might be possible from radiative semileptonic $B$ decays, for instance.

Leaving the uncertainty of $\lambda_B$ aside, the variation of the remaining parameters changes the $\bar B_d\to\gamma\gamma$ and $\bar B_s\to\gamma\gamma$ branching ratios by order $\pm 50\%$ and $\pm 35\%$, respectively. If both the $\bar B_d\to\gamma\gamma$ and $\bar B_s\to\gamma\gamma$ branching ratios were measured one could consider their ratio where the dominant uncertainties from the $\lambda_B$ and $f_B$ cancel to a large extent.

The main theoretical uncertainty for the CP asymmetries at present comes from the variation of the renormalization scale $\mu$. This could be reduced when QCD corrections are included. But as the CP asymmetries in double radiative exclusive decays are extremely hard to measure experimentally we see at present no urgent need to perform this calculation.

% =========================================================
% =     part IV: Conclusions                              =
% =========================================================
\part{Conclusions}

% ===== Conclusions =======================================
\chapter{Conclusions}
\label{ch:conclusions}

While waiting for the completion of LHC and new results from Tevatron Run II, $B$ physics is among the most active fields in recent particle physics. The main goal of $B$ physics is a precision study of the flavour sector to extract Standard Model parameters with high precision in order to reveal potential New Physics effects. Theoretically, the complication is that three fundamental scales are involved in $B$ decays: the weak interaction scale $M_W$, the $b$ quark mass $m_b$, and the QCD scale $\Lambda_{QCD}$. Via the machinery of operator product expansion and renormalization group equations in the framework of an effective theory one can factorize perturbatively calculable short distance Wilson coefficients from the long distance operator matrix elements. The latter pose the main difficulty as all low-energy contributions below the factorization scale $\mu={\cal O}(m_b)$ are collected into them. Yet, one can take advantage of the fact that $m_b\gg\Lambda_{QCD}$ and achieve a further separation of short- and long-distance part of these matrix elements. Employed for exclusive final states this application of QCD factorization amounts to a systematic expansion in $\Lambda_{QCD}/m_b$ and the strong coupling parameter $\alpha_s$.

Within this thesis we have analyzed in detail the exclusive radiative decays $B\to V\gamma$ ($V=K^*$ or $\rho$) and $B\to\gamma\gamma$ in QCD factorization. This enabled us to separate perturbatively calculable contributions from the nonperturbative form factors and universal meson light-cone distribution amplitudes. A power counting in $\Lambda_\mathrm{QCD}/m_b$ implies a hierarchy among the possible transition mechanisms and allows to identify leading and subleading contributions.

Our framework has major advantages when compared with previous approaches to exclusive radiative decays. They all had to rely on hadronic models to estimate bound state effects. These models, however, did not allow a clear separation of short- and long-distance dynamics or a clean distinction of model-dependent and model-independent features. QCD factorization is a systematic and model-independent approach based on first principles and the heavy-quark limit $m_b\gg\Lambda_\mathrm{QCD}$. It holds for $m_b\to\infty$. In reality, however, the bottom quark is not infinitely heavy so that we have to worry about $\Lambda_\mathrm{QCD}/m_b$ power corrections. For the radiative $B$ decays we considered, some of these power corrections - possibly the numerically most important ones - are calculable in QCD factorization. The remaining ones are not expected to be very large. In particular, no ``chirally enhanced'' contributions, as in non-leptonic modes with pseudoscalar mesons in the final state, can arise in our case.

Sizeable uncertainties come from the variation of the nonperturbative input parameters, in particular form factors, decay constants, and information on the $B$ meson wave function ($\lambda_B$), which are all poorly known. But we await progress on these quantities from lattice QCD and analytical methods as well as from an extraction from experimental results.

An important conceptual aspect of our analyses is the interpretation of quark loop contributions. We showed that these effects are calculable in QCD factorization. Before, such quark loops were considered as generic, uncalculable long-distance contributions. Non-factorizable long-distance corrections may still exist, but they are power-suppressed. Let us in the following address our findings for $B\to V\gamma$ and $B\to\gamma\gamma$ separately.\bigskip

The radiative decay with a vector meson $K^*$ in the final state is of particular prominence as its branching ratio was already measured. We presented in this work the first really complete calculation with next-to-leading logarithmic accuracy. Compared to our publication \cite{BBVgam} we also included the effect of QCD penguin operators and added an explicit proof, at one-loop order, for the factorizability of annihilation contributions. At next-to-leading order in QCD interesting effects occur: the operators $Q_{1\ldots 6}$ and $Q_8$ start contributing, quark loops allow for CP-violating effects, and the hard-spectator corrections appear. We have seen that weak-annihilation amplitudes are power suppressed, but calculable in QCD factorization and numerically important because they enter with large coefficients. Therefore, we included their effects in our phenomenological discussion. We estimated other power corrections to be numerically small.

Our NLO predictions for the central values of the branching ratios are
\begin{displaymath}
  \begin{array}{lclclcl}
    B(\bar B^0\to \bar K^{*0}\gamma) &=& (7.4^{+2.6}_{-2.4})\cdot 10^{-5} & \qquad & B(B^-\to K^{*-}\gamma) &=& (7.3^{+2.6}_{-2.4})\cdot 10^{-5}\\[0.2cm]
    B(\bar B^0\to\rho^0\gamma)       &=& (0.8^{+0.4}_{-0.3})\cdot 10^{-6} & \qquad & B(B^-\to\rho^-\gamma)  &=& (1.6^{+0.7}_{-0.5})\cdot 10^{-6}
  \end{array}
\end{displaymath}
They are substantially larger than the leading logarithmic values when the same form factors are used. The dominant correction comes from the hard-vertex contributions, which similarly increased the result for the inclusive $B\to X_s\gamma$ decay. Adding the effect of varying the input parameters in quadrature leads to an error estimate of ${\cal O}(35\%)$ and ${\cal O}(40\%)$ for the $B\to K^*\gamma$ and $B\to\rho\gamma$ branching ratios, respectively. For $B(B\to K^*\gamma)$ by far the dominant uncertainty comes from the $B\to K^*$ form factor, which alone already amounts to $30\%$. This situation, however, can be systematically improved. In particular, our approach allows for a consistent perturbative matching of the form factor to the short-distance part of the amplitude. The largest uncertainty in $B(B\to\rho\gamma)$ is also the form factor. Yet, variation of the hadronic parameter $\lambda_B$ and the CKM angle $\gamma$ affect the prediction for the branching ratio also considerably.

Compared to the experimental measurements of the $B\to K^*\gamma$ branching ratios our predicted central values are somewhat high. Although within the uncertainties still compatible, the discrepancy of the central values could be a hint that the form factors at $q^2 =0$ are smaller than the value obtained from QCD sum rules. At the present stage we also cannot exclude that power corrections or even a failure of the QCD factorization approach are responsible for the difference. Because the experimentally measured inclusive $b\to s\gamma$ rate is in good agreement with its theoretical prediction we do not expect sizeable New Physics contributions to the exclusive $B\to K^*\gamma$ decays either. For the $b\to d\gamma$ transitions on the other hand there is still room for physics beyond the Standard Model as experimentally at present only upper bounds for these decays exist. They are only a factor of two above our predictions so that we expect the $B\to\rho\gamma$ branching ratios to be measured soon. From such a measurement we can obtain important information on our input parameters, in particular the form factor, the hadronic parameter $\lambda_B$, and the CKM angle $\gamma$, and also on the validity of the QCD factorization approach to exclusive radiative $B$ decays.\medskip

For $B\to\rho\gamma$ both CKM sectors of the effective Hamiltonian have the same order of magnitude. This allows a sizeable CP-violating asymmetry of ${\cal A}(\rho^\pm\gamma)=10.1\%$. In $B\to K^*\gamma$, on the other hand, $|\lambda^{(s)}_u|\ll|\lambda^{(s)}_c|$ and therefore the CP asymmetry is tiny, typically $-0.5\%$. Because the CP asymmetry arises at ${\cal O}(\alpha_s)$ for the first time, we have a large dependence on the renormalization scale $\mu$. This could be reduced by going one order further in perturbation theory, which would amount to a three-loop calculation.

We included the numerically most important power corrections - weak annihilation - into our analysis.  Within our approximations weak annihilation is the only effect that is sensitive to the charge of the $B$ meson and therefore to isospin breaking. In $B\to K^*\gamma$ annihilation contributions from $Q_{1,2}$ are CKM suppressed such that their influence on the branching ratio is small but a curiosity is a rather large effect of the QCD penguin operator $Q_6$ \cite{KN}. So for both $B\to K^*\gamma$ and $B\to\rho\gamma$  the isospin asymmetry is predicted to be sizeable: $\Delta(K^*\gamma)=(-7.5^{+4.1}_{-5.9})\%$ and $\Delta(\rho\gamma)=(2.0^{+27.0}_{-15.7})\%$. The isospin breaking in $B\to\rho\gamma$ depends very sensitively on the CKM angle $\gamma$ and can thus in principle serve to constrain this quantity once measurements become available.

An explicit estimate of U-spin breaking effects in $B\to V\gamma$ decays quantified the limitations of the use of U-spin symmetry in exclusive decays as a Standard Model test.
\medskip

Our formalism can also straightforwardly be applied to other radiative rare $B$ decays, such as $B_s\to V\gamma$ where $V=\phi,\,K^*$, $B_d\to\omega\gamma$, or radiative decays to higher resonances. Some of these modes are discussed in \cite{GP,BK**gam}. Yet, for QCD factorization it is again the form factors, which pose the problem. To our knowledge they were never calculated. This would be highly desireable because for example the $B\to K_2^*(1430)\gamma$ decay was already experimentally observed \cite{CHEN,expBK**gam}.\bigskip

The double radiative decays $B_s\to\gamma\gamma$ and $B_d\to\gamma\gamma$ were not seen in experiment yet. Within QCD factorization we predict the branching ratios to leading logarithmic accuracy:
\begin{eqnarray*}
  B(\bar B_s\to\gamma\gamma) &=& (1.2^{+2.5}_{-0.7})\cdot 10^{-6}\\[0.2cm]
  B(\bar B_d\to\gamma\gamma) &=& (3.1^{+6.7}_{-2.1})\cdot 10^{-8}
\end{eqnarray*}
In the heavy quark limit the dominant contribution at leading power comes from only a single diagram, but an important class of subleading contributions can also be calculated. We used these corrections to estimate CP asymmetries in $B\to\gamma\gamma$ and one-particle irreducible two-photon emission from quark loops in $D\to\gamma\gamma$. The CP-violating ratio $r_{CP}^-(B_s)=0.4\%$ is negligible and $r_{CP}^-(B_d)=-9.7\%$, albeit not very small, will not be measured in the near future.

The absolutely dominant uncertainty in the branching ratios comes from the variation of the hadronic parameter $\lambda_B$. This universal quantity is theoretically very poorly known and enters the branching ratio quadratically. We propose to further investigate $\lambda_B$ which is of relevance also for other applications of QCD factorization to exclusive $B$ meson decays.
\bigskip

In table \ref{tab:exp}
%%%%%%%%%%%%%%%%%%%%%%%%%%%%%%%%%%%%%%%%%%%%%%%%%%%%
%%%%%%%%    table of decay reach     %%%%%%%%%%%%%%%
%%%%%%%%%%%%%%%%%%%%%%%%%%%%%%%%%%%%%%%%%%%%%%%%%%%%
\begin{table}
\renewcommand{\arraystretch}{1.2}
\begin{center}
\begin{tabular}{|l||c||c|c||c|c|}
\hline
&& \multicolumn{2}{c||}{Hadron Collider} & \multicolumn{2}{|c|}{$e^+ e^-$ $B$ Factories}\\
&& \multicolumn{2}{c||}{Experiments} & \multicolumn{2}{|c|}{}\\
\hline
Decay & Branching & CDF & BTeV  & {\sc BaBar} & Super-\\
Mode  & Fractions & D0  & LHC-b & Belle &{\sc BaBar}\\
 & & $(2\,\mathrm{fb}^{-1})$ & $(10^7 \,\mathrm{s})$ & $(0.5\,\mathrm{ab}^{-1})$ & $(10\,\mathrm{ab}^{-1})$\\
\hline\hline
$B\to K^*\gamma$       & $5\cdot 10^{-5}$ & 170 & 25K & 6K  & 120K\\
$B\to\rho\gamma$       & $2\cdot 10^{-6}$ &     &     & 300 & 6K\\
\hline
$B_d^0\to\gamma\gamma$ & $3\cdot 10^{-8}$ &     &     & 1.3 & 25\\
\hline
\end{tabular}
\end{center}
\caption{Decay reach of $B$ experiments for exclusive radiative decays \cite{SuperBaBar}.\label{tab:exp}}
\end{table}
%%%%%%%%%%%%%%%%%%%%%%%%%%%%%%%%%%%%%%%%%%%%%%%%%%%%
we summarize the decay reach of the modes we discussed. The information is taken from the Super-{\sc BaBar} proposal \cite{SuperBaBar}. Although in our estimate the branching fraction for $B_d^0\to\gamma\gamma$ is by a factor of three larger compared to the value in \cite{SuperBaBar}, these decays are indeed very rare. The $B_s^0\to\gamma\gamma$ modes are roughly 40 times more frequent, but they cannot be produced at the $\Upsilon(4s)$ resonance. At hadron colliders, on the other hand, also $B_s$ mesons are produced. But the problem with the $\rho\gamma$ and $\gamma\gamma$ final state in a hadron collider environment is that the combinatorial background is extremely large. Photons and $\rho$ mesons are produced copiously. Therefore, in table~\ref{tab:exp} no estimates exist how many of these decays could be really observed. Nevertheless we trust in the excellence of our experimental colleagues that they manage to measure the $B\to\rho\gamma$ or even $B\to\gamma\gamma$ modes at hadron collider experiments. Running a prospective high-energy $e^+ e^-$ linear collider on the $Z$ resonance could also improve the bounds on the $B\to\gamma\gamma$ branching ratios or even turn them into measurements.
\bigskip

According to Niels Bohr, the ``goal of science is to gradually reduce our prejudice'' \cite{quigg}. We think we made one step in reducing the prejudice against a theoretical calculation of exclusive decays. Many exclusive $B$ decays can be treated successfully in QCD factorization and the radiative ones are part of this set. QCD factorization has put the fear of God into hadronic matrix elements of exclusive $B$ decays. Still a lot of work has to be done, especially concerning the power corrections. But we have for the first time also for exclusive decays a tool which is based on a systematic framework, namely the heavy-quark limit. We will see if the phenomenological impact of QCD factorization is equal.

% =========================================================
% =     part V: Appendices                               =
% =========================================================
\part{Appendices}

\begin{appendix}
\chapter{Wilson Coefficients for $b\to s\gamma$}
\label{app:WC}

For completeness we want to give the explicit formulas for the Wilson coefficients of the operator bases (\ref{q1def}--\ref{q8def}) and (\ref{p1def}--\ref{p8def}) in the naive dimensional regularization and $\overline{MS}$ renormalization scheme.

\begin{figure}
   \begin{center}\psfig{figure=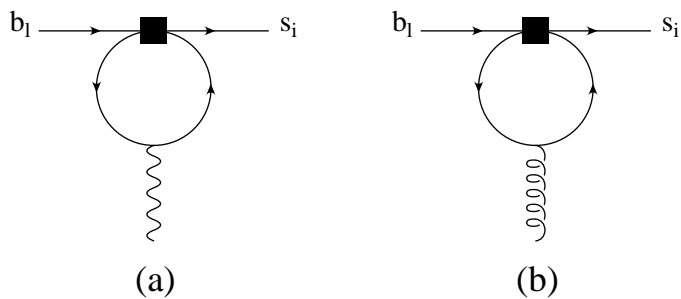}\end{center}
\caption{Magnetic photon and gluon penguin diagrams.\label{fig:magnpen}}
\end{figure}

Let us consider the magnetic penguin diagrams in Fig.~\ref{fig:magnpen}
%%%%%%%%%%%%%%%%%%%%%%%%%%%%%%%%%%%%%%%%%%%%%%%%%%%%%%%%%%%%%%%%%%%
%%%%%%%%%%%%%%%%%%%%%%%%%%%%%%%%%%%%%%%%%%%%%%%%%%%%%%%%%%%%%%%%%%%
The loop integral of these diagrams contains two $\gamma$ matrices. Therefore, insertions with the Dirac structure $\gamma_{\mu_1}\ldots\gamma_{\mu_{2n+1}}(1-\gamma_5)\otimes\gamma^{\mu_1}\ldots\gamma^{\mu_{2n+1}}(1-\gamma_5)$ vanish. This is in particular the case for $Q_{1\ldots 4}$ and is eventually a consequence of chirality conservation in the quark loop and QED gauge invariance. Non-vanishing results come from $Q_5$ and $Q_6$ only in the NDR scheme. The magnetic photon penguin diagrams gives
\begin{equation}\label{y56def}
  \begin{array}{lcrlcl}
    \displaystyle \langle Q_5\rangle^{(a)} &=& \displaystyle -\frac{1}{3} & \displaystyle \!\!\frac{e}{4\pi^2} \bar s \not\!\epsilon \not\! q m_b(1+\gamma_5) b &=:& y_5 \langle s\gamma|Q_7|b\rangle\\[0.4cm]
    \displaystyle \langle Q_6\rangle^{(a)} &=& \displaystyle -\frac{N}{3} & \displaystyle \!\!\frac{e}{4\pi^2} \bar s \not\!\epsilon \not\! q m_b(1+\gamma_5) b &=:& y_6 \langle s\gamma |Q_7|b\rangle
  \end{array}
\end{equation}
For the insertion in the magnetic gluon penguin diagram we replace the photon vertex with the gluon vertex and get
\begin{equation}\label{z56def}
  \begin{array}{lclcl}
    \displaystyle \langle Q_5\rangle^{(b)} &=& \displaystyle \frac{g_s}{4\pi^2} T^a\bar s \not\!\epsilon \not\! q m_b(1+\gamma_5) b &=:& z_5 \langle s\gamma|Q_8|b\rangle\\[0.4cm]
    \displaystyle \langle Q_6\rangle^{(b)} &=& \displaystyle 0 &=:& z_6 \langle s\gamma |Q_8|b\rangle
  \end{array}
\end{equation}
The above matrix elements vanish for an on-shell photon in any 4-dimensional regularization scheme and also the HV scheme. Such a peculiar regularization scheme dependence of leading order matrix elements has to be cancelled by a corresponding scheme dependence of the LO anomalous dimension matrix \cite{CFMRS}. This is disturbing because $\gamma^{(0)}$ is usually scheme independent. It is convenient to introduce the so-called ``effective coefficients'' for the operators $Q_7$ and $Q_8$ \cite{BMMP}
\begin{eqnarray}\label{c78effdef}
  C_7^{(0)eff}(\mu) &=& C_7^{(0)} +\sum_{i=1}^6 y_i C_i^{(0)}(\mu)\\
  C_8^{(0)eff}(\mu) &=& C_8^{(0)} +\sum_{i=1}^6 z_i C_i^{(0)}(\mu)
\end{eqnarray}
with $\vec y=(0,0,0,0,-\frac{1}{3},-\frac{N}{3})$ and $\vec z=(0,0,0,0,1,0)$ in the NDR scheme. Corresponding values for the CMM operator basis (\ref{p1def}--\ref{p8def}) are $\vec y=(0,0,-\frac{1}{3},-\frac{4}{9},-\frac{20}{3},-\frac{80}{9})$ and $\vec z=(0,0,1,-\frac{1}{6},20,-\frac{10}{3})$. The leading order effective coefficients $C^{(0)eff}_i$ with the corresponding anomalous dimension matrix $\gamma^{(0)eff}$ then are regularization and renormalization scheme independent. In addition, the leading order $b\to s\gamma$ and $b\to s \,g$ matrix elements are proportional to $C_7^{(0)eff}$ and $C_8^{(0)eff}$, respectively. In order to simplify the notation we will omit the label ``eff'' throughout this work. Whenever $C_7$ or $C_8$ appear, the effective coefficients are understood.

Let us expand the $C_i$ in powers of $\alpha_s$:
\begin{equation}\label{expci}
  C_i(\mu)=C_i^{(0)}(\mu)+\frac{\alpha_s}{4\pi}C_i^{(1)}(\mu)+\ldots
\end{equation}
The evolution of the (effective) Wilson coefficients is driven by the (effective) anomalous dimension matrix. Its leading order in the expansion (\ref{adm}) is given for our operator basis (\ref{q1def}--\ref{q8def}) by
\begin{equation}\label{adm0}
\gamma^{(0)}=\left(\begin{array}{cccccccc}
	-2 & 6  & -\frac{2}{9}   & \frac{2}{3}     & -\frac{2}{9}  & \frac{2}{3}      & \frac{416}{81}                   & \frac{70}{27}\\[0.2cm]
	6  & -2 & 0              & 0               & 0             & 0                & 0                                & 3\\[0.2cm]
	0  & 0  & -\frac{22}{9}  & \frac{22}{3}    & -\frac{4}{9}  & \frac{4}{3}      & -\frac{464}{81}                  & \frac{140}{27}+3f\\[0.2cm]
	0  & 0  & 6-\frac{2f}{9} & -2+\frac{2f}{3} & -\frac{2f}{9} & \frac{2f}{3}     & \frac{416u-232d}{81}  & 6+\frac{70f}{27}\\[0.2cm]
	0  & 0  & 0              & 0               & 2             & -6               & \frac{32}{9}                     & -\frac{14}{3}-3f\\[0.2cm]
	0  & 0  & -\frac{2f}{9}  & \frac{2f}{3}    & -\frac{2f}{9} & -16+\frac{2f}{3} & \frac{-448u+200d}{81} & -4-\frac{119f}{27}\\[0.2cm]
	0  & 0  & 0              & 0               & 0             & 0                & \frac{32}{3}                     & 0\\[0.2cm]
	0  & 0  & 0              & 0               & 0             & 0                & -\frac{32}{9}                    & \frac{28}{3}
\end{array}\right)
\end{equation}
Here we have already set the number of quark colours to 3. The number of active up- and down-type flavours is denoted by $u$ and $d$, respectively and $f=u+d$. The corresponding ADM for the Chetyrkin, Misiak, M\"unz operator basis is \cite{CMM}
\begin{equation}\label{adm0CMM}
  \gamma^{(0)}_\mathrm{CMM}=\left(\begin{array}{cccccccc}
    -4 & \frac{8}{3} &  0             & -\frac{2}{9}    & 0            &  0           & -\frac{208}{243}  &  \frac{173}{162}\\[0.2cm]
    12 & 0           &  0             &  \frac{4}{3}    & 0            &  0           &  \frac{416}{81}   &  \frac{70}{27}\\[0.2cm]
    0  & 0           &  0             & -\frac{52}{3}   & 0            &  2           & -\frac{176}{81}   &  \frac{14}{27}\\[0.2cm]
    0  & 0           & -\frac{40}{9}  & -\frac{100}{9}  & \frac{4}{9}  &  \frac{5}{6} & -\frac{152}{243}  & -\frac{587}{162}\\[0.2cm]
    0  & 0           &  0             & -\frac{256}{3}  & 0            &  20          & -\frac{6272}{81}  &  \frac{6596}{27}\\[0.2cm]
    0  & 0           & -\frac{256}{9} &  \frac{56}{9}   & \frac{40}{9} & -\frac{2}{3} &  \frac{4624}{243} &  \frac{4772}{81}\\[0.2cm]
    0  & 0           &  0             &  0              & 0            &  0           &  \frac{32}{3}     &  0\\[0.2cm]
    0  & 0           &  0             &  0              & 0            &  0           & -\frac{32}{9}     &  \frac{28}{3}\end{array}\right)
\end{equation}
These two matrices are related via the leading order expansion of equation (\ref{ADMtransform})
\begin{equation}
  \gamma^{(0)}_\mathrm{CMM} = W \gamma^{(0)} W^{-1}
\end{equation}
with $W$ given in (\ref{defW}). At next-to-leading order the anomalous dimension matrices not only look complicated but are very difficult to calculate. The complete $8\times 8$ ADM requires performing three-loop renormalization of the effective theory. This calculation was done in the CMM basis with the result
\begin{equation}\label{adm1CMM}
  \gamma^{(1)}_\mathrm{CMM}=\left(
    \begin{array}{cccccccc}
      -\frac{355}{9} & -\frac{502}{27} &  -\frac{1412}{243} &  -\frac{1369}{243} &    \frac{134}{243} &   -\frac{35}{162} &     -\frac{818}{243} &     \frac{3779}{324} \\[0.2cm]
      -\frac{35}{3}  &   -\frac{28}{3} &    -\frac{416}{81} &    \frac{1280}{81} &      \frac{56}{81} &     \frac{35}{27} &       \frac{508}{81} &     \frac{1841}{108} \\[0.2cm]
      0              &        0        &   -\frac{4468}{81} &  -\frac{31469}{81} &     \frac{400}{81} &  \frac{3373}{108} &    \frac{22348}{243} &     \frac{10178}{81} \\[0.2cm] 
      0              &        0        &  -\frac{8158}{243} & -\frac{59399}{243} &    \frac{269}{486} & \frac{12899}{648} &   -\frac{17584}{243} &  -\frac{172471}{648} \\[0.2cm]
      0              &        0        & -\frac{251680}{81} & -\frac{128648}{81} &   \frac{23836}{81} &   \frac{6106}{27} &  \frac{1183696}{729} &  \frac{2901296}{243} \\[0.2cm]
      0              &        0        &  \frac{58640}{243} & -\frac{26348}{243} & -\frac{14324}{243} & -\frac{2551}{162} & \frac{2480344}{2187} & -\frac{3296257}{729} \\[0.2cm]
      0              &        0        &         0          &         0          &         0          &        0          &      \frac{4688}{27} &          0           \\ [0.2cm]
      0              &        0        &         0          &         0          &         0          &        0          &     -\frac{2192}{81} &     \frac{4063}{27}\end{array} \right)
\end{equation}
For the standard operator basis the $6\times 6$ submatrix governing the mixing of current-current and QCD-penguin operators among each other is known to be
\begin{equation}\label{adm1}
  \gamma^{(1)}=\left(\begin{array}{cccccc}
      -\frac{21}{2}-\frac{2f}{9} & \frac{7}{2}+\frac{2f}{3}   & -\frac{202}{243}                & \frac{1354}{81}               & -\frac{1192}{243}                & \frac{904}{81}                \\[0.2cm]
      \frac{7}{2}+\frac{2f}{3}   & -\frac{21}{2}-\frac{2f}{9} & \frac{79}{9}                    & -\frac{7}{3}                  & -\frac{65}{9}                    & -\frac{7}{3}                  \\[0.2cm]
      0                          & 0                          & -\frac{5911}{486}+\frac{71f}{9} & \frac{5983}{162}+\frac{f}{3}  & -\frac{2384}{243}-\frac{71f}{9}  & \frac{1808}{81}-\frac{f}{3}   \\[0.2cm]
      0                          & 0                          & \frac{379}{18}+\frac{56f}{243}  & -\frac{91}{6}+\frac{808f}{81} & -\frac{130}{9}-\frac{502f}{243}  & -\frac{14}{3}+\frac{646f}{81} \\[0.2cm]
      0                          & 0                          & -\frac{61f}{9}                  & -\frac{11f}{3}                & \frac{71}{3}+\frac{61f}{9}       & -99+\frac{11f}{3}             \\[0.2cm]
      0                          & 0                          & -\frac{682f}{243}               & \frac{106f}{81}               & -\frac{225}{2}+\frac{1676f}{243} & -\frac{1343}{6}+\frac{1348f}{81}
  \end{array}\right)
\end{equation}
More on the transformation relations between the two operator bases can be found in the subsequent Appendix~\ref{app:opbase}.

As explained in section \ref{sec:ET}, the initial conditions for the Wilson coefficients are found by matching the effective theory amplitudes with the full Standard Model ones. At $\mu_0=M_W$ the leading-order results are
\begin{equation}\label{wcinilo}
  C_i^{(0)}(M_W)=\left\{\begin{array}{cl}
    1 & \mbox{for } i=1\\
    0 & \mbox{for } i=2\ldots 6\\
    \displaystyle \frac{3x_t^3-2x_t^2}{4(x_t-1)^4}\ln x_t+\frac{-8x_t^3-5x_t^2+7x_t}{24(x_t-1)^3} & \mbox{for } i=7\\
    \displaystyle \frac{-3x_t^2}{4(x_t-1)^4}\ln x_t+\frac{-x_t^3+5x_t^2+2x_t}{8(x_t-1)^3}       & \mbox{for } i=8\end{array}\right.
\end{equation}
where
\begin{equation}
  x_i = \frac{m_i^2}{M_W^2}
\end{equation}
Up and charm quark mass are much smaller than the top quark mass and were therefore neglected in (\ref{wcinilo}). The only difference for the initial conditions in the CMM basis is that the indices 1 and 2 are interchanged. Equating out (\ref{cevolve}) we get for our standard operator basis
\begin{eqnarray}\label{cjLO}
  C^{(0)}_j(\mu) &=& \sum_{i=1}^8 k_{ji}\eta^{a_i} \qquad \mbox{for } j=1\ldots 6\\ \label{c7LO}
  C^{(0)}_7(\mu) &=& \eta^\frac{16}{23} C^{(0)}_7(M_W) +\frac{8}{3}\left(\eta^\frac{14}{23} -\eta^\frac{16}{23}\right) C^{(0)}_8(M_W) +\sum_{i=1}^8 k_{7i}\eta^{a_i}\\ \label{c8LO}
  C^{(0)}_8(\mu) &=& C^{(0)}(M_W)\eta^\frac{14}{23} +\sum_{i=1}^8 k_{8i}\eta^{a_i}
\end{eqnarray}
and
\begin{eqnarray}\label{zjLO}
  Z^{(0)}_j(\mu) &=& \sum_{i=1}^8 h_{ji}\eta^{a_i} \qquad \mbox{for } j=1\ldots 6\\ \label{z7LO}
  Z^{(0)}_7(\mu) &=& C^{(0)}_7(\mu)\\ \label{z8LO}
  Z^{(0)}_8(\mu) &=& C^{(0)}_8(\mu)
\end{eqnarray}
for the CMM basis.

At next-to-leading order we only need the Wilson coefficient $C_7$. As explained in Appendix \ref{app:opbase} we have $C_7^{(1)}(\mu)=Z_7^{(1)}(\mu)$ and can directly take the formula from \cite{CMM}
\begin{eqnarray}\label{c7NLO}
  C_7^{(1)}(\mu)&=&\eta^\frac{39}{23}C_7^{(1)}(M_W)+\frac{8}{3}\left(\eta^\frac{37}{23}-\eta^\frac{39}{23}\right)C_8^{(1)}(M_W)\nonumber\\
	&& +\left(\frac{297664}{14283}\eta^\frac{16}{23}-\frac{7164416}{357075}\eta^\frac{14}{23}+\frac{256868}{14283}\eta^\frac{37}{23}-\frac{6698884}{357075}\eta^\frac{39}{23}\right)C_8^{(0)}(M_W)\nonumber\\
	&& +\frac{37208}{4761}\left(\eta^\frac{39}{23}-\eta^\frac{16}{23}\right)C_7^{(0)}(M_W)+\sum_{i=1}^8(e_i \eta E(x)+f_i+g_i \eta)\eta^{a_i}
\end{eqnarray}
The initial conditions for the NLO Wilson coefficients $C_7$ and $C_8$ are
\begin{eqnarray}
\label{c7ininlo}
C_7^{(1)}(M_W) &=& \frac{-16x^4-122x^3+80x^2-8x}{9(x-1)^4}L_2\!\left(1-\frac{1}{x}\right)+\frac{6x^4+46x^3-28x^2}{3(x-1)^5}\ln^2\! x\nonumber\\
	&& +\frac{-102x^5-588x^4-2262x^3+3244x^2-1364x+208}{81(x-1)^5}\ln x\nonumber\\
	&& +\frac{1646x^4+12205x^3-10740x^2+2509x-436}{486(x-1)^4}\\
\label{c8ininlo}
C_8^{(1)}(M_W) &=& \frac{-4x^4+40x^3+41x^2+x}{6(x-1)^4}L_2\!\left(1-\frac{1}{x}\right)+\frac{-17x^3-31x^2}{2(x-1)^5}\ln^2\! x\nonumber\\
	&& +\frac{-210x^5+1086x^4+4893x^3+2857x^2-1994x+280}{216(x-1)^5}\ln x\nonumber\\
	&& +\frac{737x^4-14102x^3-28209x^2+610x-508}{1296(x-1)^4}.
\end{eqnarray}
All the ``magic numbers'' for the Wilson coefficients necessary in a NLL analysis of $b\to s\gamma$ processes are summarized in table~\ref{tab:magic}.
%%%%%%%%%%%%%%%%%%%%%%%%%%%%%%%%%%%%%%%%%%%%%%%%%%%%%%%%%%%%%%%%%%%
%%%%%%%%            Table of magic numbers                 %%%%%%%%
%%%%%%%%%%%%%%%%%%%%%%%%%%%%%%%%%%%%%%%%%%%%%%%%%%%%%%%%%%%%%%%%%%%
\begin{table}[htbp]
\renewcommand{\arraystretch}{1.2}
\begin{center}
  \begin{tabular}{||c||r|r|r|r|r|r|r|r||}
    \hline\hline
    $i$      & 1                        & 2                      & 3               & 4                & 5         & 6         & 7         & 8\\ \hline\hline
    $a_i$    & $\frac{14}{23}$          & $\frac{16}{23}$        & $\frac{6}{23}$  & $-\frac{12}{23}$ & 0.4086    & $-$0.4230 & $-$0.8994 & 0.1456\\[0.1cm] \hline
    $k_{1i}$ & 0                        & 0                      & $\frac{1}{2}$   & $\frac{1}{2}$    & 0         & 0         & 0         & 0\\
    $k_{2i}$ & 0                        & 0                      & $\frac{1}{2}$   & $-\frac{1}{2}$   & 0         & 0         & 0         & 0\\
    $k_{3i}$ & 0                        & 0                      & $-\frac{1}{14}$ & $\frac{1}{6}$    & 0.0510    & $-$0.1403 & $-$0.0113 & 0.0054\\
    $k_{4i}$ & 0                        & 0                      & $-\frac{1}{14}$ & $-\frac{1}{6}$   & 0.0984    & 0.1214    & 0.0156    & 0.0026\\
    $k_{5i}$ & 0                        & 0                      & 0               & 0                & $-$0.0397 & 0.0117    & $-$0.0025 & 0.0304\\
    $k_{6i}$ & 0                        & 0                      & 0               & 0                & 0.0335    & 0.0239    & $-$0.0462 & $-$0.0112\\
    $k_{7i}$ & $\frac{626126}{272277}$  & $-\frac{56281}{51730}$ & $-\frac{3}{7}$  & $-\frac{1}{14}$  & $-$0.6494 & $-$0.0380 & $-$0.0186 & $-$0.0057\\
    $k_{8i}$ & $\frac{313063}{363036}$  & 0                      & 0               & 0                & $-$0.9135 & 0.0873    & $-$0.0571 & 0.0209\\[0.1cm] \hline
    $h_{1i}$ & 0                        & 0                      & 1               & $-$1               & 0         & 0         & 0         & 0\\
    $h_{2i}$ & 0                        & 0                      & $\frac{2}{3}$   & $\frac{1}{3}$    & 0         & 0         & 0         & 0\\
    $h_{3i}$ & 0                        & 0                      & $\frac{2}{63}$  & $-\frac{1}{27}$  & $-$0.0659 & 0.0595    & $-$0.0218 & 0.0335\\
    $h_{4i}$ & 0                        & 0                      & $\frac{1}{21}$  & $\frac{1}{9}$    & 0.0237    & $-$0.0173 & $-$0.1336 & $-$0.0316\\
    $h_{5i}$ & 0                        & 0                      & $-\frac{1}{126}$& $\frac{1}{108}$  & 0.0094    & $-$0.100  & 0.0010    & $-$0.0017\\
    $h_{6i}$ & 0                        & 0                      & $-\frac{1}{84}$ & $-\frac{1}{36}$  & 0.0108    & 0.0163    & 0.0103    & 0.0023\\[0.1cm] \hline
    $e_i$    & $\frac{4661194}{816831}$ & $-\frac{8516}{2217}$   & 0               & 0                & $-$1.9043 & $-$0.1008 & 0.1216,   & 0.0183\\
    $f_i$    & -17.3023               & 8.5027                 & 4.5508          & 0.7519           & 2.0040    & 0.7476    & $-$0.5358 & 0.0914\\
    $g_i$    & 14.8088                  & $-$10.809             & $-$0.8740       & 0.4218           & $-$2.9347 & 0.3971    & 0.1600    & 0.0225\\ \hline\hline
  \end{tabular}
  \caption{The ``magic numbers'' for the Wilson coefficients in $b\to s\gamma$. The $a_i$ are the eigenvalues of $\gamma^{(0)}$ divided by $2\beta_0$.
\label{tab:magic}}
\end{center}
\end{table}
%%%%%%%%%%%%%%%%%%%%%%%%%%%%%%%%%%%%%%%%%%%%%%%%%%%%%%%%%%%%%%%%%%%

\chapter{Operator Basis Transformations}
\label{app:opbase}

Here we address the transformation relations of the operator bases (\ref{q1def}--\ref{q8def}) and (\ref{p1def}--\ref{p8def}).
In general, if we change the operator basis from $\vec Q$ to $\vec P=W \vec Q$, to translate the scheme-dependent NLO quantities like the NLO Wilson coefficients or the NLO anomalous dimension matrix, we need the finite parts of the one-gluon exchange corrections to the operator matrix elements where the scheme dependence is manifest. If the one-loop matrix elements in the starting basis are
\begin{equation}\label{qdef}
  \langle \vec Q\rangle = \left( 1+\frac{\alpha_s}{4\pi}r\right)\langle \vec Q\rangle^{(0)}
\end{equation}
then the ones in the transformed basis can be written as
\begin{eqnarray}\label{pdef}
  \langle \vec P\rangle &=& \left( 1+\frac{\alpha_s}{4\pi}r'\right)\langle \vec P\rangle^{(0)}\\ \label{ptrans}
  &=& W\left( 1+\frac{\alpha_s}{4\pi}\Delta r\right)\langle \vec Q\rangle +{\cal O}(\alpha_s^2)
\end{eqnarray}
with
\begin{equation}\label{rdef}
  \Delta r=W^{-1} r' W -r
\end{equation}
Denoting the Wilson coefficients in the transformed basis with $\vec Z$, we can derive the transformation rules starting from the fact that the full amplitude has to be the same in both bases
\begin{equation}\label{ccond}
  \vec C^T \langle \vec Q\rangle \stackrel{!}{=} \vec Z^T\langle \vec P\rangle
\end{equation}
which leads to
\begin{equation}\label{ctransform}
  \vec Z = \left(W^{-1}\right)^T\left( 1-\frac{\alpha_s}{4\pi}\Delta r^T\right) \vec C +{\cal O}(\alpha_s^2)
\end{equation}
For the sake of completeness we also give the transformation rule for the anomalous dimension matrix $\gamma'$ in the new basis which can be derived for example via differentiation of the left and right hand side of (\ref{ctransform}) with respect to $\ln\mu$ using the definition of the anomalous dimension matrix in~(\ref{RGEC}).
\begin{equation}\label{ADMtransform}
  \gamma' =W\left\{\gamma +\frac{\alpha_s}{4\pi}[\Delta r,\gamma] +2\beta_0\left(\frac{\alpha_s}{4\pi}\right)^2 \Delta r \right\} W^{-1} +{\cal O}(\alpha_s^3)
\end{equation}
Here $\beta_0$ was defined in (\ref{betadef}) and the square brackets denote the commutator $\Delta r\,\gamma -\gamma\Delta r$.

The operator basis (\ref{q1def}--\ref{q8def}) we use mostly is apart from an overall factor of 4 and the fact that we reversed the numbering of the current-current operators $Q_{1,2}$ the same as the one of Greub, Hurth, and Wyler \cite{GHW}. Chetyrkin, Misiak, and M\"unz \cite{CMM} and also Buras, Czarnecki, Misiak, and Urban \cite{BCMU,BCMU2}, however, use another operator basis (\ref{p1def}--\ref{p8def}), which is better suited for multi-loop calculations. The transformation matrix $W$ in four space-time dimensions is given by
\begin{equation}\label{defW}
  W=\frac{1}{4}\left(\begin{array}{rrrrrrrr}
      1           & 0            & 0 & 0 & 0 & 0 & 0 & 0\\
      -\frac{1}{6} & \frac{1}{2} & 0 & 0 & 0 & 0 & 0 & 0\\
      0           & 0            & 1 & 0 & 1 & 0 & 0 & 0\\
      0           & 0            & -\frac{1}{6} & \frac{1}{2} & -\frac{1}{6} & \frac{1}{2} & 0 & 0\\
      0           & 0            & 16 & 0 & 4 & 0 & 0 & 0\\
      0           & 0            & -\frac{8}{3} & 8 & -\frac{2}{3} & 2 & 0 & 0\\
      0           & 0            & 0 & 0 & 0 & 0 & 1 & 0\\
      0           & 0            & 0 & 0 & 0 & 0 & 0 & 1\end{array}\right)
\end{equation}
and transforms our operator basis (\ref{q1def}--\ref{q8def}) into the CMM basis (\ref{p1def}--\ref{p8def})
\begin{equation}\label{PrelQ}
  P_i \stackrel{D=4}{=} \sum_{k=1}^8 W_{ik} Q_k
\end{equation}
Note the block diagonal structure of $W$. The most important consequence hereof is that $W^{-1}$ as well has block diagonal form with the inverse of the submatrices as entries.

To determine $\Delta r$ we have to insert all the operators into current-current and penguin diagrams with one additional gluon. For the $6\times 6$ submatrix of $\Delta r$ concerning the current-current and QCD penguin operators we calculated the diagrams (b), (c), (d), and (e) in Fig.~\ref{fig:opdiag} with insertions of $Q_{1\ldots 6}$ and $P_{1\ldots 6}$. We get for $N=3$
\begin{equation}\label{defDeltar}
  \Delta r^{6\times 6}=\left(\begin{array}{cccccc}
     \frac{128}{27} & \frac{52}{9} & -\frac{7}{81} & \frac{7}{27} & -\frac{7}{81} & \frac{7}{27}\\[0.2cm]
     -\frac{2311}{81} & -\frac{173}{27} & \frac{35}{243} & -\frac{35}{81} & \frac{35}{243} & -\frac{35}{81}\\[0.2cm]
     0 & 0 & \frac{178}{27} & -\frac{34}{9} & -\frac{164}{27} & \frac{20}{9}\\[0.2cm]
     0 & 0 & 1-\frac{f}{9} & \frac{-25 + f}{3} & \frac{-18 - f}{9} & 6 + \frac{f}{3}\\[0.2cm]
     0 & 0 & -\frac{160}{27} & \frac{16}{9} & \frac{146}{27} & -\frac{2}{9}\\[0.2cm]
     0 & 0 & -2+\frac{f}{9} & 6-\frac{f}{3} & 3+\frac{f}{9} & \frac{-11-f}{3}\end{array}\right)
\end{equation}
which indeed transforms the NLO anomalous dimension matrix for our operator basis (\ref{adm1}) into the $6\times 6$ submatrix of (\ref{adm1CMM}). Without calculation we already can say that the first six entries of the seventh and eighth row of $\Delta r$ vanish because of the block diagonal structure of $W$ and $W^{-1}$ and the fact that the magnetic penguin operators do not mix into the four-quark operators. Yet, determining the entries in the seventh and eighth columns of $r$ and $r'$ would amount to calculating the finite parts of the ${\cal O}(\alpha_s)$ virtual corrections to the matrix elements for $b\to s\gamma$. This challenging two-loop calculation is what Greub, Hurth, and Wyler \cite{GHW} did for $Q_{1,2}$ and $Q_{7,8}$ of our operator basis. Recently Buras, Czarnecki, Misiak, and Urban performed this calculation for the complete CMM basis (\ref{p1def}--\ref{p8def}) \cite{BCMU,BCMU2}. But the NLO matrix elements of the penguin operators in our operator basis have not yet been calculated so that we cannot determine the complete $\Delta r$. Therefore we lack the information to translate the results of \cite{BCMU,BCMU2} into our operator basis. The use of only one operator basis would perhaps be interesting from an esthetical point of view, but it is not necessary. We use the two different operator bases for physically distinct processes in $B\to V\gamma$, namely hard vertex and hard spectator corrections which is perfectly alright if the corresponding Wilson coefficients are used. For the Wilson coefficient of the electromagnetic penguin operator we need the next-to-leading value, which was calculated in~\cite{CMM} in the CMM basis.

To get $C_7$ and $C_8$ no transformation is necessary because the electro- and chromomagnetic penguin operators are the same in both bases. Furthermore $Q_7$ and $Q_8$ cannot be expressed as linear combinations of $Q_{1\ldots 6}$ or $P_{1\ldots 6}$ and they also do not mix into these operators. Using eq.~(\ref{ccond}) we therefore can identify the coefficients of $\langle Q_{7,8}\rangle$ and $\langle P_{7,8}\rangle$ to be the same. So we can directly use the NLO result for $Z_7$ from \cite{CMM} for our operator basis.

Chetyrkin, Misiak, and M\"unz use a different strategy to derive the transformation properties between the $6\times 6$ submatrices of the NLO anomalous dimension matrices \cite{CMM2}. They pass from one operator basis to the other by a series of subsequent redefinitions of physical and evanescent operators.

\chapter{Weak Annihilation in $B\to V\gamma$}
\label{app:WAproof}

In this appendix we explicitely prove to ${\cal O}(\alpha_s)$ that weak annihilation contributions to $B\to V\gamma$ decays are calculable within QCD factorization. To this end we have to show that all collinear and soft divergences cancel to leading power when one-gluon exchange to Fig.~\ref{fig:ann} is considered. By leading power we here mean the lowest nonvanishing order in the power expansion, where the entire annihilation contributions start only at subleading power (cf. section~\ref{sec:factBVgam}). Our proof here is analogous to the one given in subsection~\ref{subsec:1loopproof}. We again expect the soft divergences to cancel because of the colour transparency argument and the collinear ones via collinear Ward identities.\medskip

\begin{figure}
   \begin{center}\psfig{figure=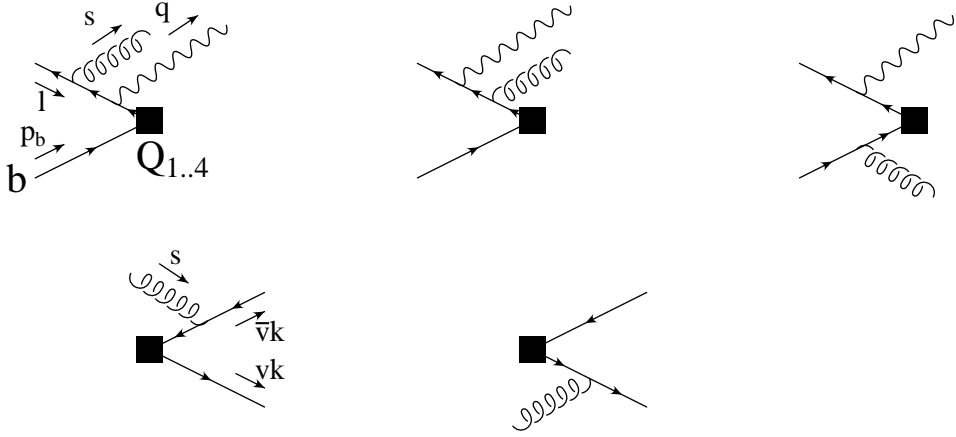}\end{center}
\caption{One-gluon exchange corrections to the annihilation contribution to $\bar B\to V\gamma$ decay from $Q_{1\ldots 4}$ operator insertions. The six possible diagrams are obtained through combining the three diagrams in the first row with each of the two in the second row.\label{fig:annNLO}}
\end{figure}

Let us first show these cancellations for the weak annihilation involving operators $Q_{1\ldots 4}$ and the case where the photon is emitted from the light quark in the $B$ meson (see Fig.~\ref{fig:annNLO}).
%%%%%%%%%%%%%%%%%%%%%%%%%%%%%%%%%%%%%%%%%%%%%%%%%%%%%%%%%%%%%%%%%%%
%%%%%%%%%%%%%%%%%%%%%%%%%%%%%%%%%%%%%%%%%%%%%%%%%%%%%%%%%%%%%%%%%%%
For the soft region all components of the gluon momentum $s$ become small simultaneously: $s\sim \lambda$. Power counting shows that every single of the six possible diagrams from Fig.~\ref{fig:annNLO} is logarithmically IR divergent. The sum of the two contributions in the second row of Fig.~\ref{fig:annNLO}, however, vanishes when the equations of motion for the vector meson constituent quarks $q$ and $q'$ are used:
\begin{eqnarray}
  \lefteqn{\bar q'\left[\frac{\gamma^\alpha (v\! \not\! k -\not\! s)\Gamma^\mu}{s^2-2v\,s\cdot k} +\frac{\Gamma^\mu (-\bar v\! \not\! k +\not\! s)\gamma^\alpha}{s^2-2\bar v\,s\cdot k}\right]q}\nonumber\\
  && =\bar q'\left[\frac{2v\, k^\alpha \Gamma^\mu}{-2v\,s\cdot k}-\frac{2\bar v\, k^\alpha \Gamma^\mu}{-2\bar v\,s\cdot k}\right]q +{\cal O}(\lambda)\nonumber\\
  && ={\cal O}(\lambda)
\end{eqnarray}
where we abbreviated $\Gamma^\mu=\gamma^\mu(1-\gamma_5)$. Hence, the soft divergences cancel to leading power in the total amplitude.

For $s$ collinear with the light-cone momentum $k$ of the vector meson, i.e. $s=\alpha k+\ldots$, we have the scaling
\begin{equation}
  s^+\sim\lambda^0,  \qquad  s_\perp\sim\lambda,  \qquad s^-\sim\lambda^2
\end{equation}
so that
\begin{equation}
  d^4 s \sim\lambda^4,  \qquad  s\cdot k \sim \lambda^2,  \qquad  s^2 \sim\lambda^2
\end{equation}
Using these scaling laws we find that the divergence of the single diagrams in the collinear region is also logarithmic. We examine again the sum of the two contributions in the second row of Fig.~\ref{fig:annNLO}:
\begin{eqnarray}
  &&\bar q'\left[\frac{\gamma^\alpha (v-\alpha) \not\! k \Gamma^\mu}{s^2 -2v\, s\cdot k} -\frac{\Gamma^\mu (\bar v-\alpha) \not\! k\gamma^\alpha}{s^2-2\bar v\, s\cdot k}\right]q \nonumber\\
  &=& \bar q'\left[\frac{2(v-\alpha) k^\alpha \Gamma^\mu}{s^2 -2v\, s\cdot k} -\frac{2(\bar v-\alpha)k^\alpha \Gamma^\mu}{s^2-2\bar v\, s\cdot k}\right]q\nonumber\\
  &\propto& k^\alpha
\end{eqnarray}
The $k^\alpha$ is contracted with the $\gamma_\alpha$ from the gluon vertices in the first row of Fig.~\ref{fig:annNLO}
\begin{eqnarray}
  k^\alpha\! \left[\frac{\gamma_\alpha\! (\not\! l-\alpha\! \not\! k)\not\! \epsilon (\not\! l-\alpha\! \not\! k-\not\! q)\Gamma_\mu}{(s^2-2 l\cdot s)(s^2-2 l\cdot s +2 s\cdot q -2 l\cdot q)} 
  -\frac{\not\! \epsilon (\not\! l-\not\! q)\gamma_\alpha\! (\not\! l -\not\! q -\alpha\! \not\! k)\Gamma_\mu}{2 l\cdot q(s^2 -2 l\cdot s +2 s\cdot q -2 l\cdot q)} \right.\nonumber\\
  \left. +\frac{\not\! \epsilon (\not\! l -\not\! q)\Gamma_\mu (\not\! p_b -\alpha\! \not\! k +m_b)\gamma_\alpha\!}{2 l\cdot q(s^2 -2 p_b \cdot s)} \right]
\end{eqnarray}
The above expression vanishes when the equations of motion for the $B$ meson quarks are used. So there are no collinear divergences at leading power either.\medskip

\begin{figure}
   \begin{center}\psfig{figure=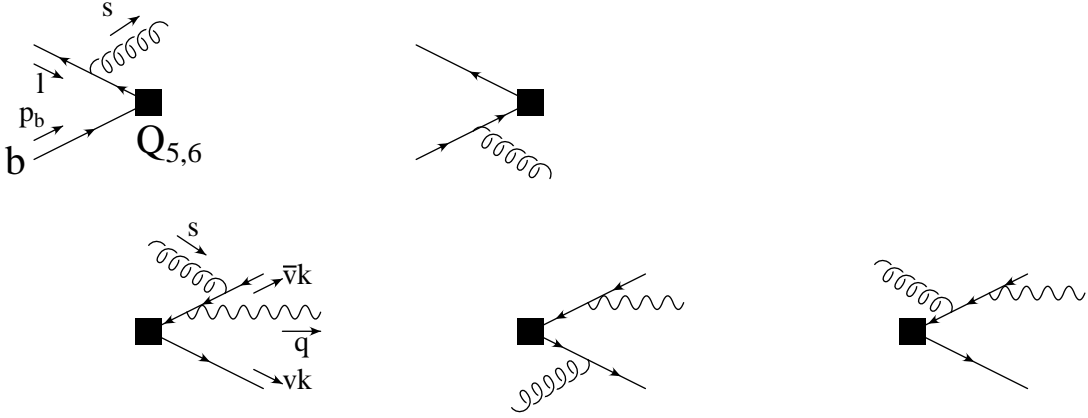}\end{center}
\caption{One-gluon exchange corrections to the annihilation contribution to $\bar B\to V\gamma$ decay from $Q_{5,6}$ operator insertions. The six possible diagrams are obtained through combining both diagrams in the first row with each of the three in the second row. Combinations with the third diagram in the second row are infrared finite. Photon emission from the other vector meson quark is not shown.\label{fig:annNLOQ56}}
\end{figure}

The other annihilation contribution which appears at the same level of power counting comes from the insertion of operators $Q_{5,6}$ and photon emission from one of the final state vector meson quarks. The one-gluon-exchange diagrams for the case where the photon is emitted from the antiquark in the vector meson are depicted in Fig.~\ref{fig:annNLOQ56}.
%%%%%%%%%%%%%%%%%%%%%%%%%%%%%%%%%%%%%%%%%%%%%%%%%%%%%%%%%%%%%%%%%%%
%%%%%%%%%%%%%%%%%%%%%%%%%%%%%%%%%%%%%%%%%%%%%%%%%%%%%%%%%%%%%%%%%%%
Both constituent quarks in the vector meson are light, collinear, and carry a momentum of the same size. Photon emission from the other vector meson quark therefore can be treated completely analogously.

A power counting as above shows that the two diagrams where the gluon hits the off-shell light antiquark in the vector meson (third diagram in second row of Fig.~\ref{fig:annNLOQ56} combined with the two diagrams in the first row) are infrared finite both in the soft and collinear region. All other combinations are logarithmically divergent.

For the soft gluon the sum of the first two contributions in the second row of Fig.~\ref{fig:annNLOQ56} again cancels to leading-power when the equations of motion for the vector meson quarks $q$ and $q'$  are used:
\begin{eqnarray}
  &&\bar q' \left[ \frac{\Gamma(-\bar v\! \not\! k +\not\! s-\not\! q) \not\! \epsilon (-\bar v\! \not\! k +\not\! s)\gamma_\alpha}{(s^2 -2\bar v s\cdot k+2\bar v k\cdot q -2s\cdot q)(s^2 -2\bar v s\cdot k)} +\frac{\gamma_\alpha (v\not\! k-\not\! s) \Gamma (-\bar v \not\! k -\not\! q) \not\! \epsilon}{(s^2 -2v s\cdot k)2\bar v k\cdot q}\right]q\nonumber\\
  && =\bar q'\left[ \frac{2\bar v k_\alpha \Gamma (-\bar v \not\! k -\not\! q)\not\! \epsilon}{4\bar v^2 k\cdot q s\cdot k} -\frac{2v k_\alpha \Gamma (-\bar v \not\! k -\not\! q)\not\! \epsilon}{4\bar v v k\cdot q s\cdot k}\right]q +{\cal O}(\lambda)\nonumber\\
  && = {\cal O}(\lambda)
\end{eqnarray}
Here $\Gamma$ represents the Dirac structure of the quark currents of $Q_{5,6}$.

Similarly, in the collinear region the sum of the same two diagrams is again proportional to $k_\alpha$, which is contracted with the $\gamma_\alpha$ from the gluon vertices in the first row of Fig.~\ref{fig:annNLOQ56}. For on-shell $B$ meson quarks this once more vanishes.\medskip

This completes our proof. For all annihilation contributions of $Q_{1\ldots 6}$, to ${\cal O}(\alpha_s)$, there are no collinear or soft infrared divergences at leading non-vanishing power in $1/m_b$. Annihilation contributions therefore are, although power suppressed, calculable in QCD factorization.
\end{appendix}
\cleardoublepage

% ===== Bibliography ======================================
\renewcommand{\chaptermark}[1]{\markboth{#1}{}}
\lhead[\fancyplain{}{\bf\thepage}]{\fancyplain{}{\em\rightmark}}
\rhead[\fancyplain{}{\em\rightmark}]{\fancyplain{}{\bf\thepage}}

%\cleardoublepage

% ===== Acknowledgements ==================================
\addcontentsline{toc}{chapter}{Acknowledgements}
\chapter*{Acknowledgements}

\noindent I want to thank everybody who helped this thesis coming into being.\medskip

\noindent In the first place I thank Gerhard Buchalla for his exclusive support. He suggested and very pedagogically introduced me to the subject of this work. At any time he was ready  to discuss my questions and ideas. I could learn a lot from him. I also want to thank him for the assistance in arranging my stay at the CERN Theory Division where our collaboration started.\medskip

\noindent I thank Andrzej Buras for supervising my thesis. I highly value his fatherly advice and the discussions with him. With all my ventures I always could rely on his quick and favourable support. I am grateful to Gerhard and Andrzej also for proofreading the manuscript.\medskip

\noindent I very much appreciate the kind hospitality of the CERN Theory Division during the ten months I stayed there and the visits afterwards. To all friends and colleagues at CERN and the Max-Planck-Institut f\"ur Physik I am grateful for stimulating discussions and creating a very pleasant social environment. In particular the lunch time conversations with Edda, Emmanuel, both Haralds, Johannes, Karsten, Martin, the two Peters, Roman, Werner, and the rest of the ``Austrian gang'' at CERN and Amol, Andr\'e, Ariel, Bj\"orn, Dima, Gennaro, Georg, Mathias, Michael, Pavel, Pushan, Ralf, Ricard, Robert, Sergio, and Thomas at the MPI will stick lively in my mind. Thanks to the members of Andrzej Buras's chair and here especially my co-postgraduates Martin, Sebastian, and Uli for the convivial receptions in Garching and helpful discussions.\medskip

\noindent A great thankyou to my parents and my whole family for all their unconditional support of any kind, the interest in my work, and for keeping me grounded. Last, but definitely not least, it is a great pleasure to express my warmest thanks to Caroline.\medskip

\vfill

\noindent I gratefully acknowledge financial support of the Max-Planck-Gesellschaft and the Stu\-dien\-stif\-tung des deutschen Volkes during my time in Munich and Geneva, respectively.\medskip

\end{document}